\documentclass[twocolumn]{aa} % for a paper on 1 column  
\usepackage{graphicx}
\usepackage{txfonts}
\usepackage{xcolor}
 
\usepackage{enumitem}
\usepackage{ulem}

\newcommand{\SC}[1]{\textcolor{black}{#1}} %Sylvie comments old: magenta
\newcommand{\SCC}[1]{\textcolor{black}{#1}} %new Sylvie changes after co-authors comments
\newcommand{\GP}[1]{\textcolor{black}{#1}} %Guillaume comments old: magenta
\newcommand{\rev}[1]{\textcolor{black}{#1}} % Sylvie revisions after refereeing
\newcommand{\MV}[1]{\textcolor{black}{#1}} %moved by Sylvie old: blue
\newcommand{\BT}[1]{\textcolor{black}{#1}} %Benoit comments old red
\newcommand{\REVbis}[1]{\textcolor{black}{#1}} 

\newcommand{\gout}[1]{} %removed
\newcommand{\kms}{km~s$^{-1}$}
\newcommand{\lbp}{$\lambda_{\rm BP}$}
\newcommand{\WW}{$\mathcal{W}$}
\newcommand{\Vk}{$V_{K}$}
\newcommand{\jobs}{$j_{\rm obs}$}
\newcommand{\Vobs}{$V_{\rm p,obs}$}

\newcommand{\robs}{$r_{\rm obs}$}
\newcommand{\lobs}{$\lambda_{\rm obs}$}
\newcommand{\jout}{$j_{\rm out}$}
\newcommand{\rout}{$r_{\rm out}$}
\newcommand{\rmax}{$r^{\infty}$}
\newcommand{\rin}{$r_{\rm in}$}
\newcommand{\zcut}{$z_{\rm cut}$}
\newcommand{\icrit}{$i_{\rm crit}$}

\newcommand{\Min}{$\dot{M}_{\rm in}$}
\newcommand{\Mout}{$\dot{M}_{\rm out}$}
\newcommand{\Mdw}{$\dot{M}_{\rm DW}$}
\newcommand{\rw}{$r_{\rm w}$}
\newcommand{\rcorr}{$r_{\rm corr}$}

\begin{document}

   \title{Constraining MHD disk winds with ALMA}
   \subtitle{Apparent rotation signatures and application to HH212}

   \author{B. Tabone 
          \inst{1,2},
          S. Cabrit \inst{2,3}, G. Pineau des For\^ ets \inst{2,4},
          J. Ferreira \inst{3}, A. Gusdorf \inst{5}, L. Podio \inst{6}, 
          E. Bianchi\inst{3}, E. Chapillon \inst{7,8}, C. Codella \inst{6,3}, F. Gueth \inst{7} 
          }
%
%   \institute{LERMA, Observatoire de Paris, Observatoire de Paris, Ecole normale sup\'erieure, PSL Research University, CNRS, Sorbonne Universit\'es, UPMC Univ. Paris 06, 75231 Paris, France \
%              \email{benoit.tabone@obspm.fr}
%         \and
%             IPAG, Universite Grenoble
%         \and
%             IAS, Orsay
%         \and
%             IRAM, Saint Martin
%         \and
%             INAF- Arceti\\
%             }
%
   \institute{Leiden Observatory, Leiden University, PO Box 9513, 2300 RA Leiden, The Netherlands \\ e-mail: {\tt tabone@strw.leidenuniv.nl}
   		\and 
	    PSL Research University, Sorbonne Universit\'es, Observatoire de Paris, LERMA, CNRS, Paris France
                       \and
             Univ. Grenoble Alpes, CNRS, IPAG, 38000 Grenoble, France 
                          \and
             Université Paris-Saclay, CNRS, Institut d’Astrophysique Spatiale, 91405, Orsay, France 
                       \and
             Laboratoire de Physique de l'Ecole Normale Sup\'erieure, ENS, PSL Research University, Sorbonne Universit\'es, CNRS, Paris, France           
                       \and
             INAF, Osservatorio Astrofisico di Arceti, Largo E. Fermi 5, 50125 Firenze, Italy
         \and
             Institut de Radioastronomie Millim\'etrique, 38406 Saint-Martin-d'H\`eres, France 
             \and
             OASU/LAB-UMR5804, CNRS, Universit\'e Bordeaux, 33615 Pessac, France}

   \date{\today}

\abstract
%%Context
{The high spectral resolution and sensitivity provided by large millimeter interferometers (ALMA, NOEMA, SMA) 
is revealing a growing number of rotating outflows, suggested to trace magneto-centrifugal disk winds (MHD DWs).
\GP{However, the angular momentum flux that they extract and its impact on disk accretion are not yet well quantified.}} 
%%Aims
{We wish to identify systematic biases in retrieving the \rev{true launch zone},  magnetic lever arm, 
\GP{and associated angular momentum flux} of an MHD DW from apparent \rev{rotation signatures,} as measured 
by observers from \rev{Position-Velocity} (PV) diagrams \GP{at ALMA-like resolution}.} 
%%Methods
{Synthetic PV cuts are constructed \BT{from} \rev{self-similar} 
MHD DW solutions over a broad range of parameters. \BT{Three methods are \rev{examined} for  estimating the specific angular momentum \jobs\ from PV cuts: the "double-peak separation" method (relevant for edge-on systems), and the "rotation curve" and "flow width" methods (applicable at any view angle).}
The launch radius and magnetic lever arm derived from \jobs \rev{with the approach} of Anderson et al. 2003 are compared to their true values \rev{on the outermost streamline}. 
\rev{Predictions for} \BT{the "double-peak separation" method} \rev{are tested on} published ALMA observations of the HH212 \rev{rotating SO wind} at resolutions from $\sim 250$~au to $\sim 18$~au.}
%%% results: 
{The "double-peak separation" method and the "flow width" method provide only a lower limit to the true outer launch radius \rout.
This bias is mostly independent of angular resolution, but increases with the wind radial extension and radial emissivity gradient
and can reach a factor 10. \BT{In contrast, the "rotation curve" method gives a good estimate of \rout\  when the flow is well resolved, and an upper limit at low angular resolution. The magnetic lever arm is always underestimated, due to invisible angular momentum stored as magnetic field torsion. ALMA data of HH212 confirm our predicted biases for the "double-peak separation" method, 
and the large \rout $\simeq$ 40 au and small magnetic lever arm first suggested by Tabone et al. 2017 from PV cut modeling.}
We also derive an exact analytical expression for the fraction of disk angular momentum extraction performed by a self-similar MHD disk wind of given radial extent, magnetic lever arm, and mass ejection/accretion ratio. The MHD DW candidate in HH212 extracts enough angular momentum to sustain steady accretion through the whole disk at the current observed rate.}
%%% Conclusions
{The launch radius estimated from observed rotation signatures in an MHD DW can markedly differ from the true outermost launch radius \rout. Similar results would apply in a wider range of flow geometries. While it is in principle possible to bracket \rout~by combining two observational methods with opposite bias, only comparison with synthetic predictions can take into account properly all observational effects, and also constrain the true magnetic lever arm. The present comparison with ALMA observations of HH212 represents the most stringent observational test of MHD DW models to date, and shows that MHD DWs are serious candidates for the angular momentum extraction process in protoplanetary disks.}


\authorrunning{B. Tabone et al.}
\titlerunning{Constraining MHD disk winds with ALMA}

   \maketitle
   
%%%%%%%%%%%%%%%%%%%%%%%%%%%%%%%%%%%
%%%%%%%%%%%%%%%%%%%%%%%%%%%%%%%%%%%
%%%%%%%%%%%%%%%%%%%%%%%%%%%%%%%%%%%
%%%%%%%%%%%%%%%%%%%%%%%%%%%%%%%%%%%

\section{Introduction} 

A major enigma in our understanding of the structure and evolution of protoplanetary disks (PPDs) is the exact mechanism by which angular momentum is extracted to allow disk accretion onto the central object at the observed rates, much larger than expected for microscopic collisional viscosity \citep[eg.][]{2016ARA&A..54..135H}. The problem is particularly acute during the early protostellar phase (so-called Class 0) where the second hydrostatic Larson's core must grow in less than $10^5$ yrs to stellar masses by accretion of disk material. An efficient mechanism, first introduced by \citet{1982MNRAS.199..883B} in the context of active galactic nuclei, is that angular momentum may be removed vertically by the twisting of large-scale poloidal magnetic field lines, and carried away in a magneto-centrifugal disk wind (hereafter MHD DW) that becomes collimated into a jet on large scale. The same process was first proposed to explain bipolar jets and outflows from young stars by \citet{1983ApJ...274..677P}, and their correlation with accretion luminosity by \citet{1989ApJ...342..208K}. The feasibility to feed a steady, super-Alfv\'enic MHD DW from a resistive Keplerian accretion flow was further demonstrated through semi-analytical works and numerical simulations  \citep[see e.g.][and references therein]{1997A&A...319..340F,2007prpl.conf..277P}. 
An alternative \rev{well-studied} mechanism able to transfer angular momentum and drive accretion through PPDs is the magneto-rotational instability \citep[MRI][]{1991ApJ...376..214B}. However, recent non-ideal MHD calculations and simulations reveal that the MRI is quenched in outer regions of PPDs 
around 1-20~au (the so-called "dead zone"), and MHD DWs are being revived as prime candidates to induce disk accretion through these outer regions \rev{\citep[see eg.][and references therein]{2014prpl.conf..411T,2017ApJ...845...75B,2017A&A...600A..75B}}.  
Therefore, robust observational tests of the presence and radial extent of MHD DWs in young stars are crucially needed to fully understand the physics of PPDs, \rev{and of planet migration inside them \citep[see eg.][]{2018A&A...615A..63O}}.

\rev{In this context, the outermost launching radius of the MHD DW (denoted as \rout\ in the following) is a particularly important parameter to determine.} A key observational diagnostic for constraining the range of launch radii is the specific angular momentum carried by the wind \citep{2002ApJ...576..222B,2003ApJ...590L.107A,2006A&A...453..785F}. In particular, \citet{2003ApJ...590L.107A} showed that in a steady, axisymmetric, and dynamically cold (negligible enthalpy) MHD DW, the launch radius $r_0$ of a given wind streamline \rev{is related to its kinematics through what we will refer to hereafter as "Anderson's relation":}
\begin{equation}
r V_{\phi} \Omega_0 = \frac{V^2}{2} + \left(\frac{3}{2}-\frac{r_0}{R} \right) \left( {GM_\star \Omega_0} \right)^{2/3}. %+ (h-h_0)
\label{eq:anderson}
\end{equation}
Here \rev{$r$ denotes the distance from the axis at the observed wind point, $V_{\phi}$ the azimuthal velocity, $V$ the total velocity modulus, $R$ the distance to the central star,} $M_\star$ the stellar mass, and
$\Omega_0 = (GM_\star / r_0^3)^{1/2}$ the Keplerian angular velocity at $r_0$. 
\rev{The term in $(r_0/R)$ accounts for gravitational potential at low altitudes, that are starting to be probed with ALMA.}
This relation further shows that a cold, steady, axisymmetric MHD DW must everywhere rotate in the same sense as the disk (ie. $V_{\phi} \Omega_0 > 0$). 
Note that an MHD DW could still be counter-rotating if it is not \rev{dynamically} cold, ie. \rev{enthalpy}-driven rather than \rev{magnetically} driven\footnote{an extra term $(h-h_0)$ must then be added to the right-hand side of Eq.~\ref{eq:anderson}, where $h$ and $h_0$ are the specific enthalpy at the observation point and at the flow base, respectively. Counter-rotation ($V_{\phi} \Omega_0 < 0$) results if $(h_0 - h) > V^2/2 + 3/2 (GM_\star \Omega_0)^{2/3}$, meaning that the enthalpy gradient dominates the flow kinematics.} \citep[see also][]{2012ApJ...759L...1S}, non-steady \citep[][]{2011ApJ...737...43F}, or non-axisymmetric \rev{\citep{2015MNRAS.446.3975S}}. However, in none of these cases would it be possible to infer $r_0$ from the above relation\footnote{in an enthalpy-driven MHD DW, the extra term $(h_0-h)$ would be too poorly known; in a non-steady or non-axisymmetric MHD DW, the MHD invariants used to derive Eq.~\ref{eq:anderson} would no longer hold.}. 

First tentative jet rotation signatures were uncovered \SCC{in optical forbidden lines} at the base of atomic T Tauri jets \SCC{thanks to the unprecedented angular resolution of} the Hubble Space Telescope (HST), in the form of centroid velocity differences $\simeq 10-20$ \kms\ between opposite edges of the flow. In two cases (DG Tau, CW Tau) the inferred jet rotation sense agrees with the disk rotation sense, as required by  Anderson's formula for a cold, steady, \rev{axisymmetric} MHD DW. The inferred values of launch radii $r_0$ range from 0.2 to 3~au, and the estimated total angular momentum flux represents 60\%-100\% of that required for accretion through the underlying disk, consistent with the MHD DW scenario \citep{2002ApJ...576..222B,2003ApJ...590L.107A,2007ApJ...663..350C}. In a more detailed modeling analysis, \citet{2004A&A...416L...9P} showed that the spatial pattern of velocity shifts along and across the DG Tau jet is in excellent agreement with synthetic predictions for an extended MHD DW launched out to 3~au. However, at flow radii smaller than the PSF diameter, velocity shifts are strongly reduced due to beam convolution. Since most atomic T Tauri jets are \rev{not well resolved across} even with HST, this effect might explain why their rotation remains so challenging to detect
at the limited spectral resolution \SC{($\simeq$ 50 \kms)} of current optical and near-infrared 2D spectro-imagers,
\SCC{or possibly contaminated by external asymmetries in the counter-rotating cases}
\citep[RW~Aur, RY~Tau, Th~28,][]{2006A&A...452..897C,2015ApJ...804....2C, 2016A&A...596A..88L}.

The \rev{unique combination of} high spectral resolution ($< 1$ \kms), sensitivity, and angular resolution 
provided by large millimeter interferometers such as PdBI/NOEMA, SMA, and ALMA is now allowing to detect much weaker rotation signatures \rev{than in the optical range}, through velocity differences of only a fraction of \kms\ between opposite sides of the flow axis. Consistent rotation signatures in the same sense as the underlying disk have thus been uncovered in a growing number of molecular jets / outflows from protostars:  CB 26 \citep{2009A&A...494..147L}, Ori-S6 \citep{2010A&A...510A...2Z}, DG Tau B \citep{2015ApJ...798..131Z}, TMC1A \citep{2016Natur.540..406B}, Orion Source~I \citep{2017NatAs...1E.146H}, HH212 \citep{2017A&A...607L...6T,2017NatAs...1E.152L,2018ApJ...856...14L}, HH211 \citep{2018ApJ...863...94L}, HH30 \citep{2018A&A...618A.120L}, IRAS4C \citep{2018ApJ...864...76Z}. Standard application of \rev{Anderson's formula} \rev{to the observed rotation signatures}
yields "observed" DW launch radii \robs\ ranging from 0.05~au to 25~au. 

\rev{When discussing the implications of these results, e.g. to favor an X-wind \citep{2000prpl.conf..789S} over an extended MHD DW, it is generally assumed that \robs\ derived in this way is close to the outermost launch radius \rout. However, it is important to realize that they are in general two different things.}

A detailed fitting of ALMA data in HH212 by MHD DW models 
required much larger outer launch radii than inferred by \rev{Anderson's} formula, \rev{namely \rout $\simeq$ 40~au instead of \robs $\simeq$ 1~au} 
for the SO-rich slow outflow, 
\rev{and \rout $\simeq 0.2-0.3$~au} instead of \robs $\simeq$ 0.05~au 
for the SiO-rich jet
\citep{2017A&A...607L...6T}.
Hence, even at the high resolution achievable with ALMA, 
it appears that application of \rev{Anderson's formula} to the "observed" angular momentum can  underestimate significantly the true 
outermost launching radius of an MHD DW, 
at least in some cases.

Another key parameter of an MHD DW 
that one wishes to estimate from observations 
is the magnetic lever arm parameter \lbp, which 
\rev{measures the total} specific angular momentum extracted by the wind \rev{in units of} the initial keplerian value \citep{1982MNRAS.199..883B}. An estimate of \lbp\ \rev{is necessary to assess angular momentum extraction by the wind.}\rev{Observational estimates of \lbp\ are generally obtained through \citep[see eg.][]{2003ApJ...590L.107A}
\begin{equation}
\lambda_{\rm obs} \simeq {r V_\phi} / \sqrt{GM_\star r_{\rm obs}}
\label{eq:lambda-anderson}
\end{equation}
where \robs\ is the launch radius inferred using Anderson' formula. However, the few detailed comparisons with MHD DW models favor \lbp\ values 2--3 times larger than this \citep{2004A&A...416L...9P, 2017A&A...607L...6T}.}

Understanding \rev{and quantifying these observational biases in \rout\ and \lbp} is crucial if we want to be able to infer robust constraints on the role of MHD DWs in sustaining accretion across PPDs. \rev{This question was first addressed by some of us in the specific case of HST optical observations of the DG Tau atomic jet  \citep{2004A&A...416L...9P}}. The goal of the present paper is to readdress this issue \rev{in the new context of ALMA-like spectral resolution, for wind parameters relevant to current molecular disk wind candidates.} \rev{We thus compute} synthetic predictions for self-similar MHD DW models at resolutions typical of current mm interferometers, \rev{and apply the same methods as observers to estimate the wind launch radius and magnetic lever arm parameter, which are then compared with the true \rout\ and \lbp\ in the model.} \rev{Quasi edge-on DWs are studied in particular detail, as their rotation shifts are maximized by projection effects, and have the interesting property of being independent of angular resolution.} 

\rev{Our main predictions in the quasi edge-on case are checked against published ALMA data of HH212 ranging in resolution from 250 au to 18 au, which represent the most stringent test of MHD DWs to date. We also derive an exact analytical expression for the fraction of disk angular momentum flux extracted by any self-similar MHD DW, that we apply to HH212 for illustration.}

The paper is layed out as follows : in Section 2, we present self-similar MHD DW solutions used for building our synthetic predictions. 
In Section 3, we describe the effect of model parameters on \rev{rotation signatures, and three methods used by observers for estimating the flow specific angular momentum
from them.} We then examine in representative cases how the launch radius and magnetic lever arm parameter deduced with \rev{Anderson's relations} \rev{differ from the true \rout\ and \lbp}.
In Section 4, \rev{we compare our predictions for the edge-on case with ALMA observations of HH212, and we examine angular momentum extraction by the proposed MHD disk wind model.} 
In Section 5, we summarize our results and their implications for ALMA-like observations of molecular MHD DW candidates in protostars.

\section{MHD disk-wind solutions}
\label{sec:MHD}
Four semi-analytical solutions of magneto-centrifugal MHD disk winds (hereafter MHD DW), of which three are new, were computed 
\SC{in order to examine their predicted rotation signatures (see Table \ref{tab:MHD-models})}. 
\SC{In this section, we briefly describe the underlying approach used, and the collimation and kinematic properties of the four chosen solutions.}
\SCC{The MHD DW models belong to the class of exact self-similar, \SC{axisymmetric}, steady-state magnetic accretion-ejection solutions
developed and described by \citet{1997A&A...319..340F, 2000A&A...353.1115C,2000A&A...361.1178C}, to which the reader is referred
for more details. 
The distributions of density, thermal pressure,  velocity, magnetic field, and electric current, 
are obtained by solving for the exact steady-state MHD fluid equations, starting from 
the Keplerian, resistive accretion disk (\SC{with $\alpha$-type prescriptions for the 
turbulent viscosity and resistivity}) and passing smoothly into the ideal-MHD disk wind regime. }
At the same time, the self-similar geometry\footnote{\rev{which assumes that the variation of a given quantity with polar angle $\theta$ is the same for all streamlines, while the variation with radius is a power law,}}
allows to solve exactly for the global 2D cross-field balance and wind collimation on scales much larger than the launching point, as required for comparing with existing observations.
\SCC{Such solutions have been shown to provide an excellent match to rotation signatures observed in the DG Tau atomic jet 
\citep{2004A&A...416L...9P} and in the HH212 molecular jet \citep{2017A&A...607L...6T}, as well as to the ubiquitous broad H$_2$O component discovered by \textit{Herschel}/HIFI towards protostars \citep{2016A&A...585A..74Y}. Hence we use the same class of models here 
to estimate observational biases on MHD DW rotation signatures observed with ALMA.}

\subsection{Relevant disk wind parameters for rotation signatures}

\SC{Two emerging global wind properties are most relevant to determine the apparent rotational signatures, and will be used to label our MHD DW solutions thereafter:}

\SC{The first key parameter, controlling the wind speed and angular momentum, is} the ``magnetic lever arm parameter" $\lambda_{BP}$ defined by \citet{1982MNRAS.199..883B} \SC{as the ratio of extracted to initial specific angular momentum,}
\begin{equation}
    \lambda_{BP} \equiv \frac{L}{\Omega_0 r_0^2}
    \label{eq:lambdaBP}
\end{equation}
where $L$ is \SC{the total} \MV{specific angular momentum carried away by the MHD DW streamline} (in the form of both matter rotation and magnetic torsion), 
and $\Omega_0$ is the Keplerian angular rotation speed at the launch point $r_0$. 
\SC{A larger/smaller value of $\lambda_{BP}$ thus corresponds to a more/less efficient extraction of angular momentum by the wind, 
and to a more/less efficient magneto-centrifugal acceleration (see next section).}

\begin{figure}
   \centering
 \includegraphics[width=0.3\textwidth]{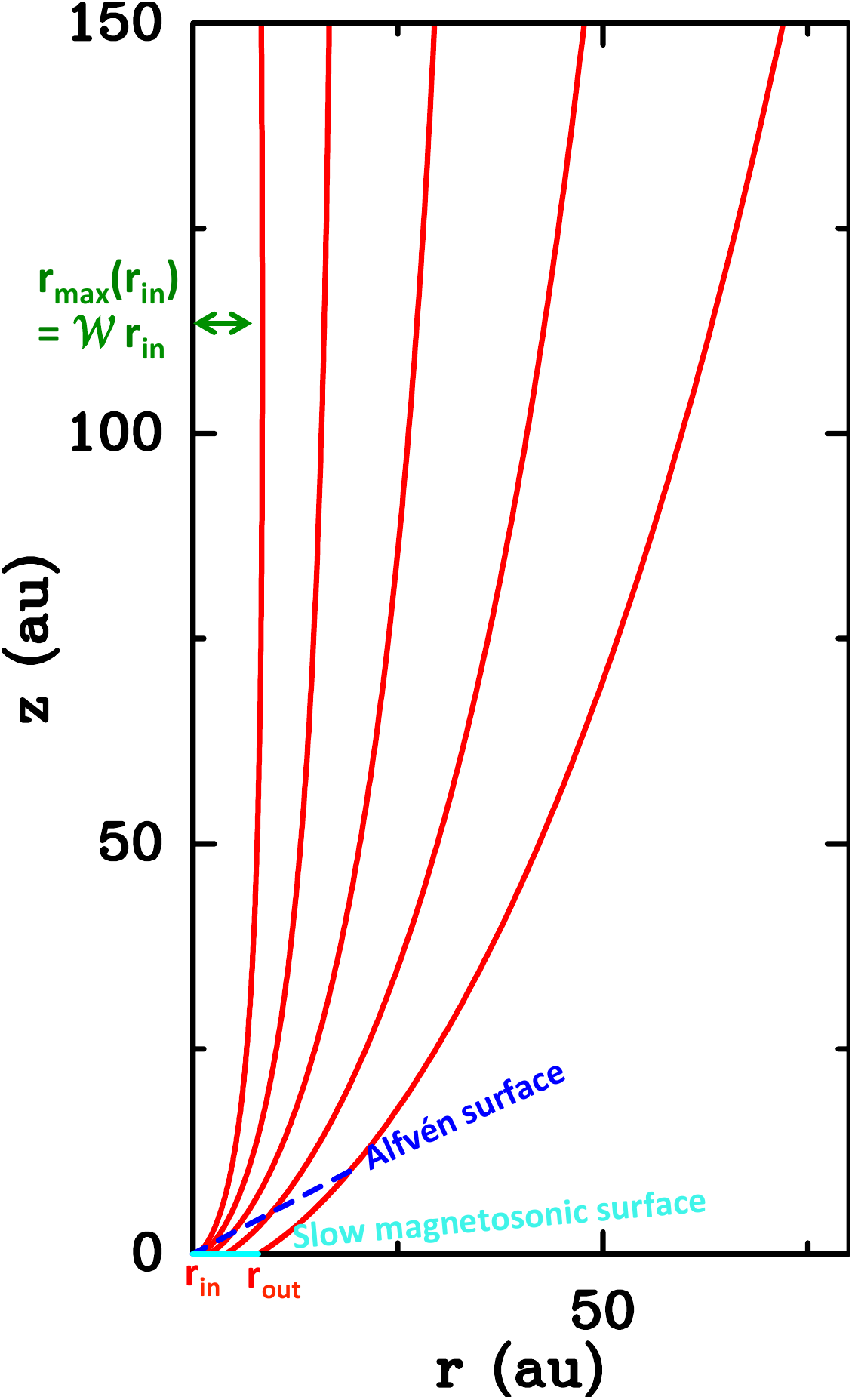}
      \caption{Poloidal cut of a self-similar, axisymmetric MHD disk-wind for our reference solution (L5W17). Selected flow surfaces are plotted in red. 
      \SC{Four} important model parameters \SC{affecting predicted rotational signatures} are illustrated here: the inner and outer launching radii, \rin\ and \rout, \SC{of the wind emitting region} \rev{(taken as 0.5~au and $8$~au in this graph)}; the magnetic lever arm parameter \lbp\ (here = 5.5) $\simeq (r_A/r_0)^2$, with \rev{$r_0$ the launch radius and} $r_A$ the cylindrical radius reached on the Alfv\'en surface (in dashed dark blue); the widening factor  $\mathcal{W} \equiv r_{\rm max}/r_0$, where $r_{\rm max}$ is the maximum radius reached by the \rev{streamline} (here \WW = 17, reached at $z/r_0 \simeq 200$). \rev{Note that due to self-similarity, all flow surfaces are homologous to each other, hence they share the same \lbp\ and \WW.}}
         \label{fig:schema}
   \end{figure}

It may be shown that $\lambda_{BP} \simeq ({r_A}/{r_0})^2$, where $r_A$ is the cylindrical radius at the Alfv\'en surface (where the gas poloidal velocity is equal to the poloidal Alv\'enic velocity \rev{$V_{A,p} = B_p/\sqrt{4 \pi \rho}$ with $B_p$ the poloidal field intensity and $\rho$ the volume density}). \SCC{The Alfv\'en  surface is illustrated in Figure~\ref{fig:schema} for our reference self-similar solution.} 

The second emerging property, \SC{affecting both the wind geometry and rotation speed,} is the wind \textit{widening factor} $\mathcal{W}$
\rev{that we define} as
\begin{equation}
    \mathcal{W}=\frac{r_{max}}{r_0},
    \label{eq:WW}
\end{equation}
where $r_{max}$ is the maximum radius reached by the streamline launched from radius $r_0$ in the disk, \SC{before it starts to (slowly) recollimate towards the axis}.
\SCC{The value of \WW\ is found by solving self-consistently for the transverse force balance between wind magnetic surfaces
\citep[see discussion in][]{1997A&A...319..340F}. This parameter is illustrated in Figure~\ref{fig:schema} for our reference solution.}

%%%%%%%%%%%%%%%%%
%%%%%%%%%%%%%%%%

\begin{table}
\caption{Wind parameters %\tablefootmark{a} of 
of MHD solutions computed in this work}              % title of Table
\label{tab:MHD-models}
\centering                                      % used for centering table
\begin{tabular}{l r r r}          % centered columns (4 columns)
\hline\hline                        % inserts double horizontal lines
Solution & $\lambda_{BP} \simeq (r_A/r_0)^2$ & $\mathcal{W} \equiv r_{max}/r_0$ & $i_{crit}$\tablefootmark{a}\\    % table heading
\hline              % inserts single horizontal line
{L13W36}\tablefootmark{b} & 13.7 & 36 & $\simeq 86^{\circ}$\\  
{L13W130} & 12.9 & 134& $< 80\degr$ \\      %
{L5W30} &   5.5 & 30 & $\simeq 86^{\circ}$\\ 
\textbf{L5W17}\tablefootmark{c} &   \textbf{5.5} & \textbf{17} & \textbf{$\simeq 84^{\circ}$}\\
\hline                           %inserts single line
\end{tabular}
\tablefoot{
\tablefoottext{a}{\rev{critical inclination below which the PV cut may be single-peaked, computed for $z_{cut}= 225$~au, \rin = 0.25~au, and \rout = 8~au.}} 
\tablefoottext{b}{solution used in the modeling of \citet{2004A&A...416L...9P, 2012A&A...538A...2P,2016A&A...585A..74Y};} 
\tablefoottext{c}{reference solution in Section \ref{sec:predict} and Figs. 3--9.}
} 

\end{table}

\SC{For easy reference, our four computed solutions are} denoted in the following as {LxWy} with x=$\lambda_{BP}$ and y=$\mathcal{W}$, and are summarized in Table~\ref{tab:MHD-models}. 

\begin{enumerate}
\item L13W36, with $\lambda_{BP} = 13.7$ and $\mathcal{W}=36$, is the solution that best fitted tentative rotation signatures across the base of the DG Tau atomic jet  \citep{2004A&A...416L...9P}; it was used by \citet{2012A&A...538A...2P} to demonstrate the molecule survival in a dusty disk wind, and by 
\citet{2016A&A...585A..74Y} to fit H$_2$O line profiles observed by \textit{Herschel} towards embedded protostars.
\item  L13W130 with $\lambda_{BP} = 12.9$ and $\mathcal{W}=134$ is a new solution with a much larger widening.
\item  L5W30 is a new, slower solution with $\lambda_{BP} = 5.5$ \SC{and} a widening factor $\mathcal{W}=30$ comparable to {L13W36}.
\item L5W17 is another new slow solution with $\lambda_{BP} = 5.5$, and an even smaller widening $\mathcal{W}=17$. This solution is our reference model in the next sections, and its geometry is illustrated in Fig.~\ref{fig:schema}.
 \end{enumerate}

 \SC{The input} physical parameters of the disk and the heating function at the \SC{disk} surface used to obtain our solutions are given in Appendix \ref{appendixA}, \SC{as well as the calculated density and magnetic field distributions along the wind streamlines.} 
 
\subsection{\SC{Collimation and} kinematics \SC{of} MHD DW solutions}

 %%%%%%%%%%%%%%%%%
 \begin{figure}
   \centering
 \includegraphics[width=0.5\textwidth]{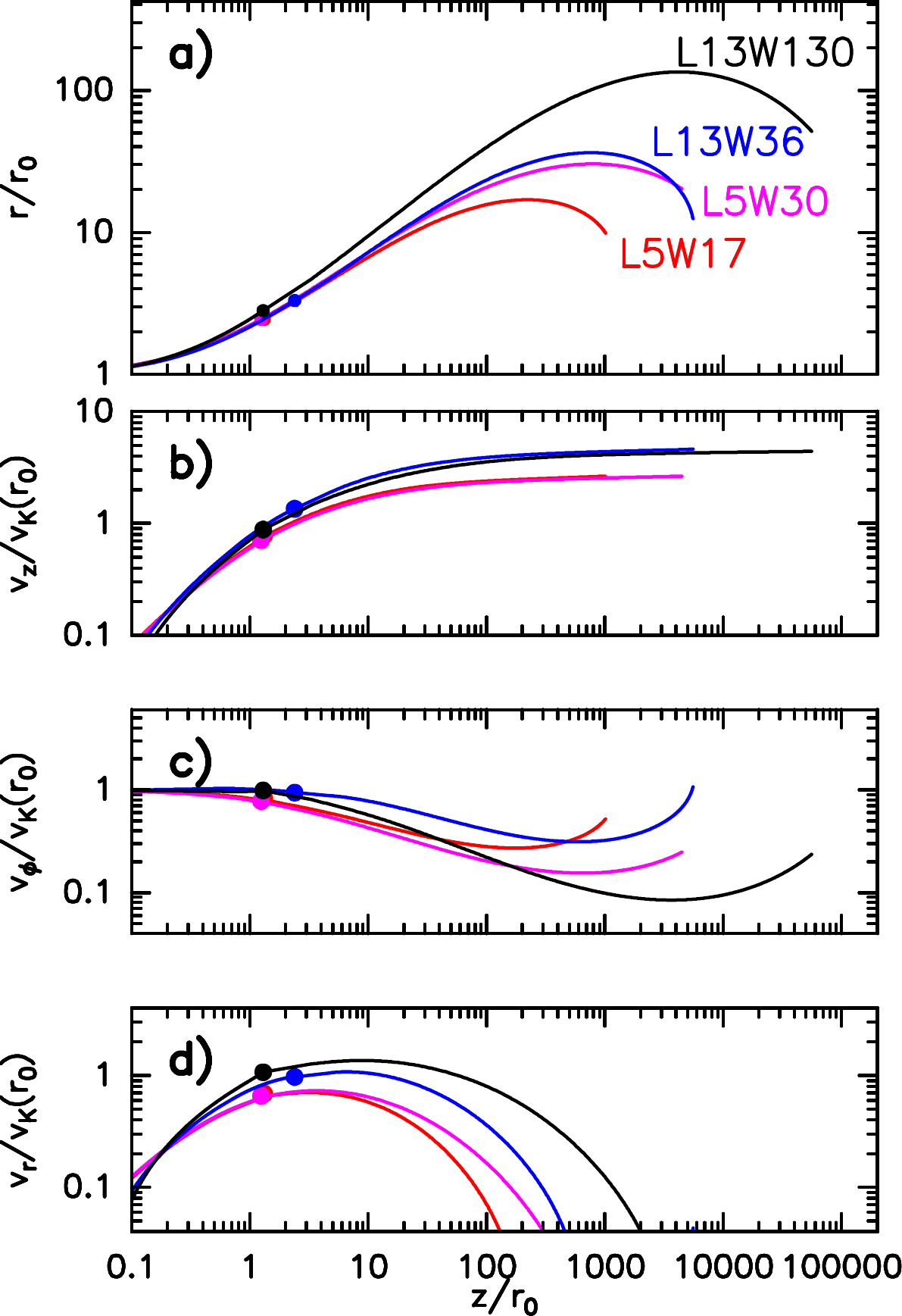}
      \caption{Shape \SC{and kinematics} of the \SC{streamlines} as a function of \SC{vertical distance $z$ above the disk midplane} for the four computed MHD DW solutions in Table~ \ref{tab:MHD-models}. \textbf{a:} \SC{cylindrical radius $r$}, \textbf{b}: velocity along the jet axis $V_z$, \textbf{c}: azimuthal rotation velocity $V_{\phi}$, \textbf{d}: radial expansion velocity $V_r$. \SC{Filled dots indicate the Alfv\'en surface.} 
     Distances are scaled by the launch radius $r_0$, and velocities by the Keplerian speed at $r_0$, \Vk($r_0$). Models are denoted as {LxWy} with x=$\lambda_{BP}$ the magnetic lever arm parameter and y=$\mathcal{W}$ = $r_{max} / r_0$ the widening factor (as defined in Eqs.\ref{eq:lambdaBP},\ref{eq:WW} and listed in Table~\ref{tab:MHD-models}).}
         \label{velocities-MHD}
   \end{figure}  
%%%%%%%%%%%%%%%%
 
\MV{Figure~\ref{velocities-MHD} compares the (self-similar) shape and velocity field of the wind streamlines for our four computed MHD DW solutions.}
\SCC{Cylindrical coordinates are adopted, and we denote hereafter $r$ the cylindrical radius, $V_z$ the velocity \SCC{component parallel} to the jet axis, $V_{\phi}$ the azimuthal (rotation) velocity, and $V_r$ the radial (sideways expansion) velocity.}

 %The shape of the streamlines is presented in 
 Figure~\ref{velocities-MHD}a \SC{shows that} the maximum radius is reached further out (i.e. at larger value of $z/r_0$) for increasing widening factor $\mathcal{W}$. After the maximum widening, the streamline slowly bends toward the axis (recollimation zone), until %the solution eventually terminates 
\SCC{refocussing becomes so strong that the steady-state solution terminates (at $z/r_0 \simeq 10^3-10^5$).} %at a recollimation shock. 
{This behavior is related to the radial distribution of physical quantities} %inherent to the self-similar ansatz, 
 \SC{and is a consequence of the dominant hoop-stress in a jet launched from a large radial extent in the disk}
 \citep[see discussion in][]{1997A&A...319..340F}. \SCC{A recollimation shock may be expected to form beyond this point.
 However, this region is not reached for the distances to the source and launch radii considered here.}
   
 \SC{Concerning kinematics, four} stages along the propagation of the jet can be distinguished in Fig.~\ref{velocities-MHD}b,c,d: \SC{below the Alfv\'en surface}, the velocity field is dominated by Keplerian rotation. At the Alfv\'en surface, the \SC{vertical, radial, and toroidal} velocities all become comparable  \SC{(and close to the initial Keplerian velocity at the launch point). Beyond this point, the jet velocity becomes dominated by $V_z$ while $V_{\phi}$, and then $V_r$, both decrease.} \SC{Finally, in} the recollimation \SC{zone where the streamline bends towards the axis},  \MV{$V_r$ becomes negative and $V_{\phi}$ increases due to conservation of angular momentum}, \SC{while $V_z$ keeps its final value}. 
 
 Figure~\ref{velocities-MHD}b shows that the \SC{increase} of $V_z$ {along a streamline} depends mainly on the magnetic lever arm $\lambda_{BP}$ with little influence of $\mathcal{W}$. 
The asymptotic \SCC{value $V_z^\infty$} is close to the maximum poloidal velocity 
predicted if all magnetic energy is transferred to the matter \citep{1982MNRAS.199..883B}: 
 %a streamline anchored at $r_0$ has a poloidal velocity given 
 \begin{equation}
V_p^{\infty}  =  V_K(r_0) \sqrt{2 \lambda_{BP} -3},
\label{eq:vp}
\end{equation}
\SCC{where $V_K(r_0)$ is the Keplerian velocity at $r_0$, and the poloidal velocity is defined as $V_p  = \sqrt{V_z^2 + V_r^2}$.}
 
\SC{In contrast},  the  expansion and rotation velocities depend on both $\lambda_{BP}$ and $\mathcal{W}$, but in different ways.  
$V_r$ increases with either of these parameters (faster and/or wider flow; cf. Fig.~\ref{velocities-MHD}d), while 
 $V_\phi$ increases with $\lambda_{BP}$ but \MV{decreases for wider solutions (Fig.~\ref{velocities-MHD}c).}
 \SC{This may be understood by noting that}, \rev{in the "asymptotic regime" ($z/r_0 \rightarrow \infty$) where} %once
the total specific angular momentum \SC{$L$} extracted by the wind magnetic torque 
has been entirely converted into matter rotation, we have
\begin{equation}
\frac{(r V_{\phi})^{\infty}}{r_0 V_K(r_0)} \simeq \frac{L}{r_0 V_K(r_0)} = \lambda_{BP},
\label{andersonvphi}
\end{equation}
\SC{where we used the definition of \lbp\ in Eq.~\ref{eq:lambdaBP}.
Combining with the definition of \WW\ in Eq.~\ref{eq:WW} we obtain that} the minimum $V_{\phi}$ on a given streamline will scale as
 \begin{equation}
 \frac{V_{\phi}^{min}} {V_{K}(r_0)} \simeq \frac{\lambda_{BP}}{\mathcal{W}}.
 \label{eq:vphimin}
 \end{equation}
 {Hence, the minimum rotation velocity reached by a streamline is smaller for wider solutions of the same $\lambda_{BP}$.}
 
 \SC{The different dependencies of $V_z$, $V_r$ and $V_\phi$ on 
 $\lambda_{BP}$ and \WW\ open the possibility to constrain these two parameters from the observed wind spatio-kinematics.}
 
\section{Observed rotation signatures, launch radius, and magnetic lever arm}
\label{sec:predict}

\begin{figure*}
   \centering
    \includegraphics[width=0.9\textwidth]{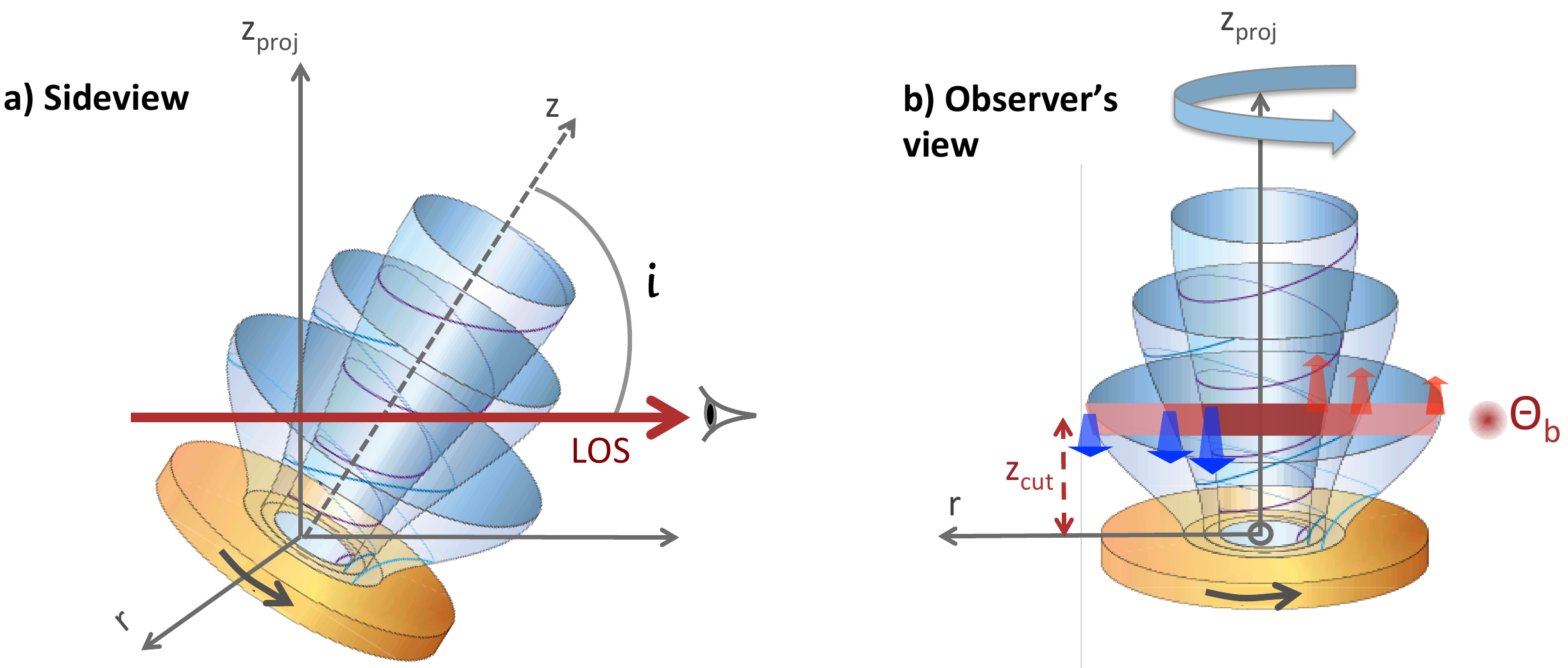} 
     \caption{\rev {\textbf{a}: Definition of inclination angle $i$ for our synthetic predictions in Section~\ref{sec:predict}. 
     \textbf{b}: Sketch illustrating the construction of the transverse Position-Velocity (PV) cut at projected altitude $z_{\rm cut}$ with a gaussian beam $\theta_b$. Wind rotation induces different line-of-sight velocities at symmetric offsets $+r$ and $-r$ on either side of the jet axis, producing a detectable "tilt" in the PV cut (see Fig.~\ref {fig:pv-perp-exemple}). This projected velocity shift is used to estimate the rotation speed and specific angular momentum of the flow (see Section \ref{sec:icrit}).} \REVbis{Adapted from Ferreira (2001)}.}
   \label{fig:DW-schema}
   \end{figure*}

\rev{In an axisymmetric wind, rotation introduces a systematic Doppler shift between spectra from symmetric positions $+r$ and $-r$ on either side of the jet axis;  this velocity shift is measured by observers using transverse Position-Velocity (PV) diagrams built perpendicular to the jet axis, as illustrated in Fig~\ref{fig:DW-schema}.}

%%%%%%%%%%%%%%%%%%%%%%%%%%%%%%%
%%%%%%%%%%%%%%%%%%%%%

In Section \ref{sec:freeparam}, we describe the range of free parameters used to compute synthetic transverse PV diagrams for our MHD DW solutions, and our choice of reference case. We then \rev{describe in Section \ref{sec:icrit} the appearance of PV cuts for radially extended disk winds, and we} introduce \rev{three methods used by observers} to estimate the "observed" specific angular momentum from the PV cuts.  
Finally, \rev{in Sections \ref{sec:doublepeak} to \ref{sec:flowwidth}}, 
we investigate \rev{for each method} how the launch radius and magnetic lever arm parameter inferred using \rev{Anderson's relations} differ from the true \rout\ and \lbp\ of the 
MHD DW model. \rev{Setting robust constraints on these two fundamental parameters is indeed crucial to assess the role of disk winds in disk accretion.}

\subsection{Free model parameters}
\label{sec:freeparam}
\MV{As shown in Fig.~\ref{velocities-MHD}, an MHD DW solution provides us with the self-similar shape of the streamline scaled by the anchor radius $r_0$ of the magnetic surface in the disk, and self-similar velocities scaled by the Keplerian velocity at $r_0$, $V_K(r_0) = \sqrt{G M_{*}/r_0}$.} 
In order to produce synthetic emission predictions comparable to observations, 
we then need \SC{to specify three dimensional} parameters to construct a wind model in physical units (see Fig.~\ref{fig:schema}):

- \textit{Mass of the central object $M_\star$}, to scale the Keplerian velocity. It has a trivial influence on line profiles and PV diagrams as it simply stretches the velocity axis by a factor $\sqrt{M_*}$. Here we set $M_\star=0.1M_{\odot}$ as a fiducial Class 0 protostellar mass, 
\SC{for consistency with} \citet{2016A&A...585A..74Y}. 

- \textit{Launch radius \rin\ of the innermost emitting wind streamline:} \rev{The value of \rin\ depends on the abundance distribution of the observed molecule, which in turn depends on the (ill-known) wind density, irradiation, and temperature. In order to limit the parameter space to explore, we will keep \rin\ constant in this Section.} Because the survival of molecules in MHD DWs has been \rev{theoretically} demonstrated so far only on dusty streamlines \citep{2012A&A...538A...2P,2016A&A...585A..74Y}, we set $r_{\rm in}$ in our models \rev{to a fiducial value of} 0.25~au, \MV{the typical dust sublimation radius in solar-mass protostars}. \SC{The resulting maximum poloidal velocity is 50--90 \kms\ for \lbp = 5.5--13 and $M_\star$ = 0.1 $M_\odot$}. 
\rev{For a given radial extent (\rout/\rin), a change in \rin\ would simply stretch the velocity axis by a factor $1/\sqrt{r_{\rm in}}$ without changing the PV shape}. 

- \textit{Launch radius \rout\ of the outermost emitting wind streamline:}  This radius, \rev{which is one key quantity that one wishes to determine from observations,} is kept as a free parameter.
We explored a range of $r_{\rm out}=0.5-32~$au, \SC{corresponding to radial extensions (\rout/\rin) of 2 -- 130.}
As a reference case, we \rev{arbitrarily} choose \rev{an intermediate value of} $r_{\rm out}=8~$au, corresponding to \rout/\rin = 32. 

\SC{Once the physical model is constructed (see Fig.~\ref{fig:schema}), we also need to specify four ``observational" parameters that 
affect the synthetic predicted PV diagrams:} 
\begin{enumerate}

\item \textit{Inclination angle $i$ of the jet axis with respect to the line of sight} \rev{(illustrated in Fig.~\ref{fig:DW-schema}a)}. 
\rev{We restrict ourselves to inclinations from $i$ = 40\degr to 90\degr, which are the most favorable to detect rotation signatures and cover 80\% of random orientations.} 
We choose $87^{\circ}$ (the inclination of HH212) as our reference model.
We show only the red lobe. PV diagrams for the blue lobe can be easily recovered by the operation $V_{proj} \rightarrow-V_{proj}$ and $r\rightarrow-r$. 

\item \textit{Power-law index $\alpha$ of the line emissivity decline with radius.} In principle, knowledge  of the emissivity function requires a full thermo-chemical and non-LTE line excitation calculations, as done by \citet{2016A&A...585A..74Y} for H$_2$O line predictions. In this work, since we aim at presenting general synthetic observations \SC{for a much broader range of} MHD DW solutions, 
a parametrized emissivity function is adopted with 
a \SC{a simple power law} radial variation\footnote{\SC{The} variation of $\epsilon$ with $z$ has little influence on transverse PV cuts as long as \SCC{this variation} is smooth over the scales \SC{probed by} the beam. Here we take a dependence of the form $e^{-z/H}$ with $H=600$~au for \SC{illustrative purposes}.}:
\begin{equation}
\epsilon(r) \propto  r^{\alpha}.
\label{eq:alpha}
\end{equation}
We choose $\alpha = -2$ as reference value \citep[based on our modeling of ALMA observations of HH212 in][and Section~\ref{sec:HH212}]{2017A&A...607L...6T}.
We also explored $\alpha= 0, -1, -3$ in our reference model. 

\item \textit{\MV{Spectral and} spatial resolutions}: a fiducial $0.44$ km s$^{-1}$ spectral sampling is adopted, \SCC{typical of what is routinely achieved with interferometric observations of faint lines.} Synthetic channel maps are then convolved by a Gaussian \SC{spatial} beam with a FWHM $\theta_b$. We choose $\theta_b = 225$~au as reference \SC{case}. It corresponds to a $0.5"$ beam for a source in \SC{the Orion molecular cloud (at $\simeq$} 450~pc). We also explored $\theta_b = 45-380$~au.

\item \textit{Position $z_{cut}$ where the transverse Position-Velocity diagram is built:} in the context of rotating disk winds from young \SC{protostars}, \SC{contamination by the} rotating \SC{infalling} envelope close to the source \SCC{(not modeled here)} \SC{has} to be \SC{minimized}. \SC{At the same time}, observations must be made sufficiently close to the source to probe a suspected pristine stationary MHD DW minimally affected by shocks or variability \citep{2018A&A...614A.119T}. A typical distance corresponding to one beam \SC{thus appears as a natural} choice for $z_{cut}$. Considering the adopted fiducial beam, we set $z_{cut}$\,=\,$\theta_b$\,= 225~au for the reference case. We \SC{will} also explore the effect of a smaller $z_{cut}=70$~au \rev{in the reference case at $i = 87^o$}.
\end{enumerate}

In summary, as a reference model, we choose the L5W17 MHD DW solution (\lbp = 5.5, \WW=17) with $M_\star=0.1M_{\odot}$, \SC{$r_{\rm in}=0.25$~au}, $r_{\rm out}=8$~au, $i=87^{\circ}$, $\alpha = -2$, and $\theta_b= 225$~au, and perform a PV cut across the redshifted lobe at $z_{cut} = 225$~au. 
\SC{At this distance from the source, $z/r_{\rm out} = 28$ and the outermost radius of the jet has thus reached $r_j \simeq 10\,r_{\rm out} \simeq$ 80 au (see Fig.~\ref{velocities-MHD}a).}
In the following, we will vary each of the above free parameters except $M_\star$ and \rin\ (which only set the velocity scale), to see how they impact apparent rotation signatures, 
and the \rev{launch radius and magnetic lever arm inferred from them using Anderson's relation.} 

% -------------------------------------------------------------------------------
\subsection{\rev{Methods for} measuring rotation from transverse PV cuts} 
\label{sec:icrit}

 %%%%%%%%%%%%%%%%%%%%%%%%%%%%%%%
%%%%%%%%%%%%%%%%%%%%%%%%%%%%%%%
   \begin{figure}
   \centering
    \includegraphics[width=0.46\textwidth]{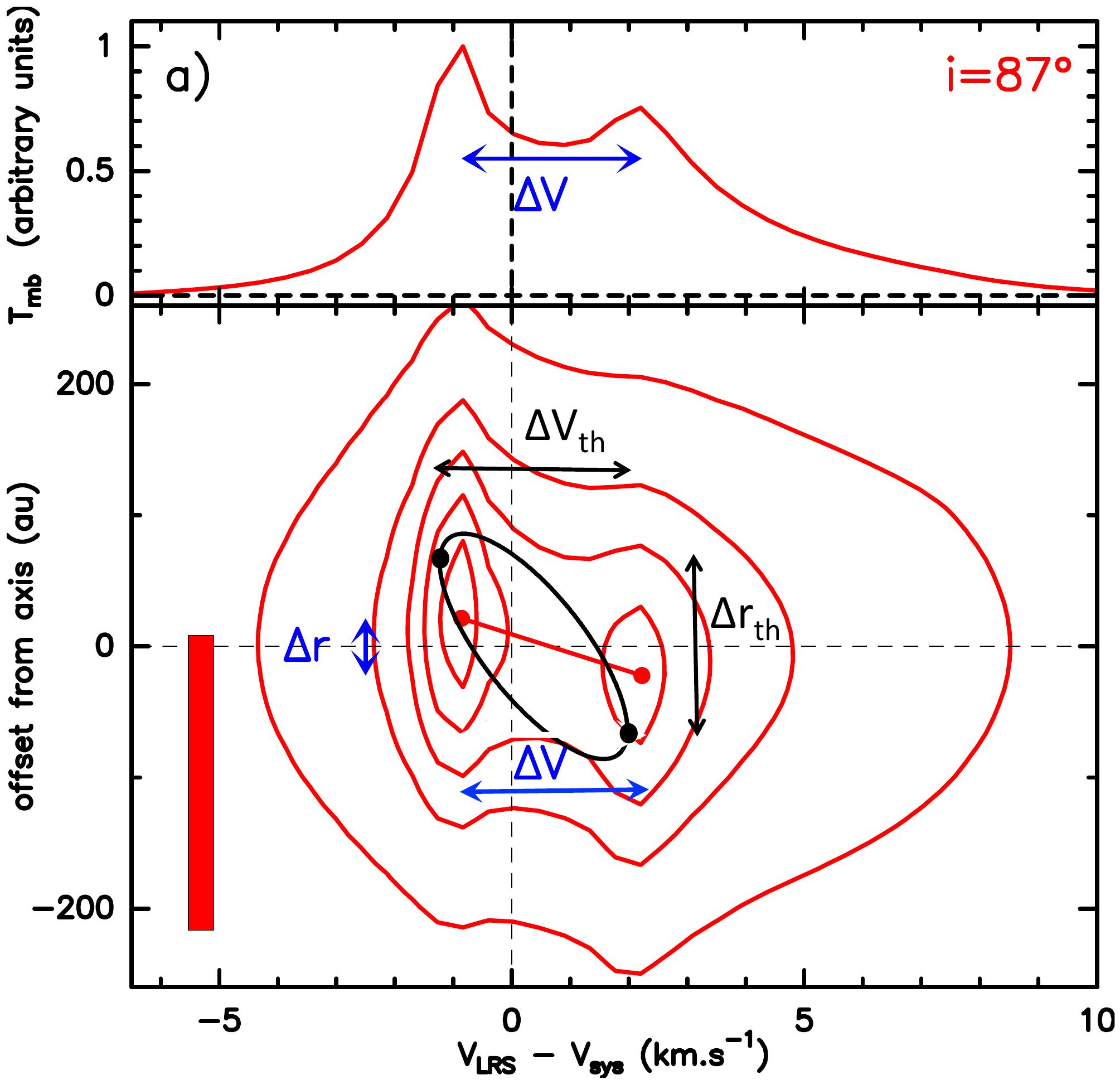} 
    \includegraphics[width=0.47\textwidth]{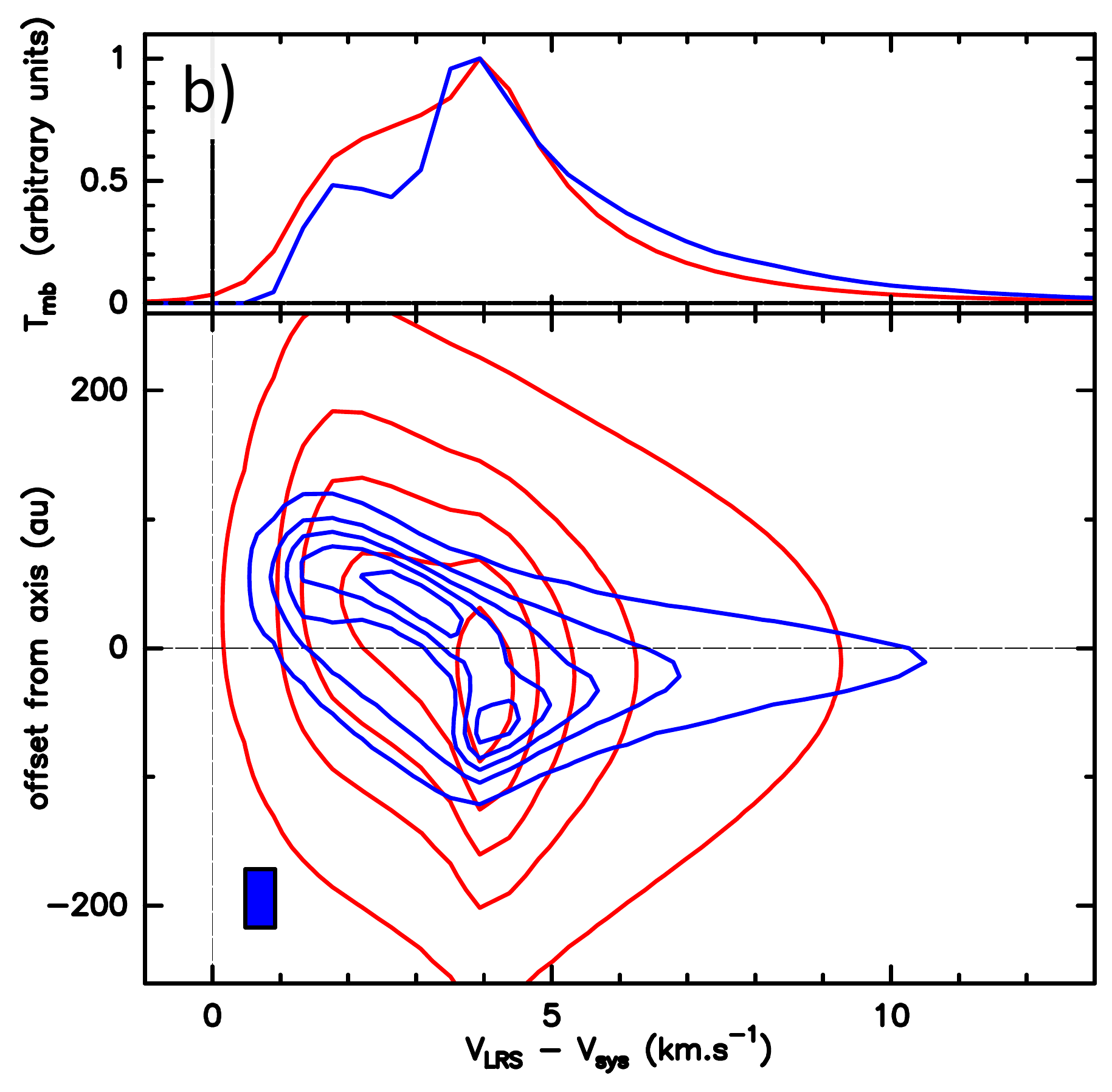}
      \caption{ \textbf{a:} On-axis spectrum and transverse PV diagram for the reference model viewed at $i=87^{\circ}$, \rev{illustrating the "edge-on" case}. The two red dots indicate the intensity peaks in the PV, \rev{which have opposite velocity signs in this configuration}; their connecting line defines the spatial and velocity separations $\Delta r$ and $\Delta V$, \rev{used to estimate the observed specific angular momentum \jobs\ in Eq.~(\ref{eq:jobs})}. We also plot in black the ellipse and peak positions contributed by the outermost streamline \rev{alone}, which predict a similar velocity shift $\Delta V_{\rm th}$ (see Eq.~(\ref{eq:deltaVth})) but a much larger spatial shift $\Delta r_{\rm th}$ (see Eq.~(\ref{eq:deltarth})). 
      Other model parameters are: $z_{cut} = 225$~au, $M_*=0.1M_{\odot}$, $r_{\rm in}=0.25$~au, $r_{\rm out}=8$~au, $\alpha = -2$, $\theta_b= 225$~au. Filled rectangles show the spectral and angular resolutions.\textbf{b:} same as \textbf{a} at lower inclination $i = 70^{\circ}$ (note the change in velocity scale) for $\alpha = 0$ and $\theta_b$ = 225 au (red) and 45 au (blue). \rev{PV double-peaks now have the same velocity sign, and can vanish at moderate angular resolution}. }
         \label{fig:pv-perp-exemple}
   \end{figure}  
%%%%%%%%%%%%%%%%%%%%%
%%%%%%%%%%%%%%%%%%%%%

\rev{Let us first consider the simple case where only a narrow rotating ring of wind material emits in the selected line tracer.
The transverse PV cut then resembles a tilted ellipse, whose major and minor axes and tilt angle 
are given in Appendix~\ref{ap:ellipse} as a function of the flow velocity field.}

\rev{Let us also assume that the ring is better resolved spectrally than spatially, as usually the case in ALMA-like observations. 
The PV ellipse then presents two emission peaks symmetrically positioned at (see Appendix~\ref{ap:peaks})}
\begin{equation}
    r_{\rm proj} = \pm r_j \left(\frac{V_\phi}{V_{\perp}}\right) 
    %\equiv \pm r_j\prime
    \label{eq:rproj-peaks}
\end{equation}
and
\begin{equation}
    V_{\rm proj} = -\cos{i} V_z \pm \sin{i} V_{\perp},
    \label{eq:Vproj-peaks}
\end{equation}
where 
\begin{equation}
    V_{\perp} = \sqrt{V_{\phi}^2+V_r^2} > V_\phi
    \label{eq:vperp}
\end{equation}
\SC{is the transverse velocity modulus (in the plane perpendicular to the jet axis),}
\MV{$r_j$ the flow radius, and $V_z$, $V_\phi$, $V_r$ are the vertical, azimuthal, and
radial expansion speeds, all measured at $z = z_{cut}$ on the outermost emitting streamline launched from \rout.}

The spatial and velocity separations between the two PV peaks, $\Delta r_{\rm th}$ and $\Delta V_{\rm th}$, are then given by:
\begin{equation}
    \Delta r_{\rm th} = 2 r_j \left( \frac{V_{\phi}}{V_{\perp}} \right),
    \label{eq:deltarth}
\end{equation}
\begin{equation}
    \Delta V_{\rm th}= 2 \sin{i} V_{\perp},
    \label{eq:deltaVth}
\end{equation}
\SC{and the true specific angular momentum on the outer streamline, \jout, is given by}
\begin{equation}
  \label{eq:jout}
j_{\rm out} \equiv r_j V_\phi = \frac{\Delta r_{\rm th}}{2} \times \frac{\Delta V_{\rm th}}{2\sin{i}}
 \end{equation}
(note that $V_{\perp}$ cancels out in the product of $\Delta r_{\rm th}$ and $\Delta V_{\rm th}$).  

{However, a \rev{narrow range of streamlines} is not the most probable case if the MHD DW dominates the extraction of angular momentum from a sizable portion of the disk, and the chosen tracer is not too chemically selective (eg. CO, SO).}

When the wind streamlines span a broad range of radii and $V_z$, 
we find that \rev{PV cuts are no longer elliptical and that two broad configurations exist, 
with a transition around a critical inclination angle 
$i_{\rm crit} \simeq \arctan(\mid{V_z}\mid/V_{\perp})$
($\simeq 84\degr$ for our models, see Table \ref{tab:MHD-models}).
At large inclinations $i >$ \icrit, which we will denote as "edge-on" in the following for brevity, PV cuts remain double-peaked regardless of model parameters, with peaks of opposite velocity signs. This is 
illustrated in Figure \ref{fig:pv-perp-exemple}a for our reference MHD DW model at $i=87^{\circ}$.}

\rev{In contrast, at moderate inclinations $i < i_{\rm crit}$, PV cuts become more complex, with curved emission ridges that stretch over a wide velocity interval. One or two main peaks may result, that rapidly shift or merge along the ridges with small changes in the parameters (spatial beam, emissivity gradient $\alpha$, wind radial extension...). An example is shown in Fig.~\ref{fig:pv-perp-exemple}b for our reference model at $i=70^{\circ}$ with $\alpha = 0$: the PV cut is double-peaked for $\theta_b = 45$~au but becomes single-peaked for $\theta_b = 225$~au. In this configuration, the PV double-peaks (of same velocity sign when present) cannot be used as reliable rotation estimators.}

\rev{In the following, we will thus consider three methods used by observers to estimate the flow specific angular momentum from PV cuts at ALMA-like resolution. They are briefly described in turn below. The resulting biases in launch radius and magnetic lever arm are discussed in Sections \ref{sec:doublepeak}, \ref{sec:single}, \ref{sec:flowwidth}.}

\subsubsection{Double-peak separation method:}
\label{sec:method1}
For a double-peaked transverse PV, it is easiest and customary in observational studies in the literature to estimate the specific angular momentum carried by the flow % the double peak separation, 
by analogy with the single annulus case (Eq.~(\ref{eq:jout})) as
\citep[eg.][]{2015ApJ...798..131Z,2016ApJ...824...72C,2018ApJ...856...14L,2018ApJ...864...76Z}
\begin{equation}
\label{eq:jobs}
j_{\rm obs} \equiv  \left( \frac{\Delta r}{2} \right) \times \left(  \frac{\Delta V}{2\sin{i}} \right),
\end{equation}
where 
$\Delta V$ is the observed \SC{velocity} shift between the two intensity peaks in the PV cut, and $\Delta r$ is their spatial centroid \rev{separation} %shift 
perpendicular to the jet axis and \rev{(see blue arrows in Figure \ref{fig:pv-perp-exemple}a).} 

\rev{In Section~\ref{sec:doublepeak}, we will investigate extensively the double-peaked method for "edge-on" inclinations ($i \ge$ \icrit), where the peaks have opposite velocity signs. We will show that the result is remarkably independent of beam size, and leads to systematically underestimate \jout, \rout, and \lbp\ (Sections~\ref{sec:jobs}, \ref{sec:robs}, and \ref{sec:lambda-obs})}.

\rev{In contrast, at lower inclinations where PV double-peaks have the same velocity sign ($i <$ \icrit), this method cannot yield robust results --- and is not recommended. The existence and positions of double-peaks are too sensitive to the exact combination of parameters (see discussion of Fig.~\ref{fig:pv-perp-exemple}b above). They would also be very sensitive to noise fluctuations along the underlying ridges. We will thus only consider the following two methods in that case.}

\subsubsection{Rotation curve method:}
\label{sec:method2}
\rev{Following optical jet rotation studies with HST
\citep{2002ApJ...576..222B,2007ApJ...663..350C}, a more generic method applicable to all inclinations and PV morphologies consists in deriving an "observed" rotation curve $V_{\phi,\rm obs}(r)$ from velocity shifts between symmetric spectra at $\pm r$ from the jet axis, through
\begin{equation}
\label{eq:vrot}
V_{\phi,\rm obs}(r) = \left[ V(r)-V(-r) \right] / 2\sin{i},
\end{equation}
from which the local specific angular momentum on each flow surface of radius $r$ may be estimated as 
\citep[][]{2003ApJ...590L.107A}
\begin{equation}
\label{eq:jrot}
j_{\rm obs}(r) = r \times V_{\phi,\rm obs}(r).
\end{equation}
This more elaborate method has only recently started to be applied to ALMA data \citep[eg.][]{2016Natur.540..406B}.
We will illustrate its typical observational biases in Section \ref{sec:single},
and show that it leads to overestimate \rout\ in our models, except when the flow is well resolved across.}

\subsubsection{Flow width method:}
\label{sec:method3}
\rev{Another generic but simpler method, mainly used when the flow is not well resolved laterally, is to take
\citep[eg.][]{2008ApJ...685.1026L}
\begin{equation}
\label{eq:jmax}
j_{\rm obs} = r_{\rm w} \times V_{\phi,\rm obs}(r^{\rm \infty}),
\end{equation}
where $V_{\phi,\rm obs}$(\rmax) is the asymptotic value of the rotation curve at large radii (assumed to trace the true rotation speed on the outer streamline) and \rw\ is the deconvolved flow radius estimated from emission maps (assumed to trace the true outer flow radius $r_j$).
In Section~\ref{sec:flowwidth}, we will show that in our models, this method systematically underestimates \rout\ and \lbp\ (by a similar amount as the double-peaked method in edge-on flows). }

\subsection{Double-peak separation: biases in edge-on flows}
\label{sec:doublepeak}

\rev{As explained in the previous section, we restrict our investigation of this method
to quasi edge-on inclinations where the PV cuts has double-peaks of opposite signs. We note that this edge-on configuration maximizes the chances of detecting rotation shifts ($\propto \sin{i}$) and minimizes contaminating shifts caused by slight asymmetries in poloidal velocity ($\propto \Delta V_p \cos{i}$).}

\rev{We will study this method in particular detail as it was used recently to measure rotation speeds in the edge-on SO outflow in HH212 \citep{2018ApJ...856...14L}. This prototypical object was studied with ALMA over a remarkably wide range of angular resolutions (factor 15), and will allow us to carry out several stringent tests of our predictions (see Section 4).}

\subsubsection{Bias in angular momentum}
\label{sec:jobs}

\rev{The outermost DW launch radius \rout\ is the most critical parameter that observers seek to estimate
in order to discriminate among disk accretion paradigms (see Introduction). Therefore, 
we will compare} the apparent specific angular momentum \jobs, measured 
from the double-peak spatial and velocity separations 
$\Delta r$ and $\Delta V$ (using Eq.~(\ref{eq:jobs})), 
\rev{with} the true specific angular momentum \jout\ along the outermost emitting DW streamline. 

%%%%%%%%%%%%%%%%%%%%%%%%%%
 \begin{figure*}%[h]
   \centering
\includegraphics[width=0.75\textwidth]{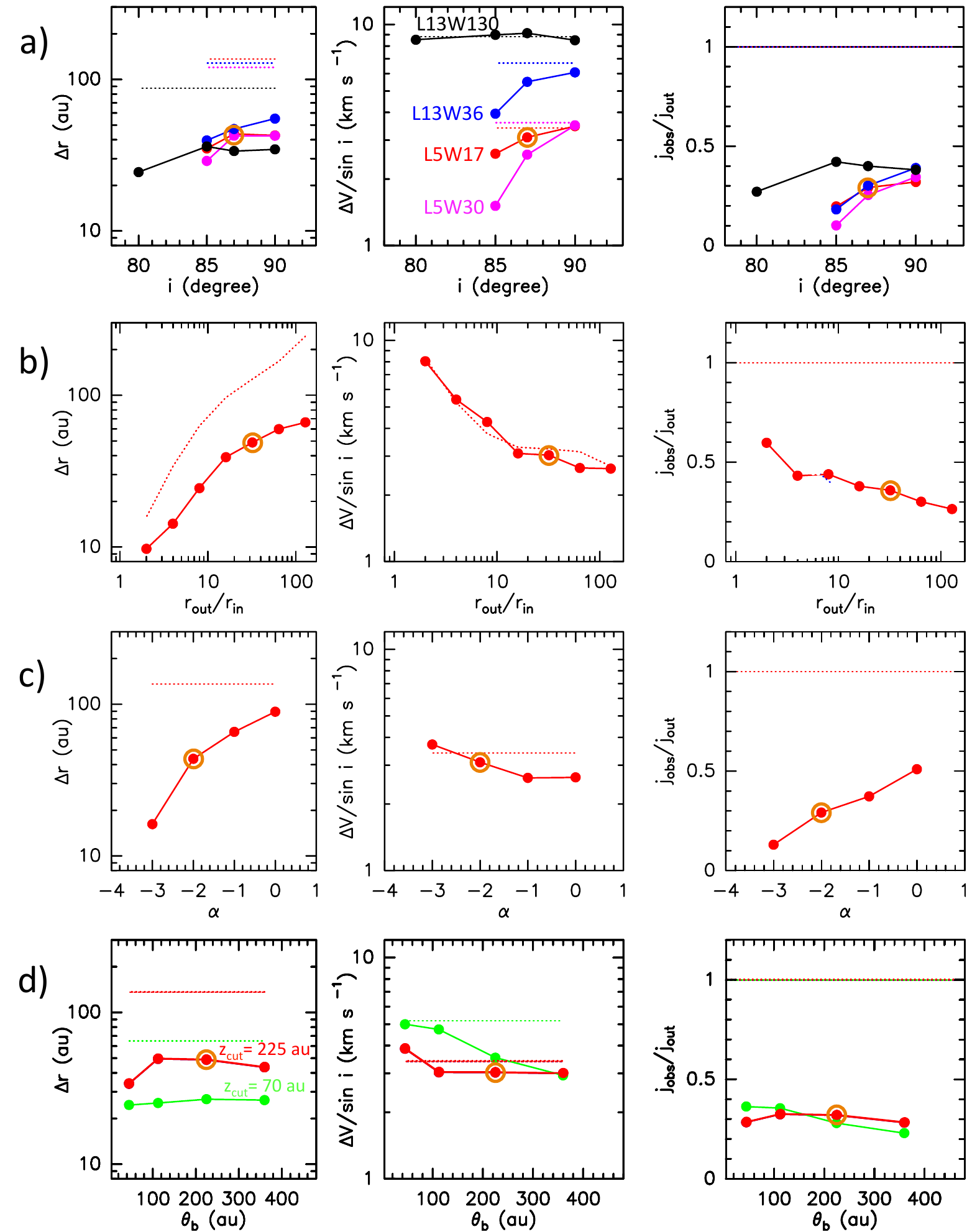}
       \caption{
\rev{Double-peak separation in edge-on PVs of radially extended DWs (connected dots) compared with theoretical value for a single wind annulus on the outermost streamline (dotted curves).} 
       \textit{Left column:} \rev{Observed spatial separation $\Delta r$ versus theoretical} $\Delta r _{\rm th}$ (from Eq.~\ref{eq:deltarth}); \textit{Middle column:} deprojected velocity shift ${\Delta V}/{\sin{i}}$ \rev{versus} theoretical value ${\Delta V_{\rm th}}/{\sin{i}}$ (from Eq.~\ref{eq:deltaVth}). \textit{Right column:} ratio of the apparent specific angular momentum ${j_{\rm obs}} = (\Delta r/2) \times (\Delta V/2\sin{i})$ to the true value on the outermost streamline, \jout\ 
       \rev{= $(\Delta r_{\rm th}/2) \times (\Delta V_{\rm th}/2\sin{i})$}.
From top to bottom, the panels show the influence of varying \textbf{a)} MHD-solution (colour-coded) and inclination angle;  \textbf{b)} outermost launching radius $r_{\rm out}$ of the emitting region of the MHD DW; \textbf{c)} index $\alpha$ of the emissivity radial power-law (see Eq.~\ref{eq:alpha});  \textbf{d)} spatial beam FWHM $\theta_b$, and PV cut position \zcut. All non-labelled model parameters are fixed at their reference value: $i = 87\degr$, MHD solution = L5W17, $r_{\rm out}=8$~au, $\alpha = -2$, $\theta_b= 225$~au, \zcut = 225~au, $M_\star = 0.1 M_\odot$, $r_{\rm in} = 0.25$~au. Datapoints for this reference case are circled in orange in each panel.}
\label{fig:peaks}
\end{figure*}  
%%%%%%%%%%%%%%%%%%%%%%%%%%
%%%%%%%%%%%%%%%%%%%%%%%%%%
%%%%%%%%%%%%%%%%%
 \begin{figure*}
   \centering
   \includegraphics[width=0.7\textwidth]{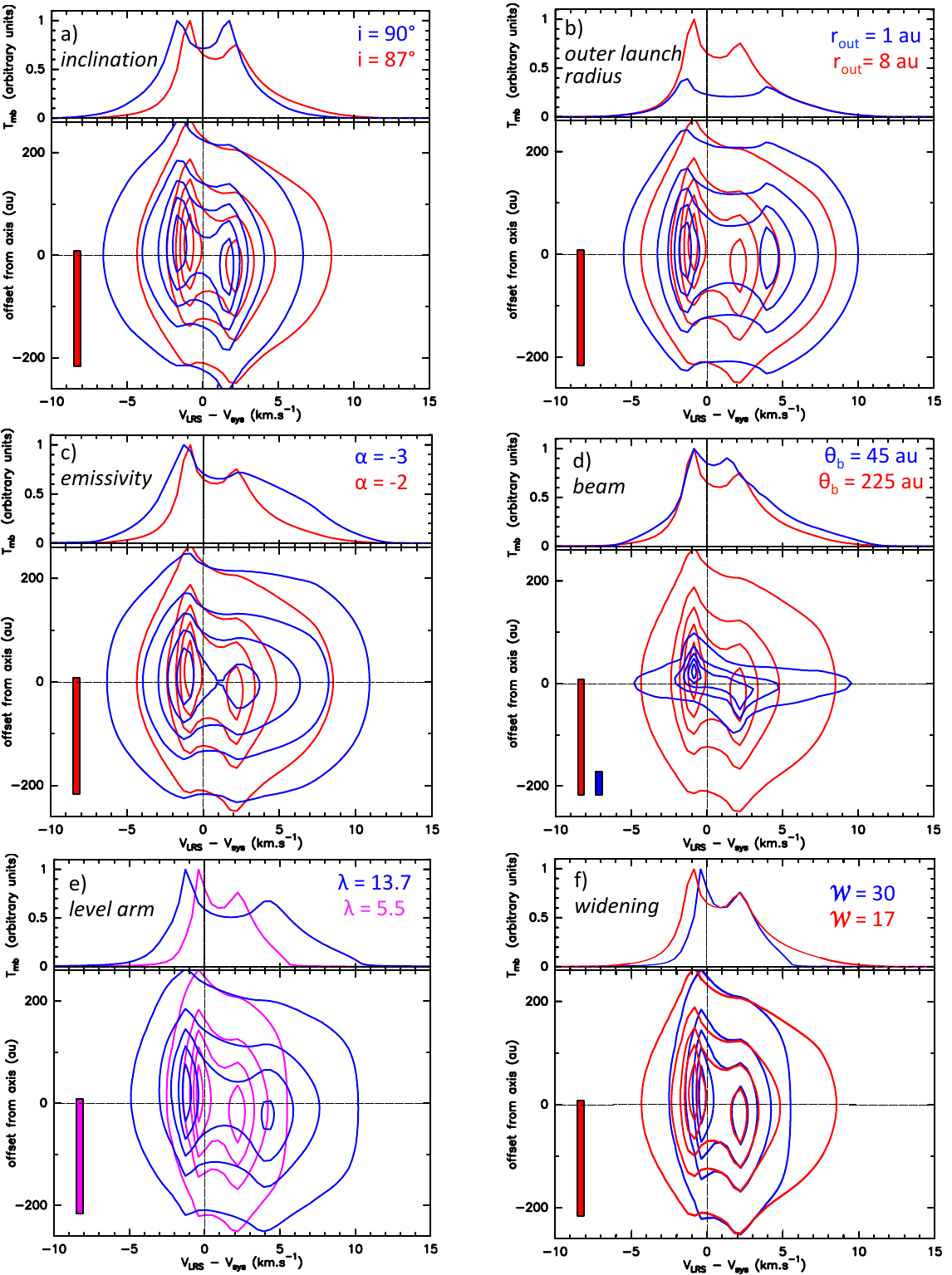}
      \caption{Comparisons of synthetic on-axis spectra and transverse PV cuts at \zcut =225~au {for selected pairs of \rev{quasi edge-on} models that differ by only one parameter at a time}. In red, the reference model with $i=87^{\circ}$, \rout = 8~au, $\alpha = -2$ (see Eq.~(\ref{eq:alpha})), $\theta_b = 225$~au, $\lambda_{BP} = 5.5$, $\mathcal{W}= 17$ (and M$_\star = 0.1 $ M$_{\odot}$, \rin = 0.25~au). \SC{In blue, the same model with only one parameter value changed  \SCC{(as labelled in each panel)}. In Panel \textbf{e}, the reference solution L5W17 is replaced by L5W30 (in magenta) so that only \lbp\ differs in the comparison (in blue: L13W30).}
     {Velocity and angular resolutions are pictured by filled rectangles}. 
     \rev{The flow is unresolved transversally except when $\theta_b = 45$ au (Panel d).}}
    \label{pv-perp-vs-r0max}
   \end{figure*}

\rev{As a first example, we consider our reference model shown in Fig.~\ref{fig:pv-perp-exemple}a.} \rev{The black ellipse and black dots show the predicted ellipse and emission peaks for
a single ring on the outermost streamline. It may be seen that}
the observed velocity separation of PV peaks, $\Delta V$, agrees with the predicted velocity separation $\Delta V_{\rm th}$. \rev{In contrast,} the spatial separation $\Delta r$ is smaller than the predicted spatial separation $\Delta r_{\rm th}$
by roughly a factor 3. As a result, the value of \jobs\ inferred from the double peak separation with Eq.~(\ref{eq:jobs}) underestimates the true \jout\ \rev{(see Eq.~(\ref{eq:jout}))} also by a factor 3, which is quite significant. 

\rev{To show that this bias is generic to the method, and how} it depends on each free parameter,
{we compare} \SC{in Fig.~\ref{fig:peaks}} 
the measured $\Delta r$, $\Delta V$, \jobs\ on PV cuts to the true values of $\Delta r_{\rm th}$, $\Delta V_{\rm th}$, \jout\ \SCC{for a series of \rev{edge-on PV} models that differ from the reference case by only one free parameter at a time.} \rev{In addition,} \SC{the detailed} effect of parameter changes on \SCC{the shape of} \SC{on-axis spectra and} PV diagrams is illustrated in Fig.~\ref{pv-perp-vs-r0max} for selected  \SC{pairs of} models.

\rev{Based on Fig.~\ref{fig:peaks}}
we find that 
\rev{\jobs\ always underestimates \jout, and that} this systematic bias is essentially due to the peak spatial separation $\Delta r$ being always much smaller than 
the predicted value $\Delta r_{\rm th}$ for the outermost streamline. 
In contrast, the velocity separation $\Delta V$ always remains close to the predicted $\Delta V_{\rm th}$
\rev{(except when $i$ approaches $i_{\rm crit}$, where $\Delta V$ drops).}

\rev{A striking result is that this bias does not improve at higher spatial resolution (Fig.~\ref{fig:peaks}d). 
It does not vary much either with position of the PV cut (Fig.~\ref{fig:peaks}d),} 
or magnetic lever arm and widening of the MHD solution (Fig.~\ref{fig:peaks}a).
In contrast, the \rev{underestimate clearly} worsens with increasing radial extension $r_{\rm out}/r_{\rm in}$ of the MHD DW \rev{(Fig.~\ref{fig:peaks}b).}
and with the slope of the radial emissivity gradient,\rev{controlled by the power-law index $\alpha$ (Fig.~\ref{fig:peaks}c).}

From this behavior, we conclude that the underestimate of \rev{$\Delta r$} is 
\rev{a contrast effect} due to the contribution of \rev{bright} nested streamlines interior to $r_{\rm out}$,
\rev{projected at low-velocity by the quasi edge-on inclination.} 
\rev{As an example, the two spectra in Fig.~\ref{pv-perp-vs-r0max}b 
show that, at the velocities of the PV peaks, inner streamlines launched within $r_0\le$ 1~au (blue curve)
contribute about $30$\% of the total line intensity integrated up to $r_{\rm out}$ = 8 au (red curve).} {This contribution of inner streamlines drags} the spatial centroids of the PV peaks closer to the axis than if emission came only from a narrow ring on the outermost streamline. The peak spatial separation $\Delta r$ is \rev{thus reduced compared to} the theoretical value $\Delta r_{\rm th}$. When the radial extension of the MHD DW grows, or when the radial gradient of emissivity steepens, the relative flux contribution of inner vs. outer streamlines automatically increases and the reduction in $\Delta r$ is more severe, reaching up to a factor 3--10 in Figs.~\ref{fig:peaks}b,c. 
%\SCC{Fig.~\ref{fig:peaks} shows that 
\rev{Of course, an even larger bias would result if both effects} (a large radial extension $\simeq 100$ and a steep emissivity gradient $\alpha = -3$) conspired together.

\subsubsection{Bias in the outer launch radius, \rout}
\label{sec:robs}

%%%%%%%%%%%%%%%%%%%%%%%%%%
%%%%%%%%%%%%%%%%%%%%%%%%%%
\begin{figure*}
\centering
\includegraphics[width=0.7\textwidth]{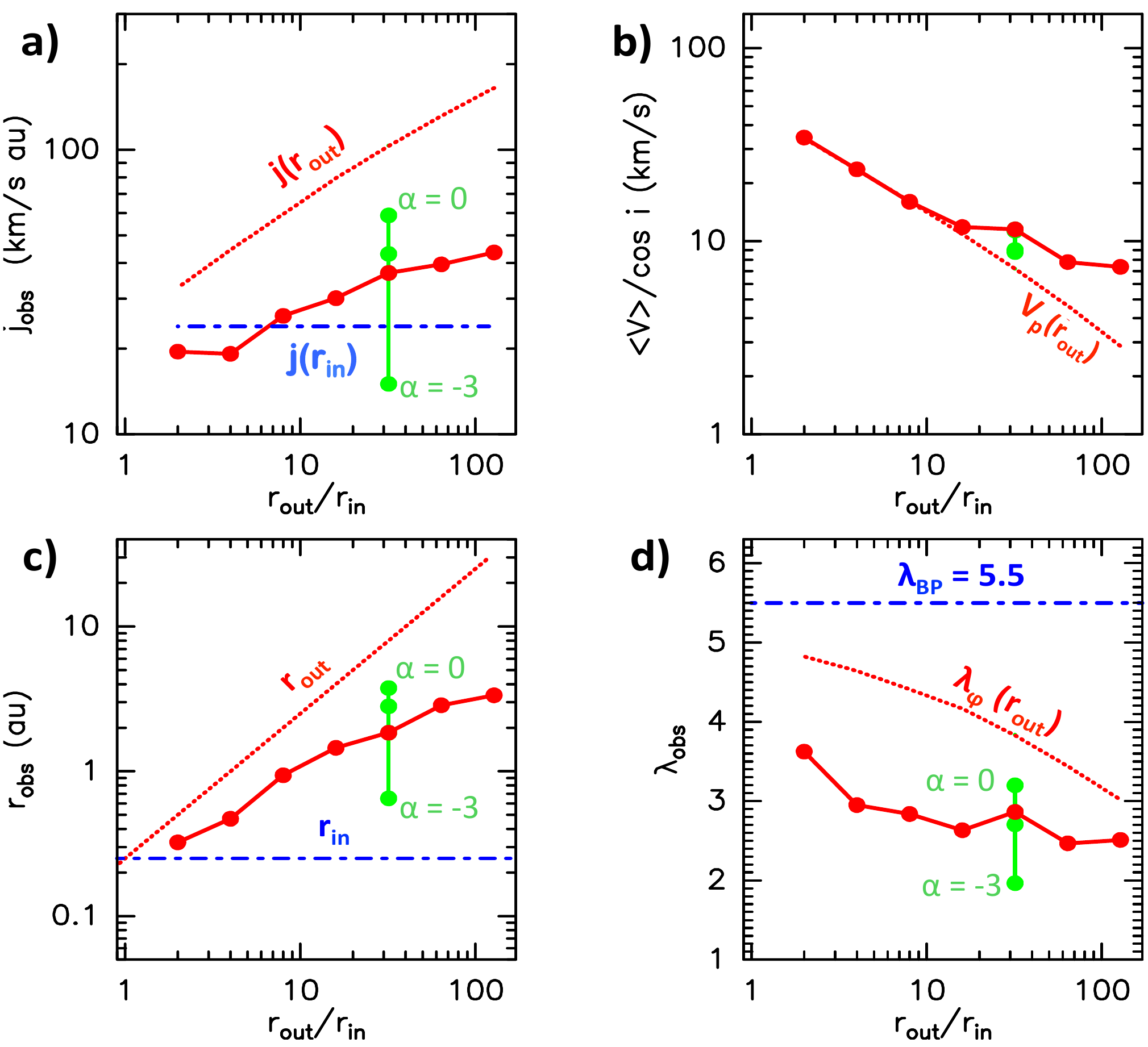}
\caption{ 
\rev{Observational biases in the PV double-peak separation method for our reference edge-on case ($i = 87\degr$), as a function of the DW radial extension (for $\alpha = -2$, connected red dots) and emissivity index $\alpha$ (for \rout/\rin = 32, connected green dots): \textbf{a,b}: specific angular momentum and poloidal velocity 
estimated from the PV double-peak separation; \textbf{c,d}; launch radius and magnetic lever arm parameter inferred from them using Anderson's relations (Eqs.~(\ref{eq:anderson}), \ref{eq:lambda-anderson}). Values that would be obtained for a single ring on the outermost wind streamline are shown in dotted red curves. The true magnetic lever arm parameter \lbp\ of the MHD solution and relevant values at \rin\ are indicated for reference in dot-dashed blue.}}
\label{fig:lambda-eff}
\end{figure*}  
%%%%%%%%%%%%%%%%%%%%%%%%%%
%%%%%%%%%%%%%%%%%%%%%%%%%% 

Since errors in \jobs\ \rev{in the edge-on configuration} do not depend much on the specific MHD solution, inclination, beam size, or position of PV cut (see  Sect.~\ref{sec:jobs}) we focus in the following on our reference \rev{edge-on model} and vary only the wind radial extension $(r_{\rm out}/r_{\rm in})$ \rev{from 2 to 130 (with $\alpha$ fixed at -2) or the emissivity index $\alpha$ from 0 to -3 (with \rout\ fixed at 8 au).}

Fig.~\ref{fig:lambda-eff}a plots the absolute value of  \jobs\ for \rev{this restricted set of models}, as a function of the radial extension. 
Fig.~\ref{fig:lambda-eff}b plots the "observed" poloidal velocity \Vobs, estimated by deprojecting the average line of sight velocity $<V>$ of the two \rev{PV peaks} 
\begin{equation}
V_{\rm p, obs} = {<V>}/{\cos{i}}.
\label{eq:Vobs}
\end{equation}

We see that \Vobs\ is close to the true $V_p(r_{\rm out})$ up to $(r_{\rm out}/r_{\rm in}) \simeq 20$, \SC{and progressively overestimates it for a more extended wind}. 

Figure~\ref{fig:lambda-eff}c plots the values of launch radii \robs\ obtained by solving Anderson's relation in Eq.~(\ref{eq:anderson})
with $r V_{\phi}$= \jobs, $V \simeq$  \Vobs\ \rev{($V_\phi$ is negligible here), and $R  \gg r_0$ (largely fulfilled at \zcut = 225 au)}.

\rev{As one might have expected, we find that \robs\ takes a value intermediate between \rin\ and \rout. It thus}
always underestimates the true outermost launching radius of the emitting disk wind. In addition, this bias worsens with the MHD DW radial extension. \rev{In our reference model, the error reaches a factor 10 for \rout = 32 au, which is a very significant effect.}

Figure~\ref{fig:lambda-eff}c also shows that for our reference emissivity index $\alpha = -2$, %used here, 
\robs\ grows roughly as the geometrical average of the innermost and outermost launch radii. \rev{One might then think of recovering the true \rout\ value as
\begin{equation}
r_{\rm corr} \simeq r_{\rm obs}^2/r_{\rm in}.
\label{eq:robs-geom}
\end{equation}}
\rev{However, the \rev{geometrical} average only holds when $\alpha = -2$.} For a steeper emissivity gradient ($\alpha = -3$) \robs\ is  closer to \rin, 
while for a shallower gradient ($\alpha = -1,0$) \robs\ is closer to \rout (see green dots in Fig.~\ref{fig:lambda-eff}c).
\rev{Since Equation~\ref{eq:robs-geom} is quadratic in \robs, an $\alpha$ value differing from -2 could introduce a large error in \rcorr\ (factor 4--9 at \rout = 8 au, cf. green dots Fig.~\ref{fig:lambda-eff}c).} 
\rev{Another problem would the relevant value of \rin\ to use. Although we fixed it for simplicity at the dust sublimation radius $\simeq 0.25$ au in this Section, \rin\ in actual disk winds will depend on the chosen chemical tracer and wind density: it could move well outside to $\ge 1$ au in evolved disk winds where FUV photodissociation is important \citep{2012A&A...538A...2P,2016A&A...585A..74Y} or well inside to \rin\ $\simeq$ 0.05--0.1 au if inner streamlines are dense enough for efficient dust-poor chemistry, as recently suggested in the dense Class 0 flow of HH212 by model fits to PV cuts \citep{2017A&A...607L...6T}. This introduces an additional uncertainty of a factor 4 either way in Eq. (\ref{eq:robs-geom}).}

\rev{We conclude that when the MHD DW is radially extended and viewed close to edge-on (ie with PV double-peaks of opposite signs), 
the launch radius \robs\ inferred from the double-peak separation using Anderson's relation only gives a {lower limit} to the true \rout. This bias cannot be accurately corrected for without additional constraints on \rin\ and the radial emissivity gradient ($\alpha$).}

\subsubsection{Bias in magnetic lever arm}
\label{sec:lambda-obs}

\rev{Figure~\ref{fig:lambda-eff}d plots the "observed" wind magnetic lever arm parameter \lobs\ inferred from the values of \jobs\ and \robs\ in Fig.~\ref{fig:lambda-eff}a,c 
following Anderson's method \citep[see eg. ][]{2003ApJ...590L.107A}:
\begin{equation}
\lambda_{\rm obs} \equiv j_{\rm obs} / \sqrt{G M_\star r_{\rm obs}}. %= j_{\rm obs} / \left({GM_\star r_{\rm obs}}\right)^{1/2}.
\label{eq:lambda_obs}
\end{equation}
For comparison, we also plot (dotted curve) $\lambda_{\phi}$(\rout), the equivalent physical quantity on the outermost streamline that would be obtained in the case of no observational bias (i.e. for a single emitting ring): 
\begin{equation}
\lambda_{\phi}(r_{\rm out})  \equiv j_{\rm out} / \sqrt{G M_\star r_{\rm out}}.
%\lambda_{\phi}(r_{\rm out})  \equiv j_{\rm out} / [r_{\rm out} V_K(r_{\rm out})] 
\label{eq:lambda_phi}
\end{equation}
Fig.~\ref{fig:lambda-eff}d shows that \lobs\ always underestimates $\lambda_{\phi}$(\rout); however, this observational bias is very mild (-20\% for $\alpha = -2$, a factor 2 for $\alpha$ =-3) and independent of the wind radial extension.}

\REVbis{This fortunate result is not a coincidence: expressing $r V_\phi$ as $\lambda_\phi \sqrt {GM_\star r_0}$ in Anderson's relation Eq.~(\ref{eq:anderson}), we see that once gravitational potential has become negligible ($R \gg r_0$), the total velocity modulus must verify\footnote{This expression is similar to Eq.~(\ref{eq:vp}) except that it involves the total velocity modulus instead of the asymptotic poloidal velocity $V_p^\infty$, and the local $\lambda_{\phi}$ instead of \lbp.}  
\begin{equation}
V = \sqrt{2 \lambda_\phi -3} \times \sqrt{G M_\star/r_0}.
\label{eq:lambda_p}
\end{equation}
Noting that our models have $V \simeq V_p$ at large distance, we obtain the following useful relation, where launch radius cancels out \citep[see Eq.~(10) in][]{2006A&A...453..785F}:
\begin{equation}
\lambda_{\phi} \sqrt{2 \lambda_\phi -3} = \frac{j_{\rm out} V_{\rm p}}{G M_\star},
\label{eq:lambda-bis}
\end{equation}
Anderson's relation for $r_0$ = \robs\ imposes the same relation between $\lambda_{\rm obs}$, \jobs, and $V_{\rm p, obs}$. For moderate magnetic lever arms $\lambda \lesssim 6$ as considered here, the function on the left-hand side of Eq.~(\ref{eq:lambda-bis}) is very steep; hence even if \jobs\ underestimates \jout\ by a large factor (Figure \ref{fig:lambda-eff}a), the bias in the inferred $\lambda_{\rm obs}$ is much smaller (Fig. \ref{fig:lambda-eff}d).}

We also observe a theoretical "MHD bias" in that  $\lambda_{\phi}$(\rout) is always smaller than the true \lbp\ in the solution. \rev{As first pointed out by \citet{2006A&A...453..785F}, this bias arises because $\lambda_{\phi}$} only measures the specific angular momentum {in the form of matter rotation}, whereas 
the total (conserved) specific angular momentum $L$ carried by the MHD DW streamline (and measured by
\lbp) also includes a contribution of magnetic field torsion.   
\rev{The dotted curve in Fig.~\ref{fig:lambda-eff}d (constructed at
\zcut = 225 au) shows that in our reference solution, $\lambda_{\phi}$ reaches
90\% of \lbp\  when $z / r_0$ = 550,
70\% when $z / r_0 \simeq 20$, and only
50\% when $z / r_0 \simeq 7$.}

\rev{In conclusion, we find that the magnetic lever arm parameter inferred with Anderson's method only gives a lower limit to the true \lbp. This is mainly caused by an MHD bias (hidden angular momentum in magnetic form), with only a minor observational bias \REVbis{for low \lbp}. \lbp\ can only be accurately estimated at high altitudes (\zcut $\ge 20$\rout\ for our self-similar models), or by modeling in detail the whole PV cut with a self-consistent MHD DW solution (see eg. Section 4 for the example of HH212).} 

%%%%% beginning of added sections %%%% 
\subsection{Rotation curve method: biases in launch radius and magnetic lever arm}
\label{sec:single}

\begin{figure}
   \centering
           \includegraphics[width=0.46\textwidth]{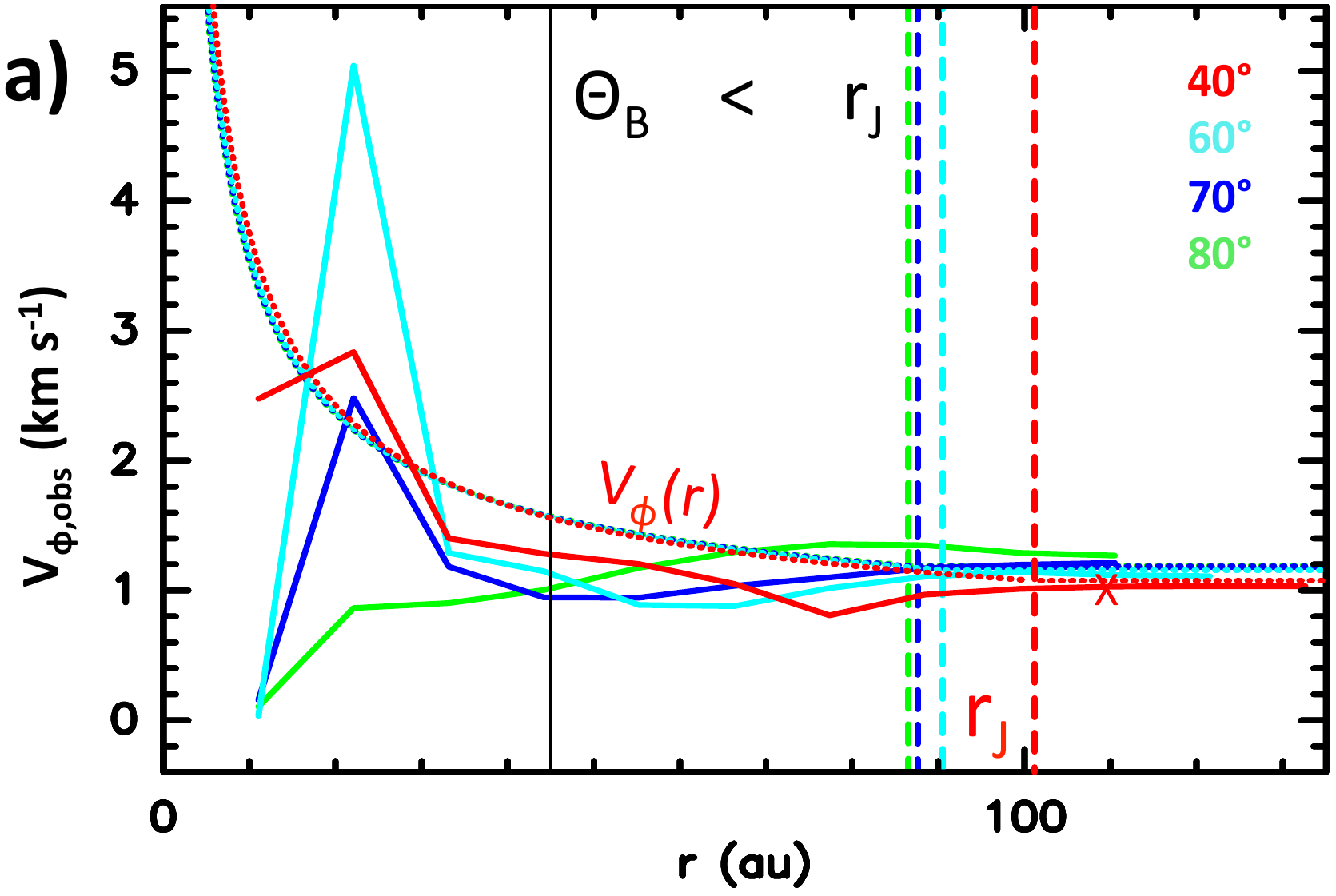}  
    \includegraphics[width=0.47\textwidth]{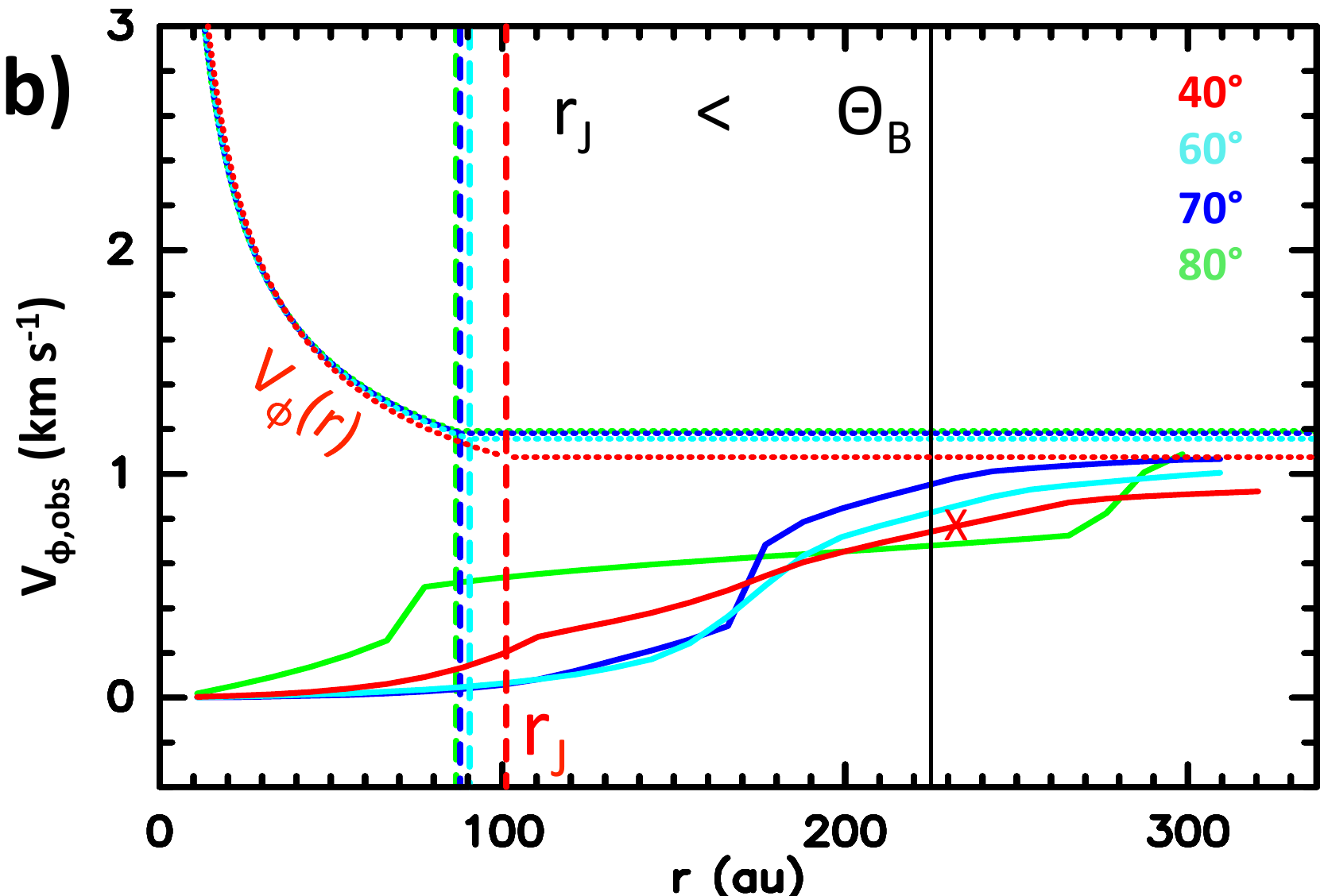}
      \caption{\rev{"Observed" rotation curve $V_{\phi,\rm obs}(r)$  obtained from the difference of peak velocity between $\pm r$ from the jet axis (Eq. \ref{eq:vrot}) for our reference model viewed at $i$ = 40\degr, 60\degr, 70\degr, 80\degr (colour-coded curves). \textbf{a:} spatially resolved regime ($\theta_b = 45$ au), \textbf{b:} spatially unresolved regime ($\theta_b$ = 225 au, note the change of scales on both axes). In both panels, a vertical black line indicates the beam FWHM, and coloured vertical dashed lines indicate the flow radius $r_j$ at \zcut = 225~au. The true wind rotation curve $V_\phi(r)$ at $z = z_{\rm cut}/\sin{i}$) is shown by dotted curves, with a constant value beyond $r_j$.  
      The cross on the red curve illustrates where the local intensity drops below 10\% of the PV peak.
     }}
      %Parameters of both models are recalled at the top right.}
         \label{fig:vshift-incl}
   \end{figure}

\rev{For consistency, we consider the same reference MHD DW parameters 
and $z_{cut} = 225$~au as in the previous section.}
\rev{We find that at moderate inclinations $i <$ \icrit, the observed rotation curves from velocity shifts (Eq.~( \ref{eq:vrot}))
depend strongly on whether the wind is laterally resolved or unresolved.
We present in Figure \ref{fig:vshift-incl}a the curves for $i$ = 40\degr\ to 80\degr 
with $\theta_b$ = 45 au $<r_j$
illustrating the well-resolved regime, and in Figure \ref{fig:vshift-incl}b the curves
with $\theta_b$ = 225 au $> r_j$, illustrating the unresolved regime.
Results for more edge-on inclinations, which are independent of beam size, will be discussed at the end of this Section.}

 %%%%%%%%%%%%%%%%%%%%%%%%%%%%%%%
%%%%%%%%%%%%%%%%%%%%%%%%%%%%%%%
%%%%%%%%%%%%%%%%%%%%%
%%%%%%%%%%%%%%%%%%%%%

\rev{Figure~ \ref{fig:vshift-incl}a shows that in the well-resolved flow regime, 
$V_{\phi,\rm obs}(r)$ follows the underlying true rotation curve $V_\phi(r)$ within a factor 2, 
until $r \le \theta_b /2$ where it falls sharply below it due to beam smearing.}
\rev{In contrast, Figure~\ref{fig:vshift-incl}b shows that in the unresolved regime, $V_{\phi,\rm obs}(r)$ 
does not follow the keplerian decline but instead increases slowly with radius. At large radii 
where emission has dropped to 10\% of the PV peak, $V_{\phi,\rm obs}(r)$ reaches about 60\%--80\% of the true rotation speed on the outermost streamline.}

 %%%%%%%%%%%%%%%%%%%%%%%%%%%%%%%
%%%%%%%%%%%%%%%%%%%%%%%%%%%%%%%
   \begin{figure*}
   \centering 
    \includegraphics[width=0.99\textwidth]{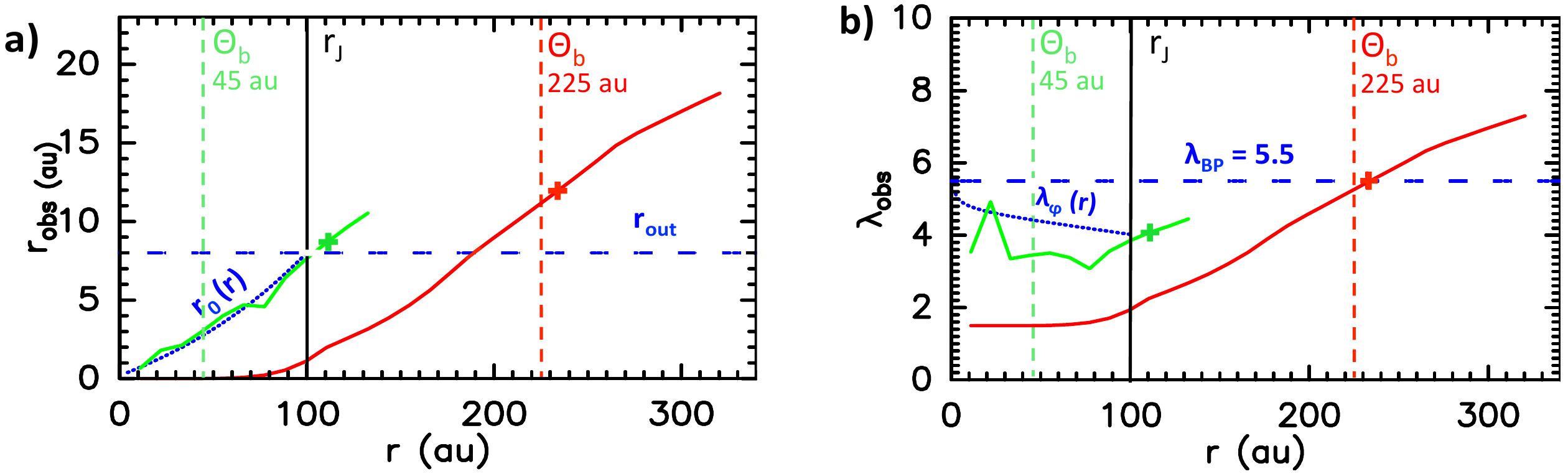}
      \caption{\rev{Examples of biases using the rotation curve method, for a representative model at $i = 40\degr$ (in red in Figure~\ref{fig:vshift-incl}): "Observed" launch radii \robs$(r)$ (\textbf{a}) and magnetic lever arms \lobs$(r)$ (\textbf{b}) inferred at each $r$ by application of Anderson's formula to \jobs$(r)$ and $V_{\rm p,obs}(r)$ (Eq.~( \ref{eq:jobsr2}), (\ref{eq:vobsr})). Green curves illustrate the spatially resolved regime ($\theta_b$ = 45 au) and red curves the unresolved regime ($\theta_b$ = 225 au).  They stop at 1\% of the PV peak intensity, with a $+$ symbol marking 10\%. The black vertical line shows the flow radius $r_j$. For comparison, dotted blue curves plot the true launch radii $r_0(r)$ (\textbf{a}) and $\lambda_\phi(r)$ (\textbf{b}) of each DW streamline tangent to $r$;  horizontal dashed blue lines mark the true outermost launch radius \rout\ = 8 au (\textbf{a}) and true \lbp = 5.5 (\textbf{b}) in the model. Other model parameters are : MHD solution L5W17, $\alpha = -2$, $M_\star=0.1M_{\odot}$, \rin = 0.25~au.}}
         \label{fig:robs-incl}
   \end{figure*}  
%%%%%%%%%%%%%%%%%%%%%
%%%%%%%%%%%%%%%%%%%%%

\rev{We note that rotation curves in Fig. \ref{fig:vshift-incl} exhibit little change with inclination, except when $i = 80\degr$ where they flatten out to become almost independent of radius, as we approach $i_{\rm crit}$ ($\simeq$ 84\degr\ for the L5W17 reference solution). Similar results were found for \rout\ = 8 au and 32~au.
In the following, we thus take $i$ = 40\degr and \rout = 8 au (red curves in Fig. \ref{fig:vshift-incl}) as our representative 
model for moderate inclinations.}

\rev{In Figure~\ref{fig:robs-incl}a, we plot for this representative model the DW launch radii
\robs$(r)$ inferred by applying Anderson's formula to the observed local specific angular momentum at each $r$
\begin{equation}
j_{\rm obs}(r) = r \times V_{\phi,\rm obs}(r),
\label{eq:jobsr2}
\end{equation}
using the corresponding observed local poloidal velocity 
\begin{equation}
V_{\rm p, obs}(r) = \left[ V(r)+V(-r) \right] / 2\cos{i}.
\label{eq:vobsr}
\end{equation}
Similarly, in Figure~\ref{fig:robs-incl}b, we plot as a function of $r$ the magnetic lever arm parameters \lobs$(r)$ inferred from \robs$(r)$ and \jobs$(r)$ using Eq.~(\ref{eq:lambda_obs}).} 

\rev{In the spatially resolved regime (green curves), 
the method performs very well, 
with little observational bias. In particular, the results at 10\% intensity level give quite accurate values of \rout\ and of $\lambda_\phi$(\rout) (defined in Eq.\ref{eq:lambda_phi}).} 

\rev{In the unresolved regime (red curves), 
\robs$(r)$ and \lobs$(r)$ suffer complex observational biases: 
They take artificially small values at radii $r \le r_j$ 
(where rotation speeds are strongly underestimated by beam smearing)
and overshoot the true \rout\ and $\lambda_\phi$(\rout) at large radii. This overshoot is caused by the
beam smearing artificially enlarging $r$ well beyond the true $r_j$, so 
that \jobs$(r)$ at 10\% intensity level exceeds \jout. This bias will of course worsen with increasing $(\theta_b/r_j)$, and provides upper limits to the true \rout\ and $\lambda_\phi$(\rout).}

\rev{Finally, we discuss the rotation curve method in the quasi edge-on case: As shown by outer contours of PV cuts in Fig.\ref{pv-perp-vs-r0max}, velocity shifts between $\pm r$ in that case are essentially constant with radius and close to $\Delta V$, the velocity separation between the PV double-peaks. The latter was found to be close to $\Delta V_{\rm th}$ (see Eq.~(\ref{eq:deltaVth}) and Fig. \ref{fig:peaks}). It follows that the (constant) value of $V_{\phi,\rm obs}(r)$ will be close to $V_\perp = \sqrt{V_\phi^2+V_r^2}$ ie. slightly larger than $V_\phi$ on the outer streamline. The inferred \jobs\ at 10\% intensity radius (where $r \ge r_j$) will thus again overestimate the true \jout\ (by an amount depending on beam smearing) and provide upper limits to \rout\ and $\lambda_\phi$(\rout).}

\subsection{"Flow width" method: biases in launch radius and magnetic lever arm}
\label{sec:flowwidth}
\rev{Here, \jobs\ is obtained from the asymptotic rotation speed at large radii and the deconvolved flow radius as
\begin{equation}
\label{eq:jmax2}
j_{\rm obs} = r_{\rm w} \times V_{\phi,\rm obs}(r^{\rm \infty}).
\end{equation}
}
\rev{To estimate \rw, observers typically measure the FWHM of the beam-convolved velocity-integrated map at \zcut\  and correct in quadrature for the gaussian beam broadening to yield an intrinsic wind FWHM, which is then assumed equal to the wind diameter so that:
\begin{equation}
r_{\rm w} = \frac{1}{2} \times \sqrt{{\rm FWHM}_{\rm obs}^2 - \theta_b^2}.
\label{eq:rw}
\end{equation}
We performed this measurement on the synthetic integrated emission maps for our reference model at $i = 40\degr$. 
%\rout = 8 au, $i = 40\degr$, $\alpha = -2$, at \zcut = 225 au; 
With $\theta_b$ = 225 au, we find \rw\ = 27 au, a factor 4 smaller than the true outer flow radius $r_j$ at that position, 
With a smaller $\theta_b$ = 45 au that fully resolves the flow across, \rw\ is almost unchanged at 20 au. Hence the fact that
\rw $\ll r_j$ is not a beam smearing effect. 
It occurs because the FWHM in emission maps is dominated by the central spine of inner bright streamlines, and does not encompass the fainter pedestal tracing the outermost streamlines. The wind thus appears much narrower than it really is ("optical illusion" effect).} %{Shang98,Cabrit99}.

\rev{It is significant that the deconvolved flow diameter 2\rw\ is of the same order as the double-peak spatial separation $\Delta r$ for the same model viewed at $i =87\degr$ (see Fig.\ref{fig:peaks}), and that both are independent of beam size. Indeed, their strong reduction compared to the true flow width has the same root cause, namely the brightness contrast between inner and outer streamlines.}

\rev{Fig. \ref{fig:vshift-incl} shows that the asymptotic rotation velocity at 10\% intensity level, $V_{\phi,{\rm obs}}(r^\infty) \simeq 0.8-1$ \kms, is also not strongly affected by beam smearing. It is relatively unaffected by inclination as well, and close to the true rotation speed on the outer streamline.}

\rev{Using Equation \ref{eq:jmax2} we thus obtain with this method 
\jobs $\simeq$ 20-22 au \kms\ for $\theta_b$ = 45-225 au.
Combining with the observed \Vobs $\simeq$ 7-8 \kms\ at 10\% intensity, we obtain with Anderson's method \robs $\simeq 1.5$ au and \lobs $\simeq 2$. These values are slightly smaller but very close to what we obtained with the double-peaked method in the same model viewed edge-on (see Fig.\ref{fig:lambda-eff} with \rout/\rin = 32). This is not surprising, since we saw that \rw\ is close to $\Delta r/2$ while $V_{\phi,{\rm obs}}(r^\infty)$ is close to $V_\phi$, which is itself slightly smaller than $V_\perp = (V_\phi^2+V_r^2)^{1/2} \simeq \Delta r/2 \sin{i}$ (see Fig. \ref{fig:peaks} and associated discussion in Section \ref{sec:jobs}).}

\rev{We conclude that the "flow width" method will underestimate \jout\ by a similar amount as the double-peak separation method in the same flow viewed edge-on, and also yield strict lower limits to the true \rout\ and \lbp.}

\section{Application to the edge-on rotating flow in HH212}
\label{sec:HH212}

\rev{The edge-on flow HH212 (viewed at $i = 87\degr$) exhibits a slow and wide rotating SO outflow first identified by \citet{2017A&A...607L...6T} as a possible MHD disk wind candidate. In this section, we use 3 sets of ALMA observations of HH212 spanning a factor 15 in angular resolution to verify our main results on the double-peak separation in edge-on PV cuts (Section \ref{sec:doublepeak}) and to test the DW model of 
\citet{2017A&A...607L...6T} down to $\simeq18$ au resolution \citep{2018ApJ...856...14L}. We also derive an exact formula for the fraction of disk angular momentum extraction performed by a self-similar MHD DW, and apply it to HH212.}

For consistency with our previous modeling work in \citet{2017A&A...607L...6T}, we adopt for HH212 \MV{a systemic velocity of $V_{\rm sys}$ = 1.7 \kms  \citep{2014ApJ...786..114L} and a distance $d$ = 450 pc}. Recent VLBI parallax measurements towards stellar members of 
\rev{the Orion B complex yield mean distances of $388\pm 10$ pc for NGC~2068  and $423\pm 15$ pc for NGC~2024 \citep{2017ApJ...834..142K}, suggesting a}
possibly closer distance $\simeq 400$ pc to HH212, located in projection between these two regions. However, \rev{the exact distance to HH212 remains uncertain; adopting 400 pc instead of 450 pc would decrease linear dimensions and mass-outflow rates by 10\%, and the mass-accretion rate by 20\%, without} altering our conclusions.

\subsection{Effect of angular resolution on apparent rotation signatures in HH212}
\label{sec:jobs-HH}

%%%%%%%%:%%%%%%%%%
 \begin{figure}
   \centering
   \includegraphics[width=0.25\textwidth]{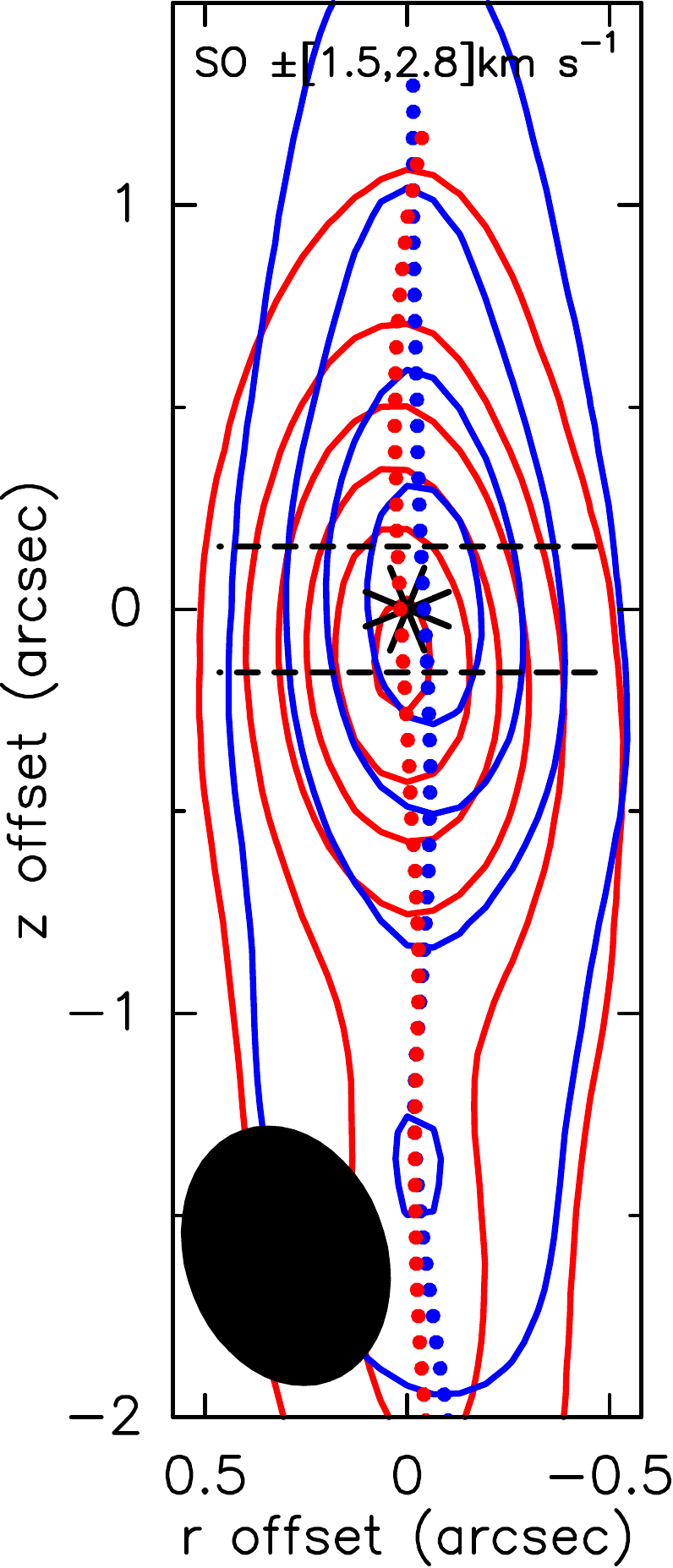} 
      \caption{\rev{Rotation signatures retrieved at 225~au resolution by spectro-astrometry towards the low-velocity HH212 outflow. Blue/red contours show SO($9_8-8_7$) emission} 
     at intermediate velocity ($1.5 $km s$^{-1}<  \mid V_{LSR}-V_{sys} \mid < 2.8 $km s$^{-1}$) mapped with ALMA Cycle 0 \citep[from][]{2015A&A...581A..85P}. 
     The blue/red \SC{dots mark} the centroid positions of the blue/red transverse intensity cuts at each altitude. The black asterisk indicates the continuum peak. 
     The jet was rotated to point upwards for clarity.      
     Horizontal black dashed lines depict the position of PV cuts \SC{at $z \pm 70$au} shown \SC{\rm in} Fig.~\ref{fig:model-70au}. 
      The clean beam FWHM of 0.65'' $\times$ 0.47'' is shown as a filled black ellipse. 
      First contours are $0.1$mJy/beam~\kms and steps are  $0.15$mJy/beam~\kms.}
         \label{channel-map-so-cs}
   \end{figure}
%%%%%%%%%%%%%%%% 

\rev{Here, we first verify on HH212 our most counter-intuitive theoretical prediction for a quasi edge-on MHD DW, namely: that the spatial shift $\Delta r$ between the redshifted and blueshifted PV emission peaks does not depend on beam size (see Section \ref{sec:jobs}).}

Figure \ref{channel-map-so-cs} shows SO %and C$^{34}$S 
blue/red channel maps \SC{of the base of the HH212 flow} obtained at $0.55" \simeq$ 250 au resolution in ALMA Cycle 0 by \citet{2015A&A...581A..85P}. \SC{They are integrated over the} intermediate velocity range $1.5 $km s$^{-1}<  \mid V_{LSR}-V_{sys} \mid < 2.8 $km s$^{-1}$, where \citet{2017A&A...607L...6T} found a clear blue/red \SC{transverse spatial shift at higher resolution of 0.15'' $\simeq 65$ au (see their Fig.~\ref{velocities-MHD}b)}.
\SC{Although no obvious shift between blue and red contours is apparent at first sight in Fig.~\ref{channel-map-so-cs}, 
the signal to noise ratio of these data is high enough that 
spatial shifts much smaller than the beam can still be detected by comparing centroid positions (the so-called ``spectro-astrometry" technique).}
\SC{At each distance $z$ along the jet axis, a transverse intensity cut is constructed across the blue and the red channel maps and the spatial centroid  measured in each cut. They are plotted in Fig.~\ref{channel-map-so-cs} as blue/red dots, respectively.} A \SC{small but significant and consistent} transverse position shift  between redshifted and blueshifted emission centroids is clearly detected, that persists out to $z \pm 0.7\arcsec$. The shift amplitude at $z \simeq 70$ au is $\Delta r \simeq 0.06"\pm 0.02"$
(27~au), \SC{in the sense of disk rotation}. The same shift is measured with this method in the higher resolution 0\farcs15 data of \citet{2017A&A...607L...6T}. 
Since the channel maps are separated by $\Delta V = 4$ \kms, the apparent specific angular momentum in both data sets is \jobs $=(\Delta r/2)(\Delta V/2)\simeq$ 27~au \kms. 

\SCC{The apparent specific angular momentum in the SO wind of HH212 was measured at yet higher angular resolution (0\farcs04) by \citet{2018ApJ...856...14L}. Their value $\simeq 30 \pm 15$ ~au~\kms\ remains remarkably similar to our results at 0\farcs55 and 0\farcs15. Hence, we verify over more than a decade in beam sizes that the apparent specific angular momentum
\rev{measured from the blue/red PV peaks separation in a quasi edge-on flow} does not depend on angular resolution, as predicted for 
an MHD DW (see Fig.~\ref{fig:peaks}d).}

\subsection{Best fitting \rout\ and MHD DW model vs. angular resolution} 

\rev{Using the apparent specific angular momentum $\simeq 30$ au \kms\ determined above, a mean deprojected poloidal speed $V_{\rm p,obs} \simeq$ 1 \kms $/\cos{i} \simeq$ 20 \kms, and $M_\star = 0.2 M_\odot$, Anderson's relation yields an estimated \robs $\simeq 1$ au.}

\rev{We show below that the true outer launch radius \rout\ of the HH212 SO wind is actually much larger than this, and close to the disk outer radius of 40 au in HH212, confirming our predicted bias that \robs $\ll$ \rout\ with the double-peak separation method (see Sect.~\ref{sec:robs}). We also show that the MHD DW model with \rout = 40 au initially proposed by \citet{2017A&A...607L...6T} remains consistent with PV cuts obtained at both 4 times lower and higher resolution.}

\begin{figure*} 
 \sidecaption
\includegraphics[width=.7\textwidth]{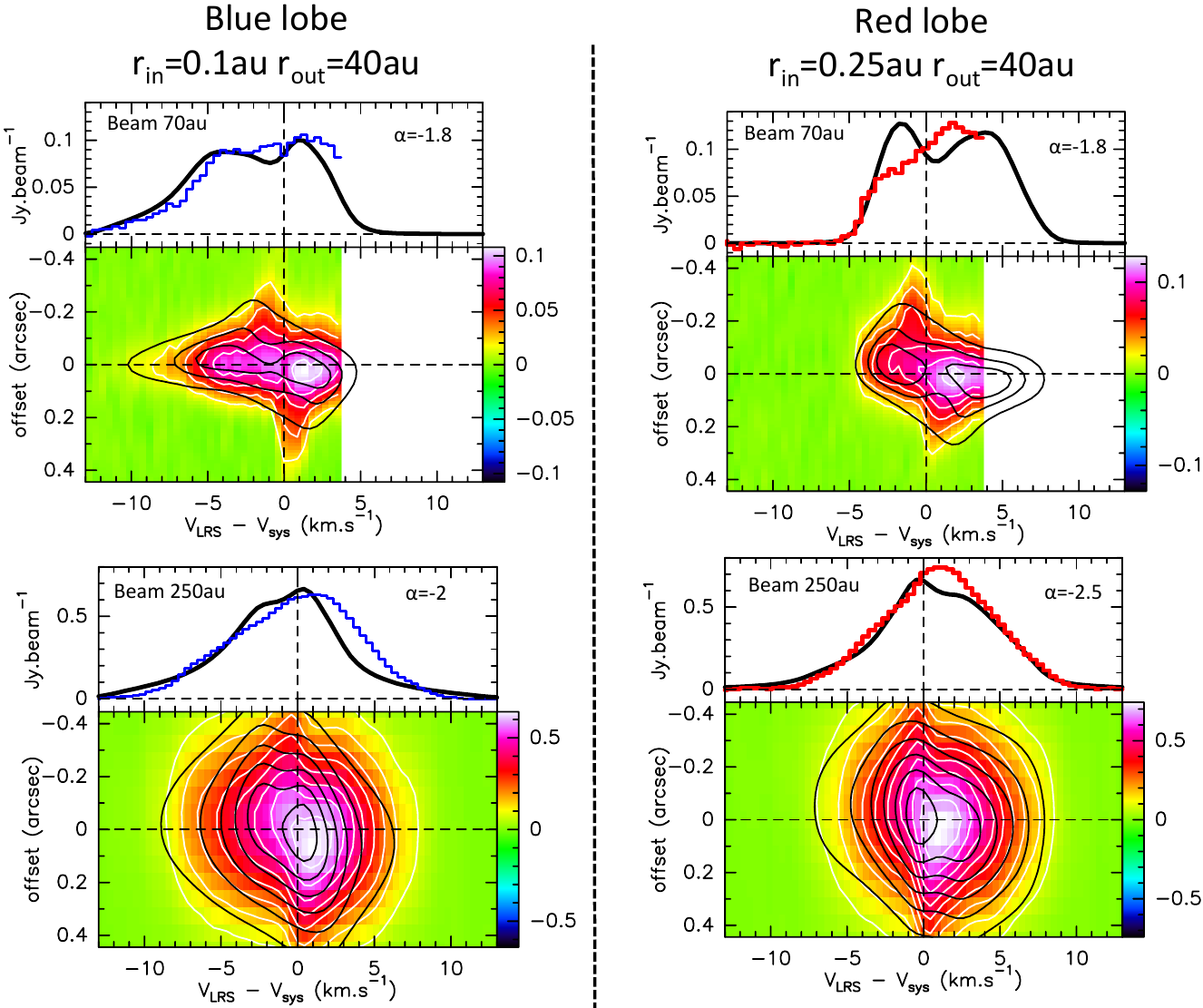}
      \caption{Comparison between \SC{observed and modeled SO} on-axis spectra and transverse position-velocity cuts taken at $\pm$ 0\farcs15 (70~au) across the blueshifted (left panels) and redshifted (right panels) lobes of the \SC{HH212} jet. \textit{Top row:} ALMA Cycle 4 data (70 au beam) from \citet{2017A&A...607L...6T}. \textit{Bottom row:} ALMA Cycle 0 data (250 au beam) from \citet{2015A&A...581A..85P}. In all panels, observed spectra are \SC{plotted as} histograms, and observed PV diagrams are pictured in color map with white contours. \SC{Synthetic predictions for the MHD DW model of \citet{2017A&A...607L...6T}, convolved by the appropriate beam size, are overplotted} in black. \SC{The model uses the MHD DW solution L5W30, $M_\star=0.2 M_{\odot}$, $i=87^{\circ}$, with the range of launch radii (\rin, \rout) indicated on top. The best fitting emissivity variation index ($\alpha$) is marked in each panel.}}
\label{fig:model-70au}
\end{figure*}

\MV{In Fig.~\ref{fig:model-70au}, we compare on-axis spectra and transverse PV diagrams of SO 
at the same $z_{\rm cut} = \pm 70$~au
\rev{for a resolution of $\simeq$ 70~au} \citep{2017A&A...607L...6T} 
and a 4 times larger beam $\simeq$ 250~au \citep{2015A&A...581A..85P}} 
\SC{(Note that we could not perform the same comparison in the SO$_2$ line, where the signal to noise in Cycle 0 was too low).}
\SC{The MHD DW model proposed by \citet{2017A&A...607L...6T}, convolved by the appropriate clean beam in each case, is superimposed in back contours.}
\SC{It was obtained with the MHD DW solution L5W30, \rout = 40 au, $M_\star=0.2 M_{\odot}$, $i=87^{\circ}$, and 
\rin = 0.1 au (blue lobe) or 0.25 au (red lobe).}

   %%%%%%%%%%%%%%%%

\SC{Fig.~\ref{fig:model-70au} shows that} the same model 
can also reproduce reasonably well the SO PV cut at a 4 times lower angular resolution, with just a slight change in radial emissivity gradient\footnote{\SCC{that could be easily produced e.g. by a slightly steeper
abundance or excitation gradient on larger scales.}} ($\alpha =$ -2 (blue lobe) or -2.5 (red lobe), instead of -1.8). 

\SC{In particular, the MHD DW model naturally explains i) the smaller peak velocity separation at lower angular resolution (cf. the drop of $\Delta V$ with beam size at $z_{cut} = 70$au visible in the green curves of Fig.~\ref{fig:peaks}d), ii) the more symmetric profile wings at lower resolution}
(in the model, this is caused by \rev{the larger beam encompassing} emission from closer to the disk surface and from the \SC{opposite} lobe). 
\MV{As a conclusion, observations at 70 au and 250 au resolution appear consistent with the same MHD DW model,}
\rev{and in particular the same large \rout\ value.} 

\rev{The agreement is of course not perfect in detail. Towards the red lobe, both datasets in Fig.~\ref{fig:model-70au} have their peak emission at redshifted velocities, while the models present a bluer peak. This is due to a global asymmetry in the HH212 SO outflow, in the sense that redshifted emission is systematically stronger than blueshifted emission in both lobes \citep[see e.g. PV cut along the flow in Figure 5 of][]{2018ApJ...856...14L}. Such behavior cannot be reproduced by an axisymmetric model like ours, where the brighter peak will necessarily switch sign between the two lobes. It could be explained by an ad-hoc non-axisymmetric emissivity distribution.}
\SC{When comparing with the 250~au resolution data, we also note that} \MV{the MHD DW model tends to predict slightly too large peak velocities further} \SC{than 0.2''} \MV{from the axis}.
This outer region might be associated with the limits of \SC{the self-similar model assumption} due to boundary effects, as discussed in \citet{2017A&A...607L...6T}. 
Alternatively, recent observations of complex organic molecules indicate temperatures $\simeq 150$~K near the disk outer edge \citep{2017ApJ...843...27L,2017A&A...606L...7B,2018A&A...617A..10C}, suggesting a sound speed \rev{in the disk atmosphere} reaching 30\% of the Keplerian speed at 40 au; hence \REVbis{"hot"} magneto-thermal DW solutions with a higher mass-loading and smaller magnetic lever arm \rev{and rotation speeds} \citep{2000A&A...361.1178C,2013ApJ...769...76B,2017A&A...600A..75B} might be more appropriate in these outermost \rev{wind} regions. Modeling such complex effects lies beyond the scope of the present paper and will be the subject of future work. 

\begin{figure*} 
\includegraphics[width=.98\textwidth]{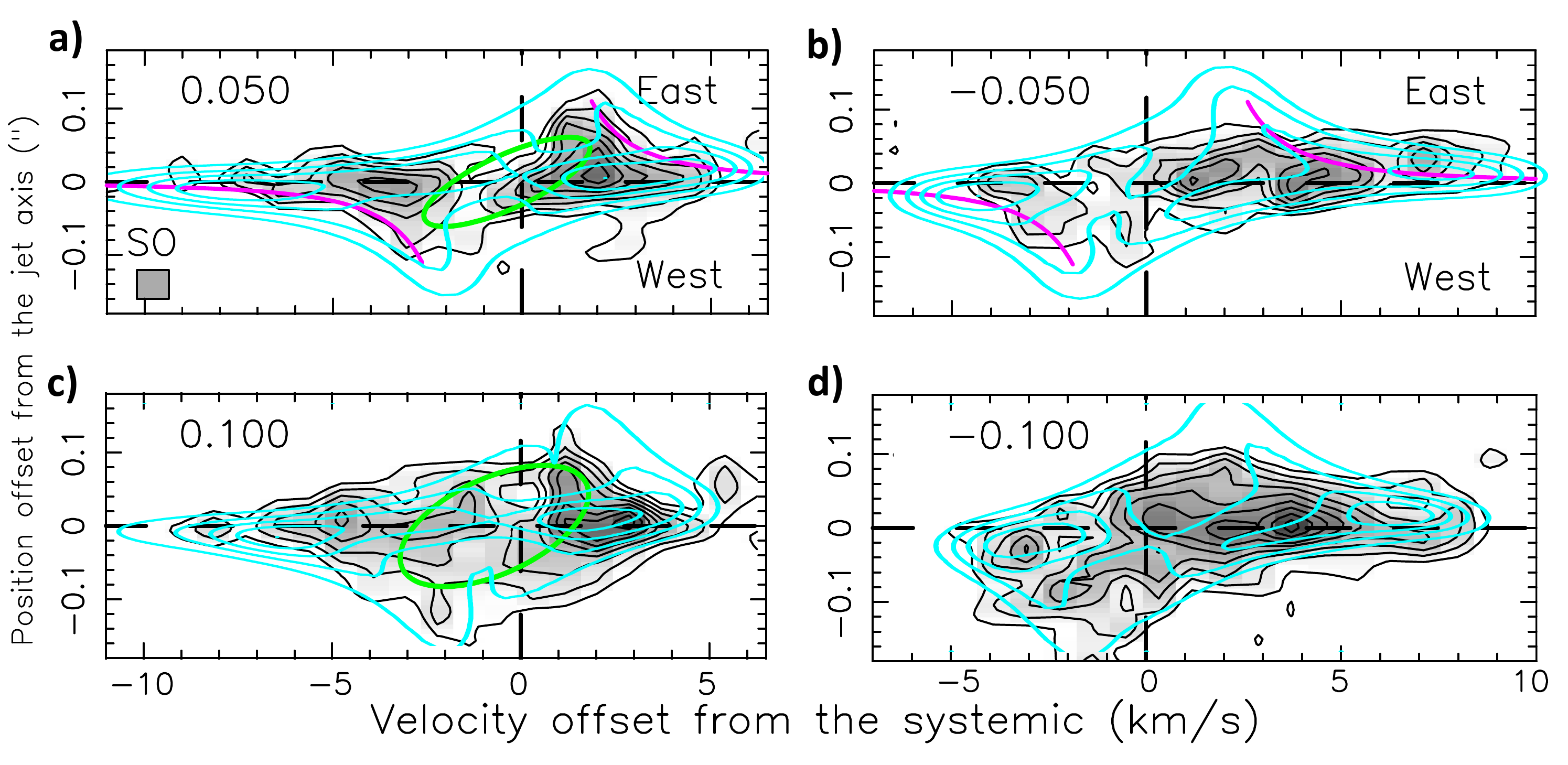}
      \caption{\rev{PV cuts at 0.04\arcsec $\simeq 18$ au resolution across the upper disk atmospheres in HH212: Black contours and greyscale show SO ALMA observations from \citet[][]{2018ApJ...856...14L}. Cyan contours show synthetic predictions for the MHD DW model of \citet[][]{2017A&A...607L...6T}. 
\zcut\ is labelled in arcsec in the upper left corner ($>0$ in the blue lobe, $<0$ in the red lobe).  
Magenta curves in \textbf{a,b} plot Keplerian curves fitted by Lee et al. to their data at $z \simeq \pm 20$ au; their blue/red velocity shift from systemic suggests an outflow from the disk atmosphere out to \rout $\simeq 0.1\arcsec \simeq 45$ au, well reproduced by the MHD DW model; green ellipses in \textbf{a},\textbf{c} show the expanding shell model fitted by Lee et al. in the blue lobe.}}
\label{fig:model-18au}
\end{figure*}

\rev{In Fig.~\ref{fig:model-18au}, we turn to smaller scales and compare the MHD DW model of \citet{2017A&A...607L...6T} with transverse PV cuts obtained by \citet{2018ApJ...856...14L} in the same SO line\footnote{\rev{The bright SO$_2$ line at 334.67335 GHz observed by \citet{2017A&A...607L...6T} was not covered by the spectral setup of  \citet{2018ApJ...856...14L}, who instead stacked 12 weak SO$_2$ lines; since stacking adds some uncertainty due to the limited spectral resolution, we focus here on the SO line common to the two studies.}} through the disk atmosphere at $z \le$ 45 au, with an unprecedented resolution of 0\farcs04 = 18 au.} A particularly noteworthy aspect is the global velocity shift observed between the two faces of the disk: Indeed, the Keplerian-like patterns fitted by \citet{2018ApJ...856...14L} \rev{at $z \simeq \pm 20$ au} (pink curves in Fig.~\ref{fig:model-18au}) are not centered on systemic velocity but shifted  \rev{globally} by $\simeq -0.3$ \kms\ \rev{to the blue} in the north (blue) lobe, and by $+0.3$ \kms\ \rev{to the red} in the south (red) lobe. This velocity shift implies that rotating disk layers probed by SO are not static but outflowing \rev{all the way out to \rout $\simeq 0.1'' \simeq 45$ au}, with a mean deprojected vertical velocity on each side $V_z \simeq 0.3 / \cos{(87\degr)} \simeq$ 6 \kms. \rev{This observation directly confirms, independently of any model, that the launch radius inferred with Anderson's relation from the PV double-peak separation (\robs $\simeq$ 1 au, see above) severely underestimates the true disk wind radial extent.}

Fig.~\ref{fig:model-18au} further shows that the MHD DW model proposed by
\citet{2017A&A...607L...6T} naturally reproduces not only the global velocity shift between the two faces of the disk, \rev{but also the overall envelope of the emission in the PV cuts at 18 resolution. The predicted regions of brightest emission (top two contour levels) also generally overlap quite well with the observed ones, although the agreement is again not perfect. The model sometimes extends to slightly higher blue velocities on axis than detected. The exact positions of emission peaks can also differ. However, observed maximum velocities and peak positions also have a component of uncertainty, due to the moderate signal-to-noise ratio and incomplete $u-v$ coverage at such high angular resolution. Moreover, our MHD DW model is probably too idealized (self-similar, steady, axisymmetric). Given these caveats, and the fact that the model was initially fitted on data at 4 times lower angular resolution (70 au), we consider the agreement to remain quite promising at this stage. }

\rev{Nevertheless, our proposed} interpretation in terms of MHD DW is not unique.
\citet{2018ApJ...856...14L} proposed an alternative model in terms of a thin swept-up shell driven by an unseen fast wide-angle X-wind \citep[see][and green ellipses in Fig.~\ref{fig:model-18au}]{2001ApJ...557..429L}.
\rev{Although this shell has a different velocity field (purely radial motion proportional to distance) than our MHD DW,} the large  
\rev{projection effect on} $V_z$
at $i = 87\degr$ make them difficult to distinguish \citep[cf. Fig.~11 in][]{2018ApJ...856...14L}. \rev{A hybrid scenario where an extended disk wind is shocked by inner jet bowshocks is also conceivable \citep[cf.][]{2018A&A...614A.119T}.} 
Indeed, wide bowshock wings are seen in HH212 in SiO at least down to $z \simeq$ 0\farcs5 $\simeq 200$~au \citep{2015A&A...581A..85P,2017NatAs...1E.152L}. 
\rev{Studies of less inclined MHD DW candidates will be crucial to constrain the $V_z$ component and discriminate between these options.}

\subsection{Role of the HH212 MHD DW candidate in disk accretion}

\rev{The disk wind streamlines used for our model PV cuts were obtained as part of a global MHD accretion-ejection solution where the wind extracts vertically most of the angular momentum flux required for steady disk accretion (see Appendix~\ref{appendixA}). However, similar emergent disk wind properties could be obtained with a dominant viscous torque in the disk, if the turbulent resistivity is highly non-isotropic \citep{2000A&A...353.1115C} \REVbis{or if the disk magnetization is low
\citep{2019MNRAS.490.3112J}}. Spiral waves could also provide extra angular momentum transfer, if the disk is perturbed by infalling material or gravitationally unstable. Therefore, observing disk wind kinematics
consistent with our MHD DW solutions does not necessarily imply that the wind performs 100\% of the disk angular momentum extraction in that system.} 

\rev{This hypothesis must be tested a posteriori, by computing the ratio $f_J$ of angular momentum flux carried off in the disk wind to that required for steady disk accretion. Below, we derive an exact expression for $f_J$ (see Eq.~(\ref{eq:fracj2}))
valid for any radially extended, self-similar, steady-state disk wind, as a function of the wind parameters and mass ejection/accretion ratio $f_M$. This expression differs from the well-known rule of thumb $f_J \simeq \lambda_{\rm BP} f_M$, valid only under specific conditions. We then apply our exact formula to the case of HH212.}

\rev{In steady-state, the rate at which angular momentum must be extracted from the disk to sustain accretion between \rin\ 
and \rout\  is given by 
\begin{equation}
\label{eq:jacc}
\dot{J}_{\rm acc} = \dot{M}_{\rm out} \sqrt{G M_\star r_{\rm out}} - \dot{M}_{\rm in} \sqrt{G M_\star r_{\rm in}},
\end{equation}
where \Min\ is the disk accretion rate at \rin,  \Mout\ is the disk accretion rate at \rout,
and \Mdw =  \Mout\ -- \Min\ is the mass-flux ejected by the disk-wind (on both sides) between \rin\ and \rout.}

\rev{The rate $\dot{J}_{\rm DW}$ at which angular momentum is extracted by the MHD disk wind between \rin\ and \rout\
depends on how the wind mass outflow rate is distributed radially across this region.  
In a self-similar system, this distribution is ruled by the "ejection efficiency" parameter $\xi$ defined by 
\citet{1995A&A...295..807F} as
\begin{equation}
\dot{M}_{\rm acc} (r) = \dot{M}_{\rm in} (r/r_{\rm in})^{\xi},
\label{eq:xi}
\end{equation}
which depends on the radial distribution of magnetic field in the disk (see Appendix A).
The wind outflow rate from an elementary disk annulus at radius $r$ is then given by mass conservation as
\begin{equation}
\frac{d \dot{M}_{\rm DW}(r)}{dr}  = \frac{d \dot{M}_{\rm acc}(r)}{dr}  = \xi \frac{\dot{M}_{\rm acc} (r)}{r}, 
\end{equation}
and the angular momentum flux extracted from the same annulus is (by definition of \lbp) 
\begin{eqnarray}
\frac{d \dot{J}_{\rm DW}(r)}{dr}  &=&  \left[\lambda_{\rm BP} r V_K(r) \right] \frac{d \dot{M}_{\rm DW}(r)}{dr} \\
						& = & \xi \lambda_{\rm BP} \dot{M}_{\rm in} \sqrt{G M_\star / r_{\rm in}} (r/r_{\rm in})^{\xi -1/2}.
\end{eqnarray}
Integration between \rin\ and \rout\ then gives
\begin{equation}
\dot{J}_{\rm DW} =  \frac{\xi \lambda_{\rm BP}}{\xi+1/2} \left[\dot{M}_{\rm out} \sqrt{G M_\star r_{\rm out}} - \dot{M}_{\rm in} \sqrt{G M_\star r_{\rm in}}\right].
\end{equation}
Comparing with Eq. \ref{eq:jacc}, we see that the fraction $f_J$ of disk angular momentum extraction performed
vertically by the MHD DW (as opposed to radially by turbulent or wave torques) is simply given by 
\begin{equation}
\label{eq:fracj}
f_J = \frac{\dot{J}_{\rm DW}}{\dot{J}_{\rm acc}} = \frac{\xi \lambda_{\rm BP}}{\xi+1/2}.
\end{equation}
}

The value of $\xi$ in a real disk is not directly measurable. However, it may be related in steady-state to the observable mass ejection to accretion ratio $f_M$ through: 
\begin{equation}
\label{eq:fm}
f_M \equiv \frac{\dot{M}_{\rm DW}}{\dot{M}_{\rm in}} = \left[ \left(\frac{\dot{M}_{\rm out}}{\dot{M}_{\rm in}}
\right) - 1\right] = \left[ \left(\frac{r_{\rm out}}{r_{\rm in}}\right)^\xi - 1\right],
\end{equation}
where we have used mass conservation (\Mdw =  \Mout\ -- \Min) and the definition of $\xi$ (Eq.~(\ref{eq:xi})).
\REVbis{Note that $f_M$ could be much larger than unity if $\xi$ large and/or \rout/\rin $\gg 1$.}

Using the above formula to eliminate $\xi$ from Eq.~(\ref{eq:fracj}), we obtain the exact expression of $f_J$ solely as a function of observable disk wind properties: 
\begin{equation}
\label{eq:fracj2}
f_J =  \lambda_{\rm BP} \times \left[ {1+ \frac{ \ln (r_{\rm out}/r_{\rm in}) }{2 \ln (1+f_M)} } \right]^{-1}.
\end{equation}

A simpler expression for $f_J$ may be obtained in the limits of small mass-fluxes $f_M \ll 1$ and large magnetic lever arms \lbp\ $\gg 1$. 
Eq.~(\ref{eq:fm}) may then be approximated as 
\begin{equation}
\label{eq:fm2}
f_M \simeq \ln(1+f_M) = \xi \ln \left(\frac{r_{\rm out}}{r_{\rm in}}\right) \ll 1,
\end{equation}
while Eq.~(\ref{eq:fracj}) may be rewritten as
\begin{equation}
\label{eq:xi2}
\xi =  \frac {f_J} {2 (\lambda_{\rm BP}-f_J)} \simeq \frac{f_J} {2 \lambda_{\rm BP}}.
\end{equation}
Combining Eq.~(\ref{eq:xi2}),(\ref{eq:fm2}) yields 
\begin{equation}
\label{eq:fracj3}
f_J \simeq  2 \xi \lambda_{\rm BP} \simeq \left[\frac{2} { \ln (r_{\rm out}/r_{\rm in}) }\right] \lambda_{\rm BP} f_M.
\end{equation}
If the wind torque dominates the disk angular momentum extraction (ie. $f_J \simeq 1$) we then recover the well-known and widely used rule of thumb $f_M \simeq 1/\lambda_{\rm BP}$  \citep{1992ApJ...394..117P,2002ApJ...576..222B,2007prpl.conf..277P}, but with an extra numerical factor in front that depends on the wind radial extent.

\rev{Unfortunately, the approximation in Eq.~(\ref{eq:fracj3}) is no longer valid for the small magnetic lever arms \lbp $\le 5$ and large $f_M \simeq 1$ favored by  ALMA-like observations, and by non-ideal MHD simulations of PPDs \citep{2017A&A...600A..75B, 2017ApJ...845...75B}. Hence, the exact Equation \ref{eq:fracj2} should be preferred to evaluate accurately $f_J$ in disk wind candidates.}

\rev{We now proceed to obtain an observational estimate of  $f_M$ in HH212.
Considering first \Mdw}, \citet{2018ApJ...856...14L} estimated the mass in the rotating SO-rich disk outflow within $z \le 0\farcs2 \simeq 90$ au from the source at $(0.4-4)\times10^{-4}M_\odot$, \rev{where the range of a factor 10 reflects the current uncertainty in SO abundance in this flow
\REVbis{\citep{2015A&A...581A..85P}}. Scaling to our adopted distance of 450 pc, and taking} the mean vertical velocity $V_z$ $\simeq$ 10 \kms\ of the best fit MHD DW model, the wind crossing time through this region is $\simeq 50$ yr, and the corresponding ejected mass-flux is \Mdw $\simeq 1-10 \times 10^{-6}M_\odot$/yr. %(where the range stems from the uncertainty in SO abundance in the wind).}

\rev{Next, we estimate the accretion rate onto the HH212 source: Since it is a young Class 0 protostar, we expect it to lie 
along the "birthline" where stellar radius grows over time \citep{1988ApJ...332..804S}. %, instead of contracting towards the main sequence
For a stellar mass $M_\star \simeq 0.25 M_\sun$ \citep{2017ApJ...843...27L}, the observed bolometric luminosity of 11 $L_\sun$ \citep[][scaled to our adopted distance of 450 pc]{
    1992A&A...265..726Z} 
is reached for an accretion rate onto the star $\dot{M}_{\star}\simeq 2 \times10^{-6}M_\odot$/yr \citep[see Fig. 9 in][]{1988ApJ...332..804S}.
The high-velocity axial SiO / CO jet, ejected from within $\simeq 0.1$ au of the source \citep{2017NatAs...1E.152L,2017A&A...607L...6T} removes
an additional $\dot{M}_{\rm jet} \simeq 10^{-6}M_\odot$/yr  from the incoming accretion flow \citep{2015ApJ...805..186L}. Therefore, in steady-state, 
the disk accretion rate at \rin = 0.1 au is \Min\ =  $\dot{M}_{\star} + \dot{M}_{\rm jet}$  $\simeq 3 \times10^{-6}M_\odot$/yr.}
%with a predicted stellar radius $R_\star \simeq 1.8R_\odot $.

\rev{From the above observational estimates of \Mdw\ and \Min\, we infer an ejection to accretion ratio $f_{M,\rm obs}$ = \Mdw/ \Min = 0.33 -- 3.3. 
Inserting these values in Eq.~(\ref{eq:fracj2}) 
and taking \rin = 0.1 au, \rout = 40 au, and \lbp = 5.5 from PV cut modeling 
\citep[see Figs.~\ref{fig:model-70au},\ref{fig:model-18au} and][]{2017A&A...607L...6T},
we obtain $f_{J,\rm obs}$ = 0.5 -- 1.8. \REVbis{Note that despite the large uncertainty on the SO abundance, we find $f_{J,\rm obs} \ge 0.5$.} In other words, the angular momentum flux extracted by the proposed MHD DW candidate in HH212 agrees within a factor 2 with that required to sustain accretion through the whole disk at the current observed rate.}

\section{Conclusions}

\rev{We studied observational biases in the rotation signatures of radially extended, MHD disk winds when observed at the typical resolution of large millimeter interferometers such as ALMA. 
We then tested our predictions in the edge-on case against published ALMA observations of HH212 covering a factor 15 in angular resolution.  
Our main results are the following.\\}

\rev{$\bullet$ The launch radius \robs\ inferred using Anderson's formula from rotation signatures in transverse PV cuts generally differs markedly from the true outermost launch radius of the MHD DW, \rout. The sign of this bias depends on the method used to estimate the flow specific angular momentum from PV cuts, opening the possibility to bracket the true value of \rout. \\}
 \rev{-- In the double-peak separation method, applied to edge-on PV cuts (Sect. \ref{sec:method1}), \robs\ always underestimates the true \rout. This bias does not improve at higher angular resolution, and worsens with the wind radial extension and emissivity gradient, reaching a factor 3--10 for typical parameters. At lower inclinations where the two PV peaks have the same velocity sign, this method becomes unreliable and should be avoided.\\} 
\rev{-- The apparent flow width method (see Sect.~\ref{sec:method3}) suffers a similar bias as the double-peak separation in edge-on PV cuts, and also provides a strict lower limit to the true \rout, available at all flow inclinations.}\\
 \rev{-- The rotation curve method (see Sect. \ref{sec:method2}) only yields a good estimate of \rout\ when the flow is well resolved across. Otherwise, it provides an upper limit to \rout, by an increasing amount for stronger beam smearing.\\}

\rev{$\bullet$ The magnetic lever arm inferred from apparent rotation signatures using Anderson's formula is not as strongly impacted by observational biases (which tend to cancel out in the calculation). However, due to unobservable angular momentum in the form of magnetic field torsion, it only gives a strict lower limit to the true Blandford \& Payne magnetic lever arm parameter \lbp\ \citep[as already pointed out in the  context of T Tauri jets by][]{2004A&A...416L...9P, 2006A&A...453..785F}. In our model, this "MHD bias" becomes significant for \zcut/\rout $< 20$, where \zcut\ is the altitude of the PV cut. The true \lbp\ can then only be constrained through detailed modeling of PV cuts.}\\

\rev{$\bullet$ While our analysis strictly pertains only to self-similar models, we expect similar biases to occur in more general geometries  \citep[see eg.][or a conical wind]{2007prpl.conf..277P} whenever the same underlying causes are present (ie. strong contrast effects between inner and outer streamlines, beam smearing). However, the biases might change if the internal velocity gradients are strongly non-keplerian. We defer the study of these more general cases to future work.\\}

\rev{$\bullet$ Our main results for the double-peaked PV method were tested against ALMA observations of the edge-on flow in HH212 at angular resolutions from 250 au to 18 au, which provide the most stringent observational test of MHD DW models to date. We verified that the PV double-peak separation indeed does not depend on beam size, and the launch radius \robs $\simeq 1$ au inferred from it using Anderson's relation does strongly underestimate the true outermost launch radius of the flow, directly resolved at \rout $\simeq 40$ au in the 18 au ALMA data.}
\rev{We also showed that the MHD DW model for HH212 proposed by \citet{2017A&A...607L...6T} still reproduces quite well (given its idealized self-similar geometry) the main features of transverse PV cuts at 4 times lower and higher resolution. However, the poloidal velocity is not well constrained in such an edge-on view, and alternative interpretations in terms of wind-driven or bowshock-driven shells are also possible \citep{2018ApJ...856...14L,2018A&A...614A.119T}.\\}

\GP{$\bullet$ The fraction of disk angular momentum flux extracted by a steady self-similar MHD DW is derived as a function of \rin, \rout, \lbp, and the observed mass ejection/accretion ratio $f_M$ (see Eq.~(\ref{eq:fracj2})). Application to HH212 supports the proposed paradigm where MHD DWs drive accretion across protoplanetary disks.\\}

$\bullet$ Observing rotating winds with less edge-on inclinations (where the distribution of $V_z$ can be better constrained) will be crucial to help discriminate between MHD DWs and alternative explanations (eg. wide-angle wind cavities) and definitely elucidate the mechanism driving disk accretion in protostars. Searches for pristine MHD DW signatures should focus on regions very close to the source with the highest possible angular resolution, in order to avoid large-scale bowshocks driven by the axial jet and interactions with the infalling envelope.

\begin{acknowledgements}
\rev{We are grateful to C. Dougados for useful suggestions, and to an anonymous referee for constructive comments that helped to improve the manuscript presentation and content.} This paper makes use of the ALMA 2012.1.00997.S and  2016.1.01475.S data (PI: C. Codella). ALMA is a partnership of ESO (representing its member states), NSF (USA), and NINS (Japan), together with NRC (Canada) and NSC and ASIAA (Taiwan), in cooperation with the Republic of Chile. The Joint ALMA Observatory is operated by ESO, AUI/NRAO, and NAOJ. This work was supported by the Programme National Physique et Chimie du Milieu Interstellaire (PCMI) of CNRS/INSU with INC/INP and co-funded by CNES, and by the Conseil Scientifique of Observatoire de Paris. \REVbis{BT acknowledges funding from the research programme Dutch Astrochemistry Network II with project number 614.001.751, which is (partly) financed by the Dutch Research Council (NWO).}
\rev{EB and CC acknowledge funding from the European Research Council (ERC) under the European Union's Horizon 2020 research and innovation programme, for the Project "The Dawn of Organic Chemistry" (DOC), grant agreement No 741002.} This research has made use of NASA's Astrophysics Data System.
\end{acknowledgements}

\bibliographystyle{aa} % style aa.bst
\bibliography{mybibli.bib} % your references Yourfile.bib

\begin{thebibliography}{56}
\expandafter\ifx\csname natexlab\endcsname\relax\def\natexlab#1{#1}\fi

\bibitem[{{Anderson} {et~al.}(2003){Anderson}, {Li}, {Krasnopolsky}, \&
  {Blandford}}]{2003ApJ...590L.107A}
{Anderson}, J.~M., {Li}, Z.-Y., {Krasnopolsky}, R., \& {Blandford}, R.~D. 2003,
  \apjl, 590, L107

\bibitem[{{Bacciotti} {et~al.}(2002){Bacciotti}, {Ray}, {Mundt},
  {Eisl{\"o}ffel}, \& {Solf}}]{2002ApJ...576..222B}
{Bacciotti}, F., {Ray}, T.~P., {Mundt}, R., {Eisl{\"o}ffel}, J., \& {Solf}, J.
  2002, \apj, 576, 222

\bibitem[{{Bai}(2017)}]{2017ApJ...845...75B}
{Bai}, X.-N. 2017, \apj, 845, 75

\bibitem[{{Bai} \& {Stone}(2013)}]{2013ApJ...769...76B}
{Bai}, X.-N. \& {Stone}, J.~M. 2013, \apj, 769, 76

\bibitem[{{Balbus} \& {Hawley}(1991)}]{1991ApJ...376..214B}
{Balbus}, S.~A. \& {Hawley}, J.~F. 1991, \apj, 376, 214

\bibitem[{{B{\'e}thune} {et~al.}(2017){B{\'e}thune}, {Lesur}, \&
  {Ferreira}}]{2017A&A...600A..75B}
{B{\'e}thune}, W., {Lesur}, G., \& {Ferreira}, J. 2017, \aap, 600, A75

\bibitem[{{Bianchi} {et~al.}(2017){Bianchi}, {Codella}, {Ceccarelli}, {Taquet},
  {Cabrit}, {Bacciotti}, {Bachiller}, {Chapillon}, {Gueth}, {Gusdorf},
  {Lefloch}, {Leurini}, {Podio}, {Rygl}, {Tabone}, \&
  {Tafalla}}]{2017A&A...606L...7B}
{Bianchi}, E., {Codella}, C., {Ceccarelli}, C., {et~al.} 2017, \aap, 606, L7

\bibitem[{{Bjerkeli} {et~al.}(2016){Bjerkeli}, {van der Wiel}, {Harsono},
  {Ramsey}, \& {J{\o}rgensen}}]{2016Natur.540..406B}
{Bjerkeli}, P., {van der Wiel}, M. H.~D., {Harsono}, D., {Ramsey}, J.~P., \&
  {J{\o}rgensen}, J.~K. 2016, \nat, 540, 406

\bibitem[{{Blandford} \& {Payne}(1982)}]{1982MNRAS.199..883B}
{Blandford}, R.~D. \& {Payne}, D.~G. 1982, \mnras, 199, 883

\bibitem[{{Cabrit} {et~al.}(2006){Cabrit}, {Pety}, {Pesenti}, \&
  {Dougados}}]{2006A&A...452..897C}
{Cabrit}, S., {Pety}, J., {Pesenti}, N., \& {Dougados}, C. 2006, \aap, 452, 897

\bibitem[{{Casse} \& {Ferreira}(2000{\natexlab{a}})}]{2000A&A...353.1115C}
{Casse}, F. \& {Ferreira}, J. 2000{\natexlab{a}}, \aap, 353, 1115

\bibitem[{{Casse} \& {Ferreira}(2000{\natexlab{b}})}]{2000A&A...361.1178C}
{Casse}, F. \& {Ferreira}, J. 2000{\natexlab{b}}, \aap, 361, 1178

\bibitem[{{Chen} {et~al.}(2016){Chen}, {Arce}, {Zhang}, {Launhardt}, \&
  {Henning}}]{2016ApJ...824...72C}
{Chen}, X., {Arce}, H.~G., {Zhang}, Q., {Launhardt}, R., \& {Henning}, T. 2016,
  \apj, 824, 72

\bibitem[{{Codella} {et~al.}(2018){Codella}, {Bianchi}, {Tabone}, {Lee},
  {Cabrit}, {Ceccarelli}, {Podio}, {Bacciotti}, {Bachiller}, {Chapillon},
  {Gueth}, {Gusdorf}, {Lefloch}, {Leurini}, {Pineau des For{\^e}ts}, {Rygl}, \&
  {Tafalla}}]{2018A&A...617A..10C}
{Codella}, C., {Bianchi}, E., {Tabone}, B., {et~al.} 2018, \aap, 617, A10

\bibitem[{{Coffey} {et~al.}(2007){Coffey}, {Bacciotti}, {Ray}, {Eisl{\"o}ffel},
  \& {Woitas}}]{2007ApJ...663..350C}
{Coffey}, D., {Bacciotti}, F., {Ray}, T.~P., {Eisl{\"o}ffel}, J., \& {Woitas},
  J. 2007, \apj, 663, 350

\bibitem[{{Coffey} {et~al.}(2015){Coffey}, {Dougados}, {Cabrit}, {Pety}, \&
  {Bacciotti}}]{2015ApJ...804....2C}
{Coffey}, D., {Dougados}, C., {Cabrit}, S., {Pety}, J., \& {Bacciotti}, F.
  2015, \apj, 804, 2

\bibitem[{{Combet} \& {Ferreira}(2008)}]{2008A&A...479..481C}
{Combet}, C. \& {Ferreira}, J. 2008, \aap, 479, 481

\bibitem[{{Fendt}(2011)}]{2011ApJ...737...43F}
{Fendt}, C. 2011, \apj, 737, 43

\bibitem[{{Ferreira}(1997)}]{1997A&A...319..340F}
{Ferreira}, J. 1997, \aap, 319, 340

\bibitem[{{Ferreira} {et~al.}(2006){Ferreira}, {Dougados}, \&
  {Cabrit}}]{2006A&A...453..785F}
{Ferreira}, J., {Dougados}, C., \& {Cabrit}, S. 2006, \aap, 453, 785

\bibitem[{{Ferreira} \& {Pelletier}(1995)}]{1995A&A...295..807F}
{Ferreira}, J. \& {Pelletier}, G. 1995, \aap, 295, 807

\bibitem[{{Hartmann} {et~al.}(2016){Hartmann}, {Herczeg}, \&
  {Calvet}}]{2016ARA&A..54..135H}
{Hartmann}, L., {Herczeg}, G., \& {Calvet}, N. 2016, \araa, 54, 135

\bibitem[{{Hirota} {et~al.}(2017){Hirota}, {Machida}, {Matsushita}, {Motogi},
  {Matsumoto}, {Kim}, {Burns}, \& {Honma}}]{2017NatAs...1E.146H}
{Hirota}, T., {Machida}, M.~N., {Matsushita}, Y., {et~al.} 2017, Nature
  Astronomy, 1, 0146

\bibitem[{{Jacquemin-Ide} {et~al.}(2019){Jacquemin-Ide}, {Ferreira}, \&
  {Lesur}}]{2019MNRAS.490.3112J}
{Jacquemin-Ide}, J., {Ferreira}, J., \& {Lesur}, G. 2019, \mnras, 490, 3112

\bibitem[{{Konigl}(1989)}]{1989ApJ...342..208K}
{Konigl}, A. 1989, \apj, 342, 208

\bibitem[{{Kounkel} {et~al.}(2017){Kounkel}, {Hartmann}, {Loinard},
  {Ortiz-Le{\'o}n}, {Mioduszewski}, {Rodr{\'\i}guez}, {Dzib}, {Torres}, {Pech},
  {Galli}, {Rivera}, {Boden}, {Evans}, {Brice{\~n}o}, \&
  {Tobin}}]{2017ApJ...834..142K}
{Kounkel}, M., {Hartmann}, L., {Loinard}, L., {et~al.} 2017, \apj, 834, 142

\bibitem[{{Launhardt} {et~al.}(2009){Launhardt}, {Pavlyuchenkov}, {Gueth},
  {Chen}, {Dutrey}, {Guilloteau}, {Henning}, {Pi{\'e}tu}, {Schreyer}, \&
  {Semenov}}]{2009A&A...494..147L}
{Launhardt}, R., {Pavlyuchenkov}, Y., {Gueth}, F., {et~al.} 2009, \aap, 494,
  147

\bibitem[{{Lee} {et~al.}(2014){Lee}, {Hirano}, {Zhang}, {Shang}, {Ho}, \&
  {Krasnopolsky}}]{2014ApJ...786..114L}
{Lee}, C.-F., {Hirano}, N., {Zhang}, Q., {et~al.} 2014, \apj, 786, 114

\bibitem[{{Lee} {et~al.}(2015){Lee}, {Hirano}, {Zhang}, {Shang}, {Ho}, \&
  {Mizuno}}]{2015ApJ...805..186L}
{Lee}, C.-F., {Hirano}, N., {Zhang}, Q., {et~al.} 2015, \apj, 805, 186

\bibitem[{{Lee} {et~al.}(2008){Lee}, {Ho}, {Bourke}, {Hirano}, {Shang}, \&
  {Zhang}}]{2008ApJ...685.1026L}
{Lee}, C.-F., {Ho}, P. T.~P., {Bourke}, T.~L., {et~al.} 2008, \apj, 685, 1026

\bibitem[{{Lee} {et~al.}(2017{\natexlab{a}}){Lee}, {Ho}, {Li}, {Hirano},
  {Zhang}, \& {Shang}}]{2017NatAs...1E.152L}
{Lee}, C.-F., {Ho}, P. T.~P., {Li}, Z.-Y., {et~al.} 2017{\natexlab{a}}, Nature
  Astronomy, 1, 0152

\bibitem[{{Lee} {et~al.}(2018{\natexlab{a}}){Lee}, {Li}, {Codella}, {Ho},
  {Podio}, {Hirano}, {Shang}, {Turner}, \& {Zhang}}]{2018ApJ...856...14L}
{Lee}, C.-F., {Li}, Z.-Y., {Codella}, C., {et~al.} 2018{\natexlab{a}}, \apj,
  856, 14

\bibitem[{{Lee} {et~al.}(2018{\natexlab{b}}){Lee}, {Li}, {Hirano}, {Shang},
  {Ho}, \& {Zhang}}]{2018ApJ...863...94L}
{Lee}, C.-F., {Li}, Z.-Y., {Hirano}, N., {et~al.} 2018{\natexlab{b}}, \apj,
  863, 94

\bibitem[{{Lee} {et~al.}(2017{\natexlab{b}}){Lee}, {Li}, {Ho}, {Hirano},
  {Zhang}, \& {Shang}}]{2017ApJ...843...27L}
{Lee}, C.-F., {Li}, Z.-Y., {Ho}, P. T.~P., {et~al.} 2017{\natexlab{b}}, \apj,
  843, 27

\bibitem[{{Lee} {et~al.}(2001){Lee}, {Stone}, {Ostriker}, \&
  {Mundy}}]{2001ApJ...557..429L}
{Lee}, C.-F., {Stone}, J.~M., {Ostriker}, E.~C., \& {Mundy}, L.~G. 2001, \apj,
  557, 429

\bibitem[{{Louvet} {et~al.}(2016){Louvet}, {Dougados}, {Cabrit}, {Hales},
  {Pinte}, {M{\'e}nard}, {Bacciotti}, {Coffey}, {Mardones}, {Bronfman}, \&
  {Gueth}}]{2016A&A...596A..88L}
{Louvet}, F., {Dougados}, C., {Cabrit}, S., {et~al.} 2016, \aap, 596, A88

\bibitem[{{Louvet} {et~al.}(2018){Louvet}, {Dougados}, {Cabrit}, {Mardones},
  {M{\'e}nard}, {Tabone}, {Pinte}, \& {Dent}}]{2018A&A...618A.120L}
{Louvet}, F., {Dougados}, C., {Cabrit}, S., {et~al.} 2018, \aap, 618, A120

\bibitem[{{Ogihara} {et~al.}(2018){Ogihara}, {Kokubo}, {Suzuki}, \&
  {Morbidelli}}]{2018A&A...615A..63O}
{Ogihara}, M., {Kokubo}, E., {Suzuki}, T.~K., \& {Morbidelli}, A. 2018, \aap,
  615, A63

\bibitem[{{Panoglou} {et~al.}(2012){Panoglou}, {Cabrit}, {Pineau Des
  For{\^e}ts}, {Garcia}, {Ferreira}, \& {Casse}}]{2012A&A...538A...2P}
{Panoglou}, D., {Cabrit}, S., {Pineau Des For{\^e}ts}, G., {et~al.} 2012, \aap,
  538, A2

\bibitem[{{Pelletier} \& {Pudritz}(1992)}]{1992ApJ...394..117P}
{Pelletier}, G. \& {Pudritz}, R.~E. 1992, \apj, 394, 117

\bibitem[{{Pesenti} {et~al.}(2004){Pesenti}, {Dougados}, {Cabrit}, {Ferreira},
  {Casse}, {Garcia}, \& {O'Brien}}]{2004A&A...416L...9P}
{Pesenti}, N., {Dougados}, C., {Cabrit}, S., {et~al.} 2004, \aap, 416, L9

\bibitem[{{Podio} {et~al.}(2015){Podio}, {Codella}, {Gueth}, {Cabrit},
  {Bachiller}, {Gusdorf}, {Lee}, {Lefloch}, {Leurini}, {Nisini}, \&
  {Tafalla}}]{2015A&A...581A..85P}
{Podio}, L., {Codella}, C., {Gueth}, F., {et~al.} 2015, \aap, 581, A85

\bibitem[{{Pudritz} \& {Norman}(1983)}]{1983ApJ...274..677P}
{Pudritz}, R.~E. \& {Norman}, C.~A. 1983, \apj, 274, 677

\bibitem[{{Pudritz} {et~al.}(2007){Pudritz}, {Ouyed}, {Fendt}, \&
  {Brandenburg}}]{2007prpl.conf..277P}
{Pudritz}, R.~E., {Ouyed}, R., {Fendt}, C., \& {Brandenburg}, A. 2007, in
  Protostars and Planets V, ed. B.~{Reipurth}, D.~{Jewitt}, \& K.~{Keil}, 277

\bibitem[{{Sauty} {et~al.}(2012){Sauty}, {Cayatte}, {Lima}, {Matsakos}, \&
  {Tsinganos}}]{2012ApJ...759L...1S}
{Sauty}, C., {Cayatte}, V., {Lima}, J.~J.~G., {Matsakos}, T., \& {Tsinganos},
  K. 2012, \apjl, 759, L1

\bibitem[{{Shu} {et~al.}(2000){Shu}, {Najita}, {Shang}, \&
  {Li}}]{2000prpl.conf..789S}
{Shu}, F.~H., {Najita}, J.~R., {Shang}, H., \& {Li}, Z.~Y. 2000, in Protostars
  and Planets IV, ed. V.~{Mannings}, A.~P. {Boss}, \& S.~S. {Russell}, 789--814

\bibitem[{{Staff} {et~al.}(2015){Staff}, {Koning}, {Ouyed}, {Thompson}, \&
  {Pudritz}}]{2015MNRAS.446.3975S}
{Staff}, J.~E., {Koning}, N., {Ouyed}, R., {Thompson}, A., \& {Pudritz}, R.~E.
  2015, \mnras, 446, 3975

\bibitem[{{Stahler}(1988)}]{1988ApJ...332..804S}
{Stahler}, S.~W. 1988, \apj, 332, 804

\bibitem[{{Tabone} {et~al.}(2017){Tabone}, {Cabrit}, {Bianchi}, {Ferreira},
  {Pineau des For{\^e}ts}, {Codella}, {Gusdorf}, {Gueth}, {Podio}, \&
  {Chapillon}}]{2017A&A...607L...6T}
{Tabone}, B., {Cabrit}, S., {Bianchi}, E., {et~al.} 2017, \aap, 607, L6

\bibitem[{{Tabone} {et~al.}(2018){Tabone}, {Raga}, {Cabrit}, \& {Pineau des
  For{\^e}ts}}]{2018A&A...614A.119T}
{Tabone}, B., {Raga}, A., {Cabrit}, S., \& {Pineau des For{\^e}ts}, G. 2018,
  \aap, 614, A119

\bibitem[{{Turner} {et~al.}(2014){Turner}, {Fromang}, {Gammie}, {Klahr},
  {Lesur}, {Wardle}, \& {Bai}}]{2014prpl.conf..411T}
{Turner}, N.~J., {Fromang}, S., {Gammie}, C., {et~al.} 2014, in Protostars and
  Planets VI, ed. H.~{Beuther}, R.~S. {Klessen}, C.~P. {Dullemond}, \&
  T.~{Henning}, 411

\bibitem[{{Yvart} {et~al.}(2016){Yvart}, {Cabrit}, {Pineau des For{\^e}ts}, \&
  {Ferreira}}]{2016A&A...585A..74Y}
{Yvart}, W., {Cabrit}, S., {Pineau des For{\^e}ts}, G., \& {Ferreira}, J. 2016,
  \aap, 585, A74

\bibitem[{{Zapata} {et~al.}(2015){Zapata}, {Lizano}, {Rodr{\'\i}guez}, {Ho},
  {Loinard}, {Fern{\'a}ndez-L{\'o}pez}, \& {Tafoya}}]{2015ApJ...798..131Z}
{Zapata}, L.~A., {Lizano}, S., {Rodr{\'\i}guez}, L.~F., {et~al.} 2015, \apj,
  798, 131

\bibitem[{{Zapata} {et~al.}(2010){Zapata}, {Schmid-Burgk}, {Muders}, {Schilke},
  {Menten}, \& {Guesten}}]{2010A&A...510A...2Z}
{Zapata}, L.~A., {Schmid-Burgk}, J., {Muders}, D., {et~al.} 2010, \aap, 510, A2

\bibitem[{{Zhang} {et~al.}(2018){Zhang}, {Higuchi}, {Sakai}, {Oya},
  {L{\'o}pez-Sepulcre}, {Imai}, {Sakai}, {Watanabe}, {Ceccarelli}, {Lefloch},
  \& {Yamamoto}}]{2018ApJ...864...76Z}
{Zhang}, Y., {Higuchi}, A.~E., {Sakai}, N., {et~al.} 2018, \apj, 864, 76

\bibitem[{{Zinnecker} {et~al.}(1992){Zinnecker}, {Bastien}, {Arcoragi}, \&
  {Yorke}}]{1992A&A...265..726Z}
{Zinnecker}, H., {Bastien}, P., {Arcoragi}, J.-P., \& {Yorke}, H.~W. 1992,
  \aap, 265, 726

\end{thebibliography}

\begin{appendix}

\section{Parameters and properties of MHD DW solutions}
\label{appendixA}

\begin{table*}
\caption{Disk parameters and emerging wind properties of the calculated self-similar MHD solutions} % title of Table
\label{tab:disk-param}      % is used to refer this table in the text
\centering                                      % used for centering table
\begin{tabular}{| c | c c  c c c | c c c | c c|}% centered columns (4 columns)
\hline 
\hline 
&   \multicolumn{5}{c |}{input disk parameters$^{a}$} &  \multicolumn{3}{c |}{solved parameters$^{a}$} &  \multicolumn{2}{c |}{wind properties$^{c}$}  \\                 
\hline
Model name  & $\xi$ &  $\epsilon$ & $\alpha_v^{(b)}$ & $\alpha_m$ & $\chi_m$ & $\mu$  & $p$ & $q$ & $\lambda_{BP}$ & $\mathcal{W}$ \\  
\hline                               
\textbf{L13W36} & $3.8 \times 10^{-2}$ & $3.2 \times 10^{-2}$ & 1.2 & 2 & 1.42 & 0.35 & 1.1 & 1.8 & 13.7 & 36  \\
\textbf{L13W130} & $4 \times 10^{-2}$ & $5 \times 10^{-2}$ & 0.9 & 1.9 & 1.38 & 0.21 & 1.2 & 2.3 & 12.9 & 134 \\ 
\textbf{L5W30}  & 0.11 & $1\times 10^{-2}$ & 1.1 &  2.3  & 4.2 & 0.22 & 2  & 4.9 & 5.5 & 30\\ 
\textbf{L5W17} &  0.11  & $1\times 10^{-2}$ & 1.2 &  2.35  & 4.1 & 0.25 & 1.85 & 4.33& 5.5 & 17 \\ 
\hline  %inserts single line
\end{tabular}
\tablefoot{
\tablefoottext{a}{See Appendix A for parameter definitions;} \tablefoottext{b}{we take $\alpha_v = \alpha_m \sqrt{\mu}$, 
corresponding to $\nu_v = \nu_m$ (magnetic Prandl number = 1).}
\tablefoottext{c}{\lbp\ is the wind Blandford \& Payne magnetic lever arm parameter, defined in Eq.~(\ref{eq:lambdaBP}). \WW\ is the wind maximum widening factor $r_{\rm max}/r_0$ (see Eq.~(\ref{eq:WW}) and Fig. \ref{fig:rho-bfield}).}
}
\label{tab:MHD-models-bis}
\end{table*}

\begin{figure}
   \centering
   \includegraphics[width=0.45\textwidth]{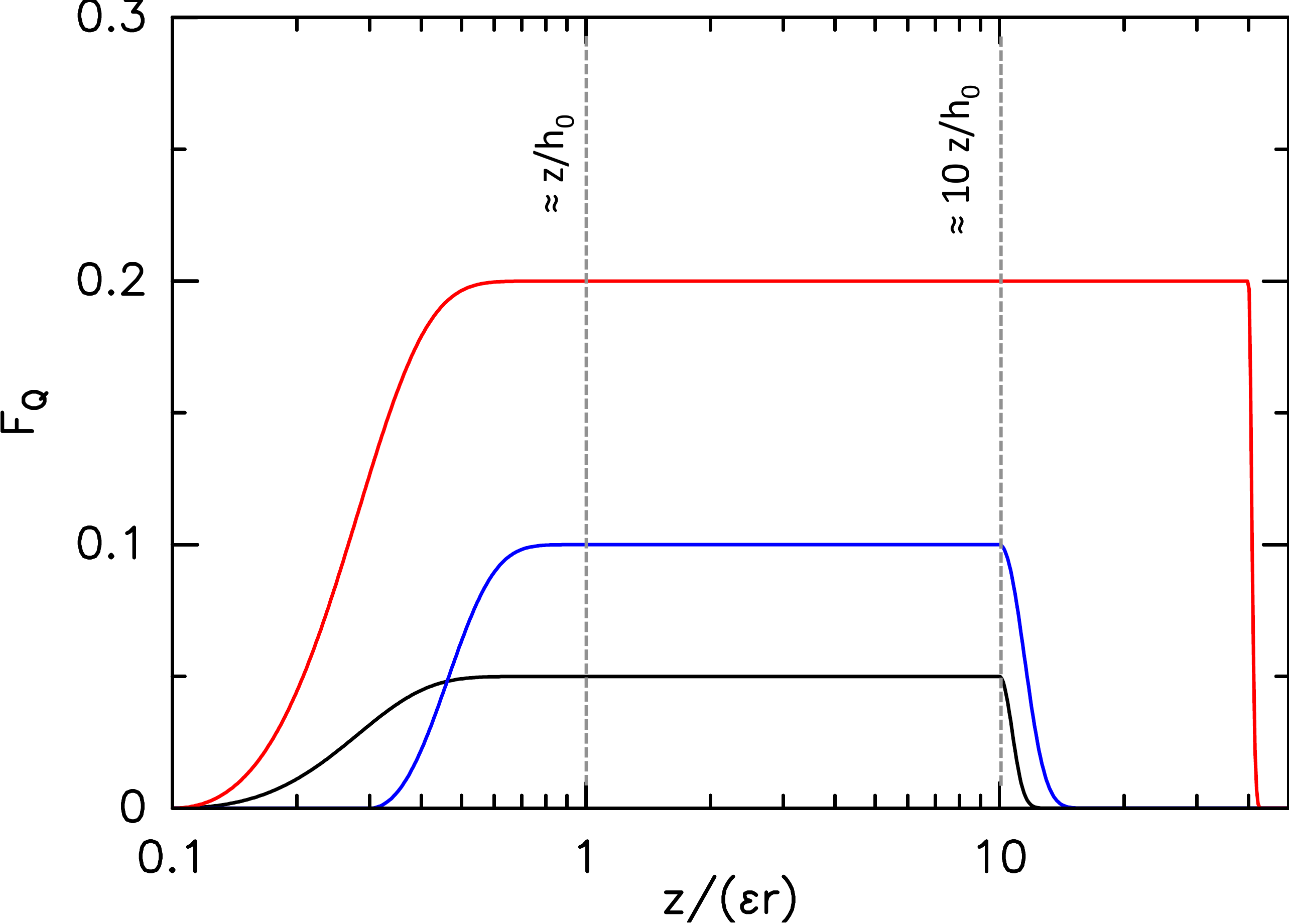}
      \caption{Vertical profile of the normalized entropy generation function $F_Q({z}/{\epsilon r})$ in the four computed MHD disk wind solutions. In blue L13W36, in black L13W120, in red the (identical) function for L5W17 and L5W30. \SCC{$F_Q$ is related to the entropy source function along the magnetic flow surface through $\Gamma-\Lambda = Q_0 \left({r}/{r_0}\right)^{\xi-4} F_Q\left({z}/{\epsilon r} \right)$ where $Q_0= \left( {P u_r}/{r}\right) |_{z=0}$.}
      }
      \label{fig:heating}
   \end{figure}

\begin{figure}
   \centering
    \includegraphics[width=0.45\textwidth]{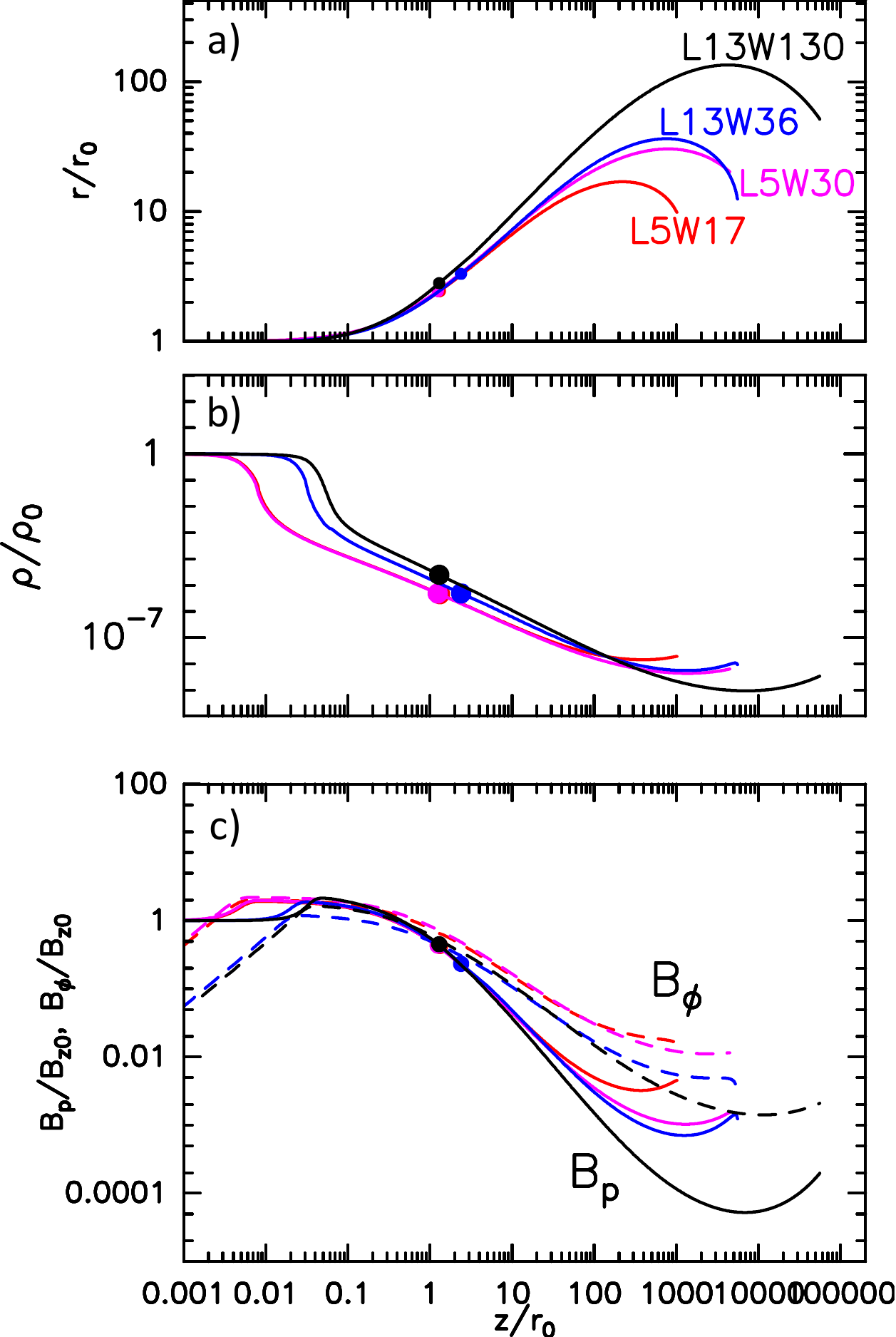}
    \caption{Variation of \textbf{a}: radius, \textbf{b}: density $\rho$,  and \textbf{c}: poloidal and toroidal magnetic field (in solid and dashed curves respectively)  along a magnetic wind surface as a function of $z/r_0$, for the four computed MHD DW solutions. The Alfv\'{e}n point for each solution is plotted as a \SC{filled dot}. Quantities are normalized by the mid-plane value of radius, density, and vertical magnetic field $B_{z,0}$ at $r_0$. \SCC{The latter are related to the dimensional parameters of the system through $\rho_0 \simeq \dot{M}_{acc}(r_0) / (4\pi r_0^2 V_{K,0} \epsilon^2 m_s)$ and $B_{z,0}^2 \simeq {\mu} \dot{M}_{acc}(r_0) V_{K,0} / (r_0^2 m_s)$, where $m_s \simeq 1-2$ is the Mach number of the accretion flow \citep{2008A&A...479..481C}.} }
    \label{fig:rho-bfield}
   \end{figure}

The self-similar, stationary and axisymmetric MHD disk-wind solutions computed for this work treat in an exact and consistent manner the disk accretion and ejection process\SC{es}, using the formalism and method described in \citet{1997A&A...319..340F,2000A&A...353.1115C}. The vertical, \SC{radial}, and rotational balance equations of the resistive, \rev{viscous}, magnetized accreting disk are integrated through a set of coupled ODEs and determine the emerging jet parameters and the large-scale collimation. Consequently, in order to compute an MHD disk wind solution, \SC{several free input} disk parameters have to be set, \SC{that describe the disk thermal scale height, viscosity, resistivity, and magnetic field structure}. For  information and reference purposes, the disk parameter values \SC{used to obtain the four solutions used in this work} are listed in Table \ref{tab:MHD-models-bis}, and the definition of each parameter is given at the end of this section.  Following \citet{2000A&A...361.1178C}, \SC{a (self-similar) heating function in the disk atmosphere is  \MV{implemented to mimic a coronal heating along magnetic surfaces,} which allows to enhance the wind mass-loading compared to isothermal or adiabatic solutions.} The heating function adopted for each of the four MHD DW solution is \SC{plotted} in Fig.~\ref{fig:heating}. The variations of density and magnetic field along magnetic surfaces in our four solutions are presented in Fig.~\ref{fig:rho-bfield}. 

\SC{We consider here solutions where the MHD DW extracts most of the angular momentum required for disk accretion, with negligible contribution from the viscous torque. However, \citet{2000A&A...353.1115C} showed that it is possible to obtain identical emerging jet configurations with a smaller ratio of wind vs. turbulent torques. Hence, the emerging wind properties in our solutions are actually relevant for a much broader range of situations.} 

The non-dimensional parameters describing the self-similar disk structure, listed in Table \ref{tab:MHD-models-bis}, are defined as follows \citep[see][for further details]{2000A&A...353.1115C}:
\begin{enumerate}
\item The ejection efficiency $\xi \equiv {d ln  \dot{M_a}(r)}/{d ln r}$ is related to the radial scaling of the mid-plane magnetic field $B_z \propto  r_0^{\alpha_B}$ through \rev{$\alpha_B = -5/4+\xi/2$}. 
\item The disk \SC{thermal} aspect ratio $\epsilon \equiv h(r_0)/r_0 \equiv c_S/V_K(r_0)$ where $h(r_0)$ is the disk vertical pressure scale height, \SC{$V_K$ the Keplerian rotation speed, and $c_S$ the sound speed at radius $r_0$ in the disk midplane.}  
\item The Shakura-Sunyaev {} \SC{viscosity} parameter in the disk midplane, $\alpha_v \equiv \left. {\nu_v}/{c_s h}\right.|_{z=0}$ with $\nu_v$ the \SC{(anomalous) effective viscosity},
\item The \SC{poloidal magnetic diffusivity} parameter at the disk mid-plane, $\alpha_m \equiv \left. {\nu_m}/{V_{A} h}\right|_{z=0}$ with $\nu_m$ the (anomalous) poloidal magnetic diffusivity and $V_A$ the Alfven speed,
\item The magnetic diffusivity anisotropy $\chi_m \equiv {\nu_m}/{\nu_m'}$ where $\nu_m'$ is the toroidal magnetic diffusivity.
\item The disk \rev{mid-plane} magnetization $\mu \equiv \left. {B_z^2}/{(4\pi P)}\right|_{z=0}$ = $\left. ({V_A}/{c_s})^2 \right|_{z=0}$ where $P$ is the thermal pressure and $B_z$ the magnetic field in the disk mid-plane \SC{($2/\mu$ is the usual plasma parameter $\beta$)}, 
\item The inclination of the magnetic field at the disk surface $p \equiv R_m \epsilon \sim {B_r^{+}}/{B_z^{+}}$, where $R_m= \left.{r u_{r}}/{\nu_m}\right|_{z=0}$ is the magnetic Reynolds \SC{number} in the mid-plane and $B_r^{+}$ and $B_z^{+}$ are the radial and vertical magnetic fields at the disk surface,
\item The magnetic shear $q \equiv - \left.\frac{h}{B_{z}} \frac{\partial B_{\phi}}{\partial z}\right|_{z=0} \sim {-B_{\phi}^{+}}/{B_z^{+}}$ where $B_{\phi}^{+}$ is the toroidal magnetic field at the disk surface. 
\end{enumerate}
For computational reasons, $\xi$, $\epsilon$, $\alpha_v$, $\alpha_m$ and $\chi_m$ are taken as free \SC{input} parameters, 
while $\mu$ and $p$ are numerically adjusted in order to cross \SC{the} slow and Alfv\'en critical points, respectively. 
\SC{The value of $q$ is inferred from the other disk parameters and is only given in Table \ref{tab:MHD-models-bis} for the sake of completeness.}
\SC{To limit the number of free parameters, we assumed that the viscosity $\nu_v$ and resistivity $\nu_m$ are equal 
(ie. a magnetic Prandl number $\simeq 1$), since they are both ``anomalous" transport coefficients arising presumably from the same turbulence. This assumption corresponds to $\alpha_v = \alpha_m \sqrt{\mu}$. It does not affect our solutions, where the turbulent viscous torque (included in our equations) is much smaller than the dominant disk wind torque.}

 %%%%%%%%%%%%%%%%%
%%%%%%%%%%%%%%%%

\section{\SC{Transverse} Position-Velocity diagram for a rotating and expanding wind annulus}
\label{appendixB}
 
In this appendix, we  show that the \SC{transverse} Position-Velocity (PV) diagram produced by \SC{a single axisymmetric wind} annulus observed at inclination $i$ to the line of sight is an ellipse. We describe extremal points of interest on the ellipse, and relate the parameters of the ellipse \SC{(semi-major and semi-minor axes, PA, central velocity)} to the {physical parameters} of the annulus (radius, $V_r, V_{\phi}, V_z$). In Appendix \ref{ap:peaks} we then {investigate the effect of spatial and velocity smearing on} the location of intensity peaks in the PV diagram.

\subsection{Ellipse shape and tilt for a single wind annulus}
\label{ap:ellipse}
%%%%%%%%%%%%%%%%%
 \begin{figure}
   \centering
   \includegraphics[width=0.4\textwidth]{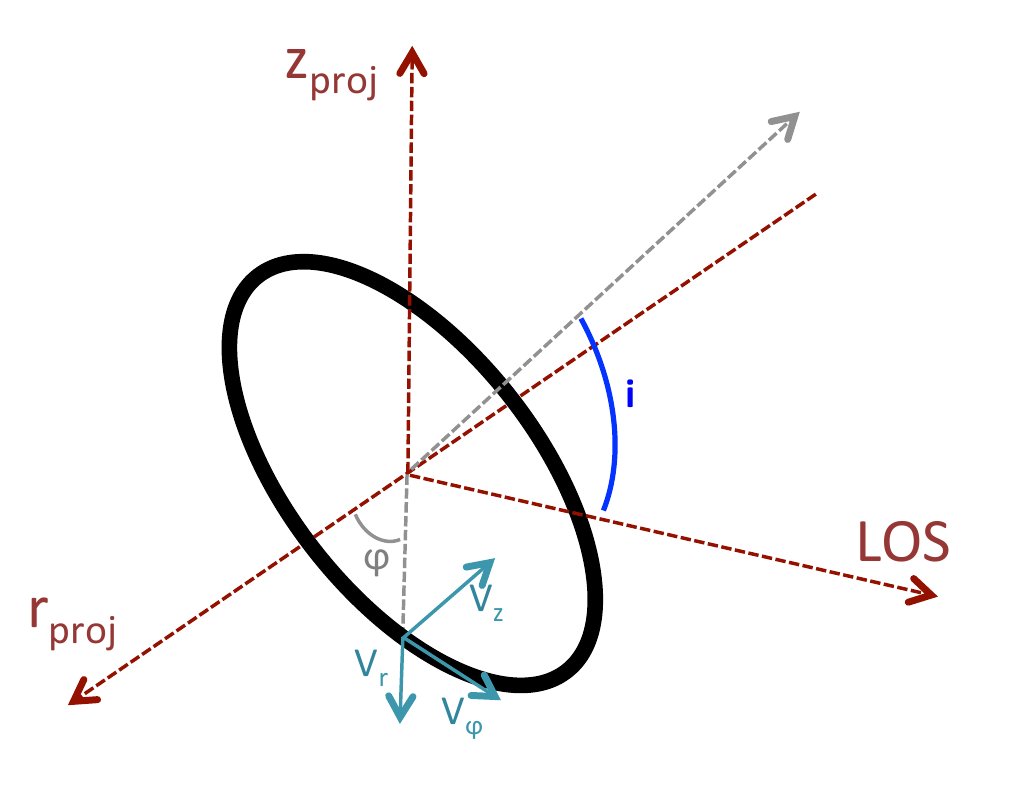}
   \includegraphics[width=0.45\textwidth]{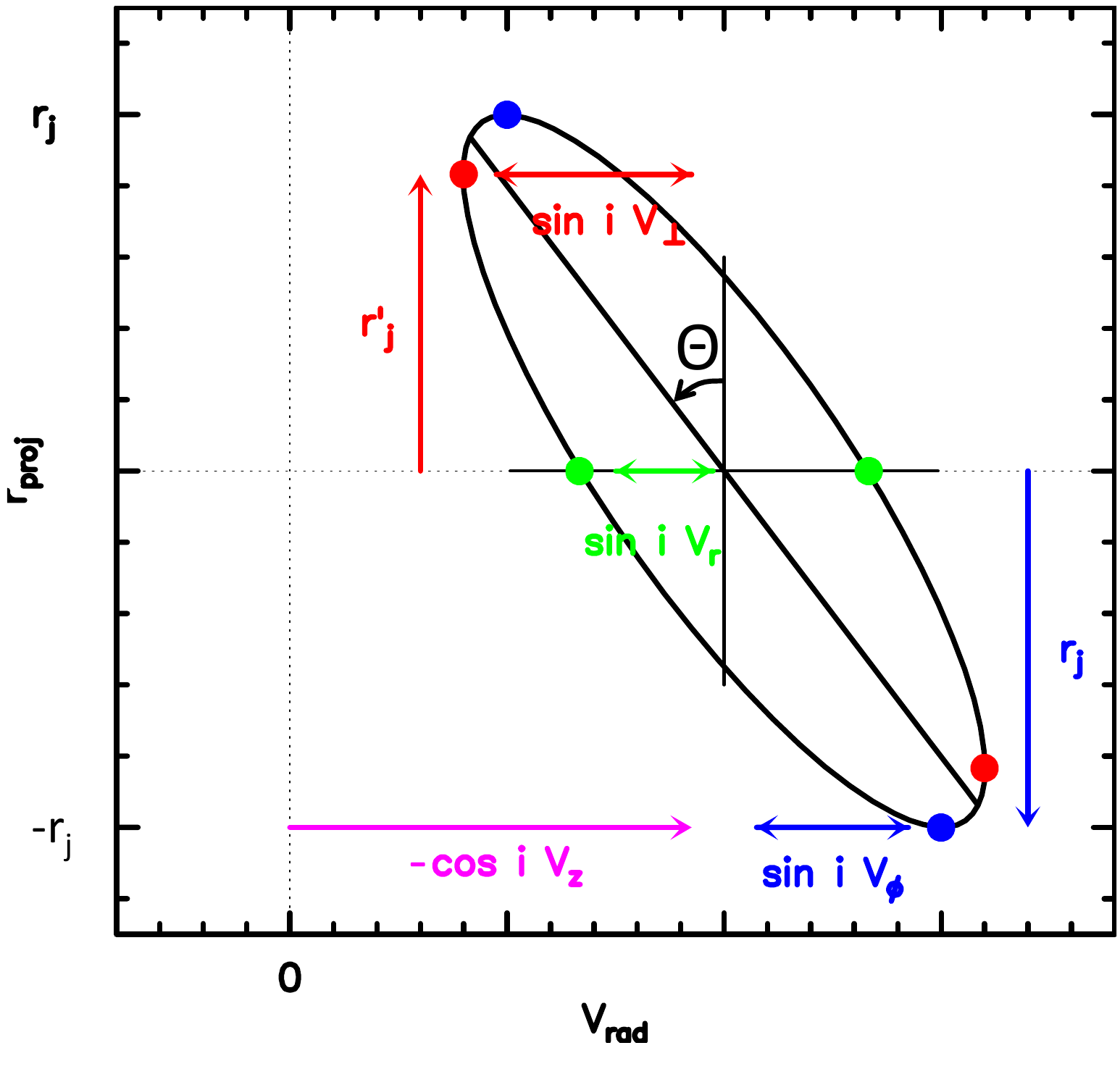} 
      \caption{ \textit{Top:} schematic view of a thin wind annulus of radius $r_j$, expansion velocity $V_r$, azimuthal velocity $V_{\phi}$, and vertical velocity $V_z$ along the jet axis, inclined by an angle $i$ from the line of sight. A point on this annulus is located by the azimuthal angle $\phi$. \textit{Bottom:} The transverse position-velocity diagram for the wind annulus is an ellipse (in black) centered on $r_{\rm proj} = 0$ and $V_{\rm rad} = -\cos{i}V_z$, of minor and major axes and tilt angle $\Theta$ given by equations \ref{velotoellipse} and \ref{eq:pa-ellipse}. Three pairs of points of interest are indicated: the on-axis points (in green) giving the projected $V_r$; the points of maximum radii (in blue) giving the projected $V_{\phi}$; and the points of maximum radial velocity (in red) \SC{with $V_{\perp} = \sqrt{V_{\phi}^2+V_r^2}$ and $r^\prime_j = ({V_{\phi}}/{V_{\perp}}) r_j$}, where intensity peaks are located at high spectral resolution (see Appendix~\ref{ap:peaks} \rev{and Fig.\ref{pv-perp-model}}). 
}
         \label{pv}
   \end{figure}
%%%%%%%%%%%%%%%%

 \SC{We consider a thin axisymmetric} annulus of radius $r_j$ with a \SC{vertical} outflow velocity $V_z$, rotation velocity $V_{\phi}$, and \SC{radial} expansion velocity $V_r$, observed \SC{at} an inclination angle $i$ with respect to the line of sight
 \SC{($i$ = 0 corresponds to a face-on ring)}. We show in Fig.~\ref{pv}(top) a schematic view of the annulus \SC{and the adopted coordinate system}. The projected position $r_{proj}(\phi)$ and \SC{projected} radial velocity $V_{rad}(\phi)$ of an elementary segment located at azimuthal angle $\phi$ are given by
\begin{eqnarray}
r_{proj}(\phi)  &=& \cos{\phi} ~ r_j, \\
V_{rad}(\phi) &=& -\left(\cos{i}~V_z +\sin{i}\sin{\phi}~V_r +\sin{i}\cos{\phi}V_{\phi}\right),
\end{eqnarray}
where we adopt the usual astrophysical convention of negative radial velocity for approaching material.
Denoting % $\tilde{r} = \frac{r}{r_j}$ 
the magnitude of the transverse velocity (in the plane perpendicular to the flow axis) 
as
  \begin{equation}
  V_{\perp} = \sqrt{V_{\phi}^2+V_r^2},
  \label{eq:Vperp}
  \end{equation}
and making the change of variable \SC{$V = {(V_{rad}+\cos{i}~V_z)}/\sin{i}$}, 
we \SC{obtain a} quadratic equation in $V$ and $r_{proj}$
\begin{equation}
\begin{split}
%\tilde{V}^2 + (1+\eta^2) \tilde{r}^2 - 2 \tilde{V}\tilde{r} = \eta^2,
\SC{V^2 + 2 V V_{\phi} \left( \frac{r_{proj}}{r_j} \right) + V_{\perp}^2 \left(\frac{r_{proj}}{r_j}\right)^2= V_r^2}.
\end{split}
\label{projectedposvel}
\end{equation}
\SC{This equation} defines an ellipse in the position-velocity diagram, centered on 
$r_{\rm proj}=0$ and $V_{\rm rad}= -\cos{i}V_z $. 
\SC{This ellipse is shown in Fig.~\ref{pv}(bottom) for our reference case where $V_z < 0$ (redshifted lobe of the jet).} Three pairs of point on this ellipse are of particular interest. 
\begin{itemize}
\item \SC{First, towards} the projected jet axis at $r_{\rm proj}=0$ (i.e. for $\phi = \pm \pi/2$), the line-of-sight component of the rotation velocity cancels out and $V_{rad}= -V_z \cos{i} \pm  V_r \sin{i}$ (green dots in Fig.~\ref{pv}). 
\item \SC{Second, towards the} maximum projected radius $r_{proj}=r_j$  where the line of sight is tangent to the annulus
(i.e. for $\phi = 0$ or $\phi=\pi$) the line-of-sight component of the expansion velocity $V_r$ vanishes and $V_{\rm rad}= -V_z \cos{i} \pm V_{\phi} \sin{i}$ (blue dots in Fig.~\ref{pv}). 
\item Third, and most importantly, the points where the ellipse reaches its minimum and maximum projected velocities (red dots in Fig.~\ref{pv}) are located at $r_{\rm proj}= \pm ({V_{\phi}}/{V_{\perp}}) r_j$ and $V_{\rm rad}= -V_z \cos{i} \pm V_{\perp} \sin{i}$.
\end{itemize}

\SC{In the following, we assume for simplicity an edge-on flow with $\sin{i} = 1$.}
\SCC{Denoting as $\bar{r_j}$, $\bar{V_r}$, and $\bar{V_\phi}$, the (dimensionless) numerical values of jet radius and velocities in the chosen units of 
the graph axes (eg. au and \kms),} and diagonalizing equation (\ref{projectedposvel}), we obtain that
the PV ellipse has a major axis $a$ and minor axis $b$ (in dimensionless graph units) given by:
\begin{eqnarray}
a^2 &=& \frac{K}{1-\sqrt{\Delta}}, \\
b^2 &=& \frac{K}{1+\sqrt{\Delta}}, {\rm with} \\
K &=& \frac{2 \bar{r_j}^2 \bar{V_r}^2}{\bar{r_j}^2+\bar{V_{\perp}}^2}, {\rm and}\\ 
\Delta &=& 1- \frac{2K}{\bar{r_j}^2+\bar{V_{\perp}}^2} =  \frac{ \left(\bar{r_j}^2-\bar{V_{\perp}}^2\right)^2 + 4\bar{r_j}^2 \bar{V_\phi}^2}{\left(\bar{r_j}^2+
\bar{V_{\perp}}^2\right)^2}
\label{velotoellipse}
\end{eqnarray}
The tilt angle $\Theta$ {of the ellipse} in the PV diagram, measured from the increasing $r_{proj}$ axis
towards the decreasing radial velocity axis (see Fig.~\ref{pv}), is such that
%\begin{eqnarray}
\begin{equation}
%\begin{split}
%\tan{(PA)} = \frac{1}{2} \left(\sqrt{\eta^2+4}-\eta^2\right) \\
% \left(V_{\perp}^2 - r_j^2 -\sqrt{(r_j^2+V_{\perp}^2)^2-4 r_j^2 V_r^2}-\right).
%\tan$\Theta$ = \frac{1}{2 r_j V_\phi} \left(r_j^2+\sqrt{(r_j^2+V_{\perp}^2)^2-4 r_j^2 V_r^2}-V_{\perp}^2 \right)\\
%\end{split}
\tan \Theta = \frac{\bar{r_j} \bar{V_\phi}} {\bar{r_j}^2-b^2}.
\label{eq:pa-ellipse}
\end{equation}
The general expressions for random inclinations can be recovered by multiplying $\bar{V_r}$, $\bar{V_\phi}$, and $\bar{V_\perp}$ by $\sin{i}$ in these expressions.
\SCC{The reverse relations allowing to calculate $r_j$, $V_r$, and $V_\phi$ from the ellipse parameters 
$a$,$b$, and $\Theta$ may be found in Appendix~B of \citet{2018A&A...618A.120L}, where $\Theta$ is denoted as $PA$.}

 %%%%%%%%%%%%%%%%%
 \begin{figure*}[!ht]
   \centering
  \includegraphics[width=0.45\textwidth]{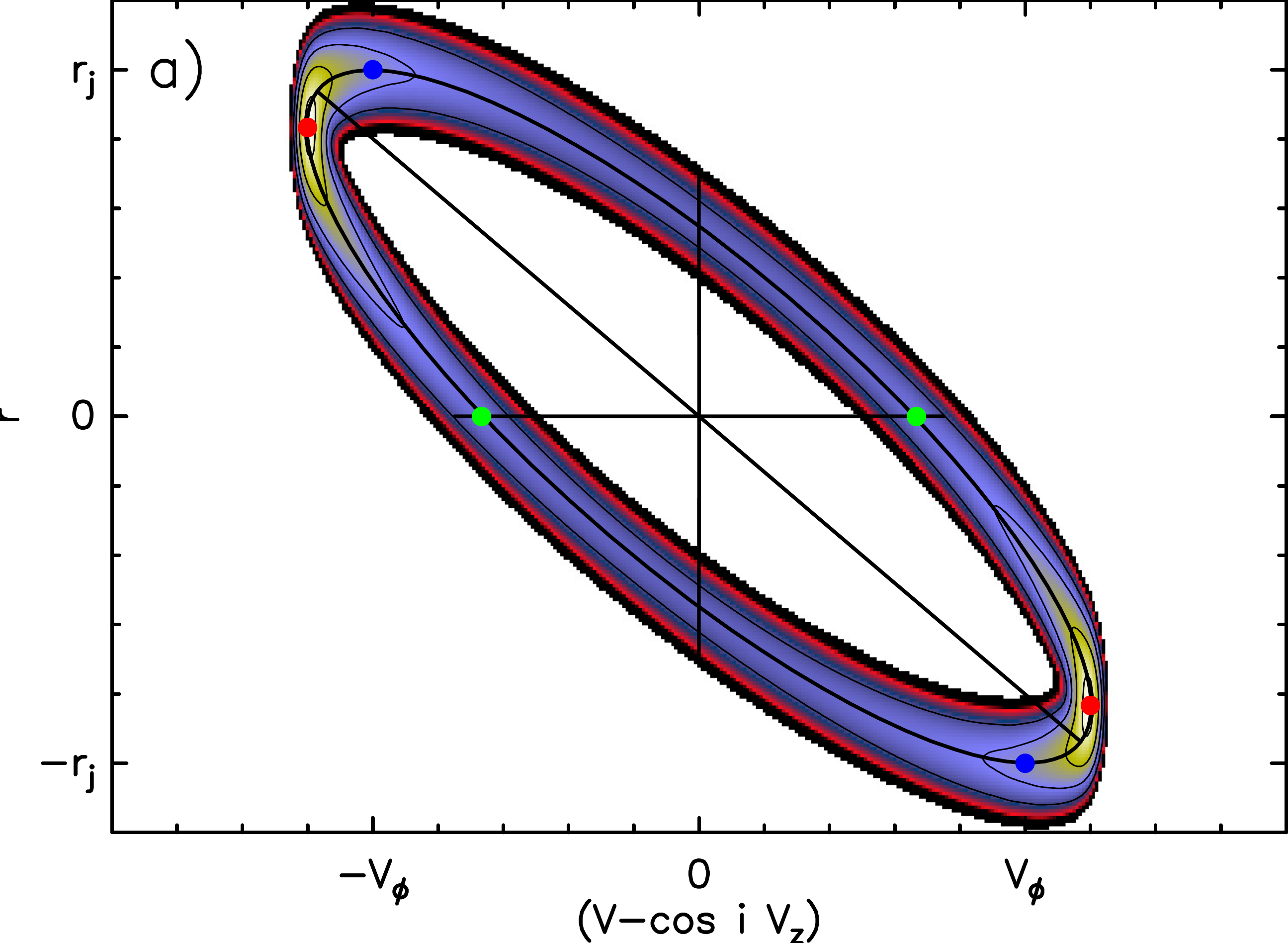} 
   \includegraphics[width=0.45\textwidth]{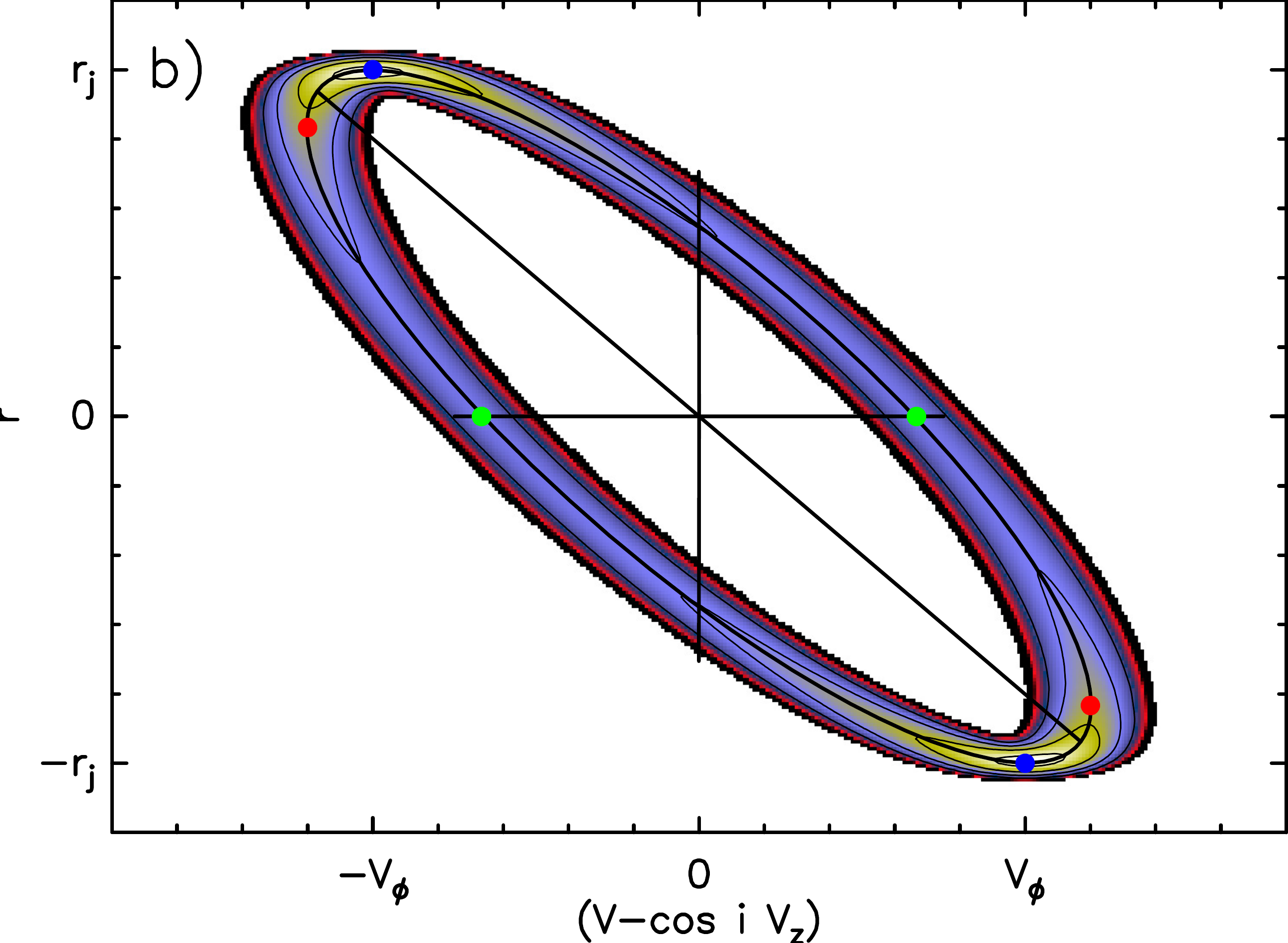}  \\
      \includegraphics[width=0.45\textwidth]{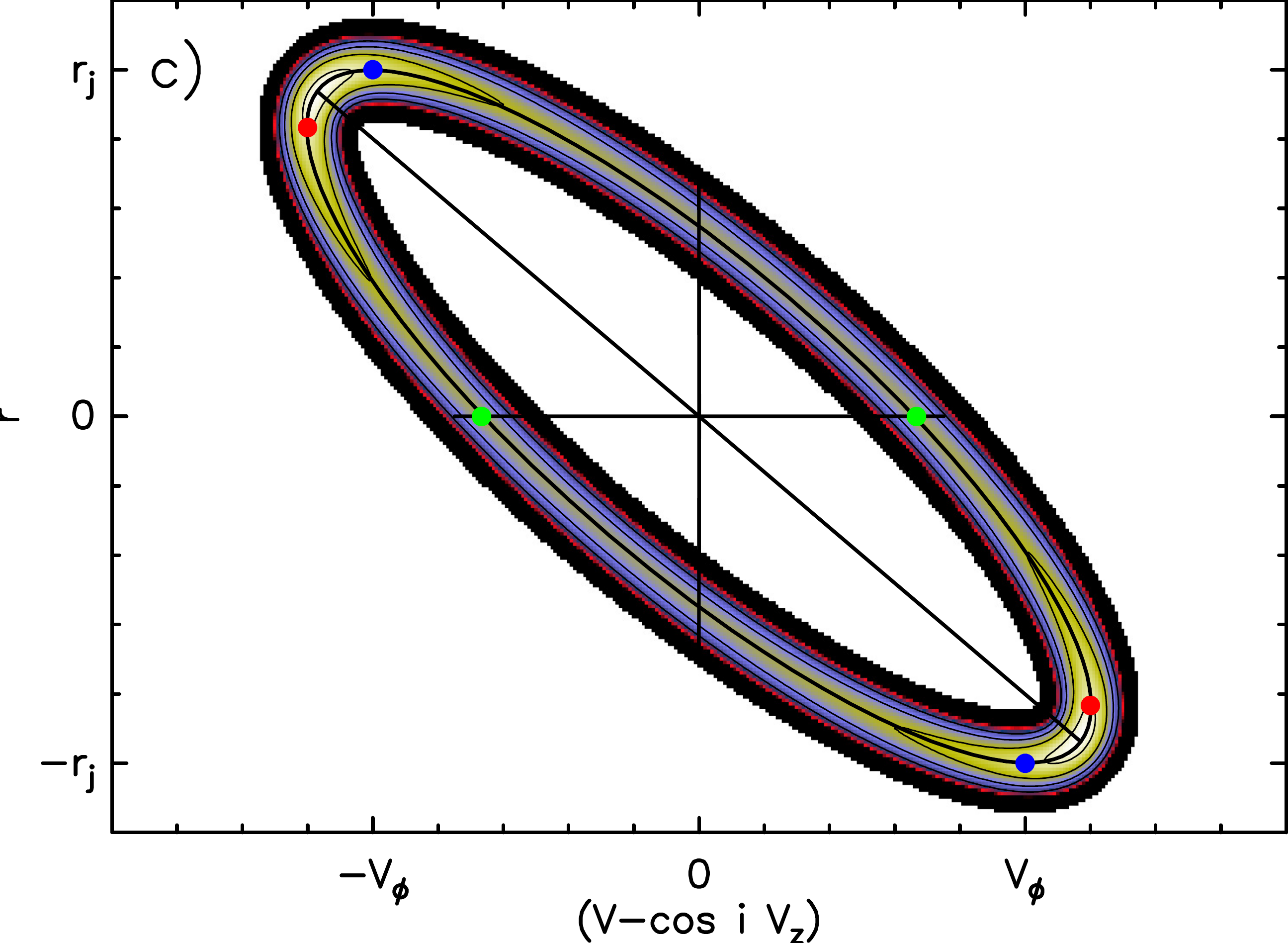} 
            \includegraphics[width=0.45\textwidth]{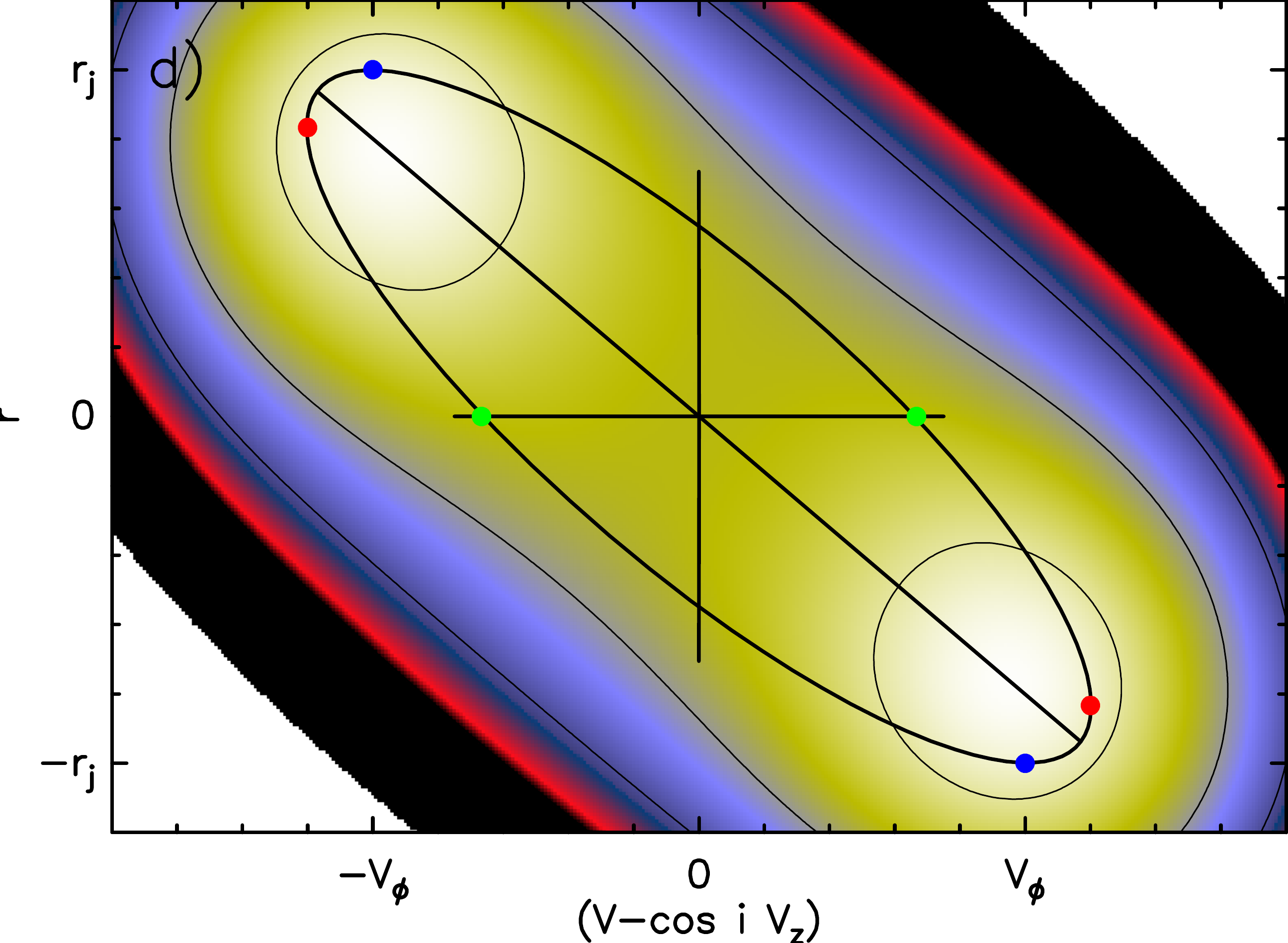}
      \caption{\SC{Effect of spatial beam ($\theta_b$) and spectral broadening ($\delta V$) on peak separation in} synthetic transverse PV diagrams for \SC{an expanding and rotating} annulus of radius $r_j$ and $V_{\phi}= 3 \times V_{r}$. \textbf{a)} significant beam smearing $\theta_b = 0.2 \times r_j$ but negligible line broadening $\delta V \ll V_{r}$: \SC{peaks on velocity extrema (red dots)}; this is the case expected for ALMA observations of cool molecular disk winds. \textbf{b)} significant line broadening $\delta V=0.2 \times V_{\phi}$ but negligible beam convolution $\theta_b \ll r_j$: \SC{peaks on  spatial extrema (blue dots)};  \textbf{c)}  high spatial and velocity resolution ($\theta_b=0.1 \times r_j$, $\delta V=0.1 \times V_{\phi}$): \SC{peaks between red and blue dots, close to the ellipse major axis.}; \textit{d)} low spatial and spectral resolution $\theta_b/r_j = \delta V/V_{\phi} = 1$: peaks move closer in, along the ellipse PA.}
         \label{pv-perp-model}
   \end{figure*}  
%%%%%%%%%%%%%%%%
 
\subsection{Impact of spatial and velocity smearing on emission peak positions} 
\label{ap:peaks}
% In the previous geometrical approach, velocities and positions of each elements of the annulus was projected in the PV diagram. 
\SC{We now examine the influence of spatial and velocity smearing on the separation of emission peaks in the elliptical PV of a single annulus,} assuming an 
optically thin line. 
We simply add up the emission contributions of \SC{each elementary arc $d\phi$} to the corresponding velocity/position bins in the PV diagram. \SC{The PV is then smoothed} by a gaussian beam of FWHM $= \theta_b$ and by a velocity \SC{broadening of} FWHM $= \delta V$. 
\SC{We find that the synthetic PV diagrams always present two symmetric peaks, whose} positions ($V_1, r_1$) and ($V_2,r_2$) depend on \gout{both $\theta_b$ and $\delta V$} \SC{the ratios $\theta_b/r_j$ and $\delta V/V_{\perp}$. 
Four cases are identified, which are displayed in Fig.~\ref{pv-perp-model}}:
\begin{itemize}
\item a) $\delta V/V_{\perp}<<\theta_b/r_j$ (Fig.~\ref{pv-perp-model}a): \SC{When the jet is better resolved spectrally than spatially (as usually the case for 
millimeter interferometric observations)} the intensity peaks are located {at the points of extremum velocities along the ellipse} (red dots in Fig.~\ref{pv-perp-model}): 
\begin{eqnarray}
r_{\{1,2\}} &=& \pm  r_j \left(\frac{V_{\phi}}{V_{\perp}}\right)\\
V_{\{1,2\}} &=& -\cos{i} V_z \pm \sin{i} V_{\perp}.
\end{eqnarray}
In that case, the spatial shift \SC{between emission peaks}, $\Delta r \equiv r_1 - r_2$ = $2  r_j ({V_{\phi}}/{V_{\perp}})$ \gout{gives only a lower limit on the actual ring radius} \SC{is smaller than the true ring diameter $2r_j$}, while the velocity shift \SC{between emission peaks} $\Delta V = 2 \sin{i}V_\perp$ \gout{gives an upper limit to $V_{\phi}$} \SC{is larger than $2V_{\phi}\sin{i}$}. Yet, the (inclination-corrected) specific angular momentum that one would estimate from the spatial and velocity separations of 
PV peaks, 
\begin{equation}
j_{\rm obs} = (\Delta r/2)(\Delta V/2\sin{i}),
\label{eq:jobs-appendix}
\end{equation} 
is still equal to the true specific angular momentum of the elementary rotating annulus, \jout\ = $r_j V_{\phi}$,
since the terms in $V_\perp$ cancel out in the product.

 \item b) $\theta_b/r_j << \delta V/V_{\perp}$ (Fig.~\ref{pv-perp-model}b): \gout{conversely, when a non-zero velocity dispersion is taken into  account, the beam size being small} \SC{Conversely, when the jet is better resolved spatially than spectrally} (as e.g. in optical/near-IR observations of the DG Tau jet), the peaks are located at the \textit{points of extremum radii} on the ellipse (blue dots in Fig.~\ref{pv-perp-model}) at
 \begin{eqnarray}
r_{\{1,2\}}& = &\pm r_j, \\
V_{\{1,2\}} &= & - \cos{i} V_z \pm \sin{i} V_{\phi}.
\end{eqnarray}
In that (simpler) case, the spatial shift  between peaks $\Delta r$ gives directly the true \SC{diameter} of the annulus, the velocity shift $\Delta V$ gives directly the true projected rotation velocity $2\sin{i} V_{\phi}$. Like in case a), the apparent (inclination corrected) specific angular momentum \jobs\ is equal to the true value in the ring.

\item c) $\theta_b/r_j \sim \delta V/V_{\perp}<<1$ (Fig.~\ref{pv-perp-model}c): \gout{when the two ratio are comparable and when the beam and the velocity dispersion are much smaller than the outflow width and the rotation velocity $V_{\phi}$,}\SC{when the jet is similarly well resolved spatially and spectrally, the two emission} peaks lie on the ellipse \SC{somewhere between the extremal velocity and radial points (ie. between the red and blue dots). In this case as well, \jobs\  will be close to the true \jout.}. 

\item d) $\theta_b/r_j \sim \delta V/V_{\perp}\sim 1$ (Fig.~\ref{pv-perp-model}d): \gout{when the beam size and the velocity dispersion are of the same order of magnitude as the radius $r_j$ and the rotation velocity} \SC{when the jet is under-resolved both spatially and spectrally, the two emission} peaks \gout{are located} \SC{migrate inwards,} \SC{roughly along the ellipse P.A. angle. In this last case, the apparent specific angular momentum \jobs\ will underestimate the true value in the ring}. 

\end{itemize}

In summary, \rev{as long as} the emitting ring is narrow and well resolved in \textit{at least} one dimension (spatial or spectral), and signal to noise is high enough to measure the centroid shift below the beam scale in the other dimension (using e.g. spectro-astrometric techniques or cross-correlation),  \jobs\ estimated from PV peak separations using Eq.~(\ref{eq:jobs-appendix}) gives a good estimate of the true specific angular momentum in the ring, \jout. 

However, this conclusion is only valid for a narrow emitting ring. For a radially extended disk wind, where we observe the summed contribution of a broad range of nested rings, \rev{the situation is more complex. When the flow is close to edge-on,}. we find that \jobs\ estimated from PV double peak separations always significantly underestimates \jout\ even at high spectral and angular resolutions (see Section~\ref{sec:jobs}). \rev{Below some critical inclination angle \icrit\ $\simeq \arctan \mid {V_\perp} \mid / V_\phi$ the PV is not systematically double-peaked anymore and other methods must be used to estimate the flow specific angular momentum (see Section~\ref{sec:icrit}).}
\end{appendix}
\end{document}